\documentclass[%
 aip,
 amsmath,amssymb,
 reprint,%
]{revtex4-1}

\usepackage{graphicx}
\usepackage{dcolumn}
\usepackage{bm}
\usepackage[mathlines]{lineno}
\usepackage[utf8]{inputenc}
\usepackage[T1]{fontenc}
\usepackage{mathptmx}
\usepackage{etoolbox}

\usepackage{xcolor}
\usepackage{subfigure}
\usepackage{amssymb}
\usepackage{tikz}
\usepackage{amsmath}
\usepackage{mathrsfs}
\usepackage{graphicx}
\usepackage{pifont}
\usepackage{subfigure}
\usepackage{mathtools}
\usepackage{mathcomp}
\usepackage{mathdots}
\usepackage{gensymb}

\usepackage[thmmarks,amsmath]{ntheorem}

\usepackage{lineno}

\usepackage{algorithm}
\usepackage{algpseudocode}
\usetikzlibrary{calc}

\newcommand{\cI}{{\cal I}}
\newcommand{\bV}{{\bf V}}

\newcommand{\enma}[1]   {\ensuremath{#1}}

\newcommand{\beq}{\begin{equation}}
\newcommand{\eeq}{\end{equation}}
\newcommand{\bseq}{\begin{subequations}}
\newcommand{\eseq}{\end{subequations}}
\newcommand{\beqn}{\begin{eqnarray}}
\newcommand{\eeqn}{\end{eqnarray}}
\newcommand{\ba}{\begin{array}}
\newcommand{\ea}{\end{array}}
\newcommand{\bct}{\begin{center}}
\newcommand{\ect}{\end{center}}
\newcommand{\btmz}{\begin{itemize}}
\newcommand{\etmz}{\end{itemize}}
\newcommand{\benum}{\begin{enumerate}}
\newcommand{\eenum}{\end{enumerate}}

\newcommand{\cH}{\enma{\mathcal H}}

\newcommand{\norm}[1]{\| #1 \|}                 

\newcommand{\trace}     {\enma{\mathrm{trace}}}

\newcommand{\bv}{{\bf v}}

\newcommand{\matbegin}{
        \left[
}
\newcommand{\matend}{
        \right]
}

\newcommand{\tbo}[2]{
  \matbegin \begin{array}{c}
       #1 \\ #2
       \end{array} \matend }

\newcommand{\tbt}[4]{
  \matbegin \begin{array}{cc}
       #1 & #2 \\ #3 & #4
       \end{array} \matend }

\newcommand{\be}{\begin{equation}}
\newcommand{\ee}{\end{equation}}

\newcommand{\cplxs}{ C\kern -.35em \rule{0.03 em}{.7 ex}~   }

\def\complex{\hbox{C\kern -.45em \rule{0.03 em}{1.5 ex}}~}

\newcommand{\bi}{\begin{itemize}}
\newcommand{\ei}{\end{itemize}}

\newcommand{\bu}{{\bf u}}

\newcommand{\bw}{{\bf w}}

\newcommand{\bbR}{\mathbb{R}}

\newcommand{\btab}{\begin{tabular}}
\newcommand{\etab}{\end{tabular}}
\newcommand{\bd}{{\bf d}}

\newcommand{\bx}{{\bf x}}

\newcommand{\non}{\nonumber}
\newcommand{\mrd}{\mathrm{d}}
\newcommand{\mre}{\mathrm{e}}

\newcommand{\ds}{\displaystyle}

\newcommand{\DefinedAs}[0]{\mathrel{\mathop:}=}
\newcommand{\vsp}{\vspace*{0.15cm}}

\DeclareMathOperator*{\logdet}{log\,det}
\DeclareMathOperator*{\minimize}{minimize}
\DeclareMathOperator*{\subject}{subject~to}

\definecolor{bgblue}{rgb}{0.04,0.19,0.53}
\definecolor{dblue1}{rgb}{0,0.3,0.7}
\definecolor{dred}{rgb}{0.4,0.2,0}

\makeatletter
\def\@email#1#2{%
 \endgroup
 \patchcmd{\titleblock@produce}
  {\frontmatter@RRAPformat}
  {\frontmatter@RRAPformat{\produce@RRAP{*#1\href{mailto:#2}{#2}}}\frontmatter@RRAPformat}
  {}{}
}%
\makeatletter
\def\@email#1#2{%
 \endgroup
 \patchcmd{\titleblock@produce}
  {\frontmatter@RRAPformat}
  {\frontmatter@RRAPformat{\produce@RRAP{*#1\href{mailto:#2}{#2}}}\frontmatter@RRAPformat}
  {}{}
}%
\makeatother
\begin{document}

\newtheorem*{remark}{Remark}

\preprint{AIP/123-QED}

\title{Short-term wind forecasting via surface pressure measurements: stochastic modeling and sensor placement}
\author{Seyedalireza Abootorabi}
 \affiliation{UTD Wind Energy Center and the Department of Mechanical Engineering,
University of Texas at Dallas, Richardson, TX, 75080, USA}
\author{Stefano Leonardi}%
\affiliation{UTD Wind Energy Center and the Department of Mechanical Engineering,
University of Texas at Dallas, Richardson, TX, 75080, USA}%

\author{Mario Rotea}
\affiliation{UTD Wind Energy Center and the Department of Mechanical Engineering,
University of Texas at Dallas, Richardson, TX, 75080, USA}

\author{Armin Zare}
\email{armin.zare@utdallas.edu}
\affiliation{UTD Wind Energy Center and the Department of Mechanical Engineering,
University of Texas at Dallas, Richardson, TX, 75080, USA}
 

\begin{abstract}
We propose a short-term wind forecasting framework for predicting real-time variations in atmospheric turbulence based on nacelle-mounted anemometer and ground-level air-pressure measurements. Our approach combines linear stochastic estimation and Kalman filtering algorithms to assimilate and process real-time field measurements with the predictions of a stochastic reduced-order model that is confined to a two-dimensional plane at the hub height of turbines. We bridge the vertical gap between the computational plane of the model at hub height and the measurement plane on the ground using a projection technique that allows us to infer the pressure in one plane from the other. Depending on the quality of this inference, we show that customized variants of the extended and ensemble Kalman filters can be tuned to balance estimation quality and computational speed 1--1.5 diameters ahead and behind leading turbines. In particular, we show how synchronizing the sign of estimates with that of velocity fluctuations recorded at the nacelle can significantly improve the ability to follow temporal variations upwind of the leading turbine. We also propose a convex optimization-based framework for selecting a subset of pressure sensors that achieve a desired level of accuracy relative to the optimal Kalman filter that uses all sensing capabilities.
\end{abstract}

\maketitle

\section{Introduction}
Adjustments to turbine blade pitch, generator toque, and nacelle direction (yaw) have shown promise in increasing energy production, lowering operational costs, and stabilizing the power distribution grid~\cite{aho12,fagmorrotste13,knubaksve15,ahmismshasulami20,shastagay22}. However, in practice, the effectiveness of feedback control strategies can be challenged by atmospheric variability ranging from meandering wind patterns to gusts and squalls resulting in achieved gains that fall short of expectations set by ideal scenarios that entail steady winds. In fact, in the absence of effective forecasting tools, almost all modern-day plants rely on data collected at or just behind the wind turbine to adjust their settings, and can consequently lag optimal operation conditions. This motivates the development of short-term forecasting tools for estimating changes in the speed of wind seconds to minutes ahead of impact, thereby enabling model-based control systems to preemptively adapt to improve turbine efficiency and reduce structural loads and failures~\cite{towjon16,doehoewin20,knubaksol11}.

In order to estimate changes in the incoming wind velocity, forecasting tools rely on the ability of models in capturing the evolution of dominant coherent motions that affect the performance and structural durability of wind turbines. While high-fidelity models, such as those that are based on large-eddy simulations 
(LES), have been shown to successfully capture the vast range of spatio-temporal scales in wind farm turbulence, their computational cost precludes their utility in developing forecasting tools that can inform real-time turbine control. To address this challenge, various studies have sought lower-fidelity alternatives by constraining the computational domain of LES ~\cite{solwisbra14,rotboevankuh17,boedoevalmeywin18}, employing the Reynolds-averaged Navier–Stokes (RANS) in conjunction with simple turbulence models~\cite{iunviocirleorot16,boedoevalmeyvan18,letiun22}, developing reduced-order models based on data-driven methods such as proper orthogonal decomposition~\cite{anngebsei16,hamvigcaltutcal18} and dynamic mode decomposition~\cite{iunsanabkporrotleo15}, and more recently, capturing the dynamics of velocity fluctuations around the time-averaged flow using the stochastically forced linearized Navier-Stokes (NS) equations~\cite{bharodberleozarEnergies23,rodburbhaleozarACC23}. Together with data-assimilation strategies such as Kalman filtering reduced-order models can adapt to atmospheric variations that affect the velocity field impinging a wind farm. Kalman filtering algorithms were initially used to improve meteorological models that provide long-term forecasts, i.e., predictions of daily-averaged wind speeds over wind plants and the resulting energy production~\cite{lou08,casbur12}. However, the focus of more recent studies has shifted toward short-term forecasting of the direction and speed of  wind to enable real-time turbine control (see, e.g., Ref.~[\onlinecite{doeboepaoknuvan18}]). 

\subsection{Real-time flow estimation for wind farms}

Wind velocity measurements from light detection and ranging (LiDAR) scanners, nacelle-mounted anemometers, and meteorological towers have provided invaluable insights into the dynamics of wind turbine wakes (see, e.g., Refs.~[\onlinecite{iunwupor13,iun16}]) in addition to opportunities for estimation and control. The early work of Schlipf et al. demonstrated the utility of LiDAR-based wind measurements in the context of preview control~\cite{schkuh08} and nonlinear model predictive control~\cite{schschkuh13}. 

Iungo et al.~\cite{iunsanabkporrotleo15} proposed Kalman filtering as a means to dynamically update predictions of wind turbine wakes obtained by a DMD-based reduced-order model that is trained using LES data. In this work, the authors used LES data throughout the entire 3D domain to correct model-based predictions at each time step. Sinner et al.~\cite{sinpaokin20} also adopted Kalman filtering to capture velocity variations at turbines' hub height based on noisy anemometer readings. In contrast to prior efforts, they proposed the use of a polynomial-based class of data-driven prior models that are trained to match the direction and magnitude of velocity at the hub-height of turbines across a wind farm.

Doekemeijer et al.~\cite{doevanboapao16} used ensemble and approximate Kalman filters to assimilate longitudinal and lateral wind speed measurements collected from a coarse-grained mesh with the predictions of a 2D LES-based model to estimate the mean centerline velocity impacting wind turbines.
Similar Kalman filters were later shown to estimate longitudinal and lateral velocity fields using only power production measurements~\cite{doeboepaowin17}. This estimation framework was further extended to incorporate both turbine power and velocity measurements in the context of extended, ensemble, and unscented Kalman filters that predict the velocity field at hub height~\cite{doeboepaoknuvan18}. The authors demonstrated improved estimation accuracy for scenarios in which a row of measurement sensors (which could be potentially provided by a LiDAR scanner) are placed downstream of wind turbines. The dynamic wake meandering model has also been used as a prior for Kalman filtering. In particular, Kalman filtering algorithms were used to assimilate power production~\cite{shastamengay19} and LiDAR~\cite{liolartho21} measurements with the predictions of this model.
While most studies assume precalculated power coefficients, Lio et al.~[\onlinecite{liolimen21}] estimated this parameter by incorporating measurements of the rotational speed of turbine rotors into their Kalman filtering algorithm.

\subsection{Motivation, contribution, and preview of results}

In this paper, we propose an estimation framework that utilizes an alternative, significantly cheaper, sensing technology relative to Doppler LiDAR. Our framework relies on the sequential self-correcting property of the Kalman filter in assimilating measurements from ground-level air-pressure sensors and nacelle-mounted anemometers with model-based predictions of hub-height velocity fluctuations. To facilitate real-time estimation, we limit the dimensional complexity of our models by confining their computational domains to the hub height of wind turbines. We bridge the vertical gap between the domain of our 2D models of hub-height velocity and the measurement plane on the ground by leveraging a data-driven projection scheme for inferring ground-level pressure from hub-height pressure.
Our contributions can be listed as follows:
\begin{itemize}
    \item Stochastic dynamical modeling of the fluctuating velocity field at the hub height of a wind farm;
    \item Devising customized linear and nonlinear Kalman filtering algorithms for assimilating ground-level pressure and nacelle velocity measurements with model-based prediction of the hub-height velocity field;
    \item Comparing the performance of various Kalman filters based on statistical and instantaneous error metrics;
    \item Proposing a sensor selection strategy for identifying subsets of ground-level pressure sensors that yield the least possible estimation error.
\end{itemize}

Our forecasting tool aims to detect atmospheric variations around the mean wind speed. We capture such variations using the linearized NS equations around analytical profiles of waked wind farm flows obtained from conventional engineering wake models. Stochastic processes are used to excite a statistical response from the linearized dynamics that is comparable with the result of high-fidelity LES. We first propose a white-in-time forcing model that is scaled to ensure the output velocity field matches the total kinetic energy of the flow. We then follow Bhatt et al.~\cite{bhazarACC22,bharodberleozarEnergies23} in adopting a convex optimization framework that shapes the spectral content of colored-in-time stochastic forcing to achieve second-order statistical consistency (i.e., matching normal and shear stresses) 
with LES.

\begin{figure}
\centering
\includegraphics[width=.32\textwidth]{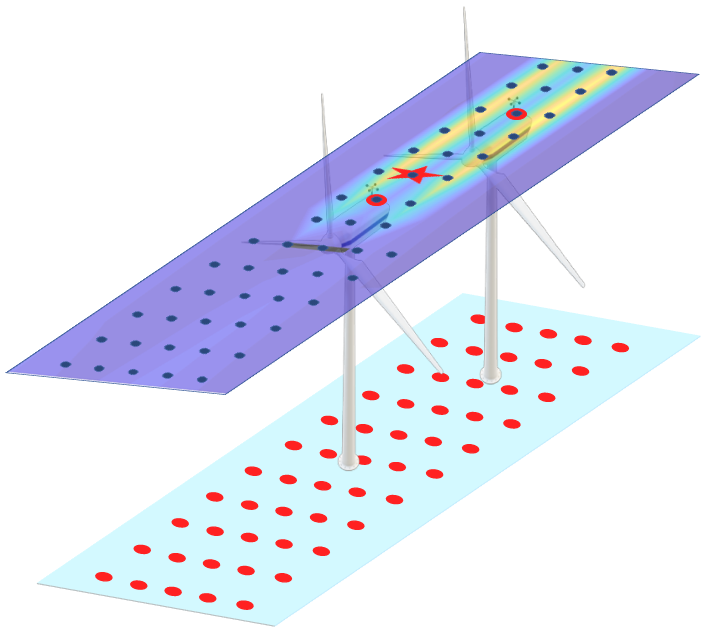}
\vspace{-.1cm}
\caption{Computational (top) and sensing (bottom) planes in our estimation framework. Blue and red dots mark the locations of training points for stochastic modeling and sensing, respectively, and the red star marks an estimation point in the wake of the leading turbine.}
\label{fig.hubheight_config}
\end{figure}

While the 2D stochastic model of hub-height velocity fluctuations can ensure statistical consistency with LES data, it does not warrant real-time tracking of wind variations. To address this issue, we invoke a Kalman filtering approach in correcting model-based predictions using information provided by ground-level pressure sensors that are distributed across the farm. To project quantities between the model's computational domain at hub height and pressure measurements on the ground, we utilize a data-driven projection scheme that relies on strong two-point correlations of the pressure field between these two planes (Fig.~\ref{fig.hubheight_config}). 
{We also show that a similar correlation-based heuristic allows us to synchronize the sign of streamwise velocity estimates ahead of the leading turbine with that of the streamwise velocity recorded by the nacelle-mounted anemometer.}

We tailor a range of linear and nonlinear Kalman filtering algorithms namely, the linearized Kalman filter (LKF), the extended Kalman filter (EKF), the ensemble Kalman filter (EnKF), and the unscented Kalman filter (UKF) to assimilate real-time pressure measurements with model predictions and consider two error metrics to compare their performance. One error metric quantifies the relative error in matching the variance of streamwise velocity at hub height with the result of LES and the other quantifies the normalized error in matching the instantaneous streamwise velocity. We demonstrate that besides the LKF, other variants achieved good accuracy in predicting the turbulent flow in the near wake region behind the first row of turbines, which provides approximately $50$ sec of preview time for the operator to adapt turbine settings ahead of variations in the direction and speed of incoming wind. This accuracy does, however, come at the cost of computational complexity for the ensemble and unscented Kalman filters. 

While our results provide a proof of concept for the deployment of inexpensive pressure sensors for wind monitoring across wind farms, structural constraints and practical issues with maintaining a large number of sensors necessitate a strategy for their optimal placement. To address this issue, we employ a sensor selection strategy to determine regions of the farm that are most important in achieving desirable estimation. This is based on a convex formulation of the optimal sensor selection problem that identifies subsets of sensors that strike a balance between estimation accuracy and sparsity of the sensing architecture. 

\subsection{Paper outline}
The remainder of the paper is organized as follows. In Sec.~\ref{sec.problem_formulation}, we formulate the problem and present our approach. In Sec.~\ref{sec.stochastic_dynamical_modeling}, we summarize our method in constructing a prior linear-time invariant (LTI) model of velocity fluctuations that is statistically consistent with high-fidelity LES. In Sec.~\ref{KF_algorithm}, we provide details of Kalman filtering algorithms including the measurement equation. In Sec.~\ref{sec.KF_comparison}, we compare the performance of various Kalman filters in estimating hub-height velocity variations based on ground-level air-pressure measurements. In Sec.~\ref{sec.sensel}, we present a convex optimization for selecting subsets of pressure sensors and examine the effect of sparsifying a uniform distribution of sensors on the performance of our forecasting tool. Finally, we conclude with a brief summary of results and future directions in Sec.~\ref{sec.concluding_remark}.

\section{Problem Formulation and approach}
\label{sec.problem_formulation}

The dynamics of the flow field impinging on a wind farm is given by the nonlinear NS and continuity equations
\be
    \label{eq.org-dyn}
    \ba{rcl}
    \partial_t\, \bu &\!\!=\!\!& f(\bu,P)
    \\[.15cm]
    0 &\!\!=\!\!& \nabla \cdot \bu
    \ea
\ee
where $t$ is time, $\bu(\bx,t)$ consists of the three components of the velocity field with $\bx = [\,x\,~y\,~z\,]^T$ denoting the vector of spatial coordinates in the streamwise ($x$), wall-normal ($y$), and spanwise ($z$) directions, $P$ is the pressure, and $\nabla$ is the gradient operator. Given a set of partially available steady-state correlations of the velocity field, we are interested in constructing an estimator that: (i) tracks velocity variations at the hub-height of wind turbines based on real-time measurements from ground-level air-pressure sensors and nacelle-mounted anemometers; and (ii) in the absence of filtering innovations, i.e., information on the difference between model predictions and current observations, provides statistical consistency with the original nonlinear model~\eqref{eq.org-dyn}. The first feature aims to support a recently proposed sensing technology~\cite{WindscapeAI} for short-term wind forecasting that is cheaper than Doppler LiDAR and the second is motivated by the desire to account for colored-in-time process noise that represents the intricacies of atmospheric turbulence beyond model predictions. To these ends, we propose a Kalman filter-based estimation framework that relies on two pillars:
\bi
    \item[$\bullet$] Stochastic dynamical modeling of the hub-height velocity field using the linearized NS equations;
    \item[$\bullet$] Data-driven inference of hub-height pressure from ground-level pressure measurements.
\ei
The first addresses the necessity for models of reduced complexity for short-term forecasting and is based on the predictive capability of the stochastically forced linearized NS equations around static solutions of engineering wake models~\cite{rodburbhaleozarACC23,bharodberleozarEnergies23}. 
The stochastic process noise that drives the linearized dynamic is identified via convex optimization to ensure second-order statistical consistency with high-fidelity LES of the wind farm flow~\cite{zarCDC21}. The second pillar is based on the construction of a data-driven kernel transfer function for projecting the hub-height pressure field to the ground (Fig.~\ref{fig.hubheight_config}). Building on these elements, we compare the performance of the LKF, EKF, EnKF, and UKF in providing a preview of the hub-height velocity field upwind of turbine rotors. 

\section{Stochastic dynamical wake modeling}
\label{sec.stochastic_dynamical_modeling}

In this section, we summarize key aspects of the 2D dynamical model we use for flow estimation. An in-depth presentation of the stochastically forced linearized NS equations for wind farm turbulence modeling is provided in Ref.~[\onlinecite{bharodberleozarEnergies23}]. Our modeling framework relies on the availability of a structured set of second-order statistics of wind farm flow together with an initial prediction of the mean wind velocity from a low-fidelity engineering wake model, which we cover next. 

\subsection{Base flow}
\label{sec.baseflow}

\begin{figure}
    \centering
    \begin{tabular}{cc}
    \begin{tabular}{c}
        \rotatebox{90}{$z$}
    \end{tabular}
    &
    \hspace{0.15cm}
    \begin{tabular}{c}
         \includegraphics[width=0.4\textwidth]{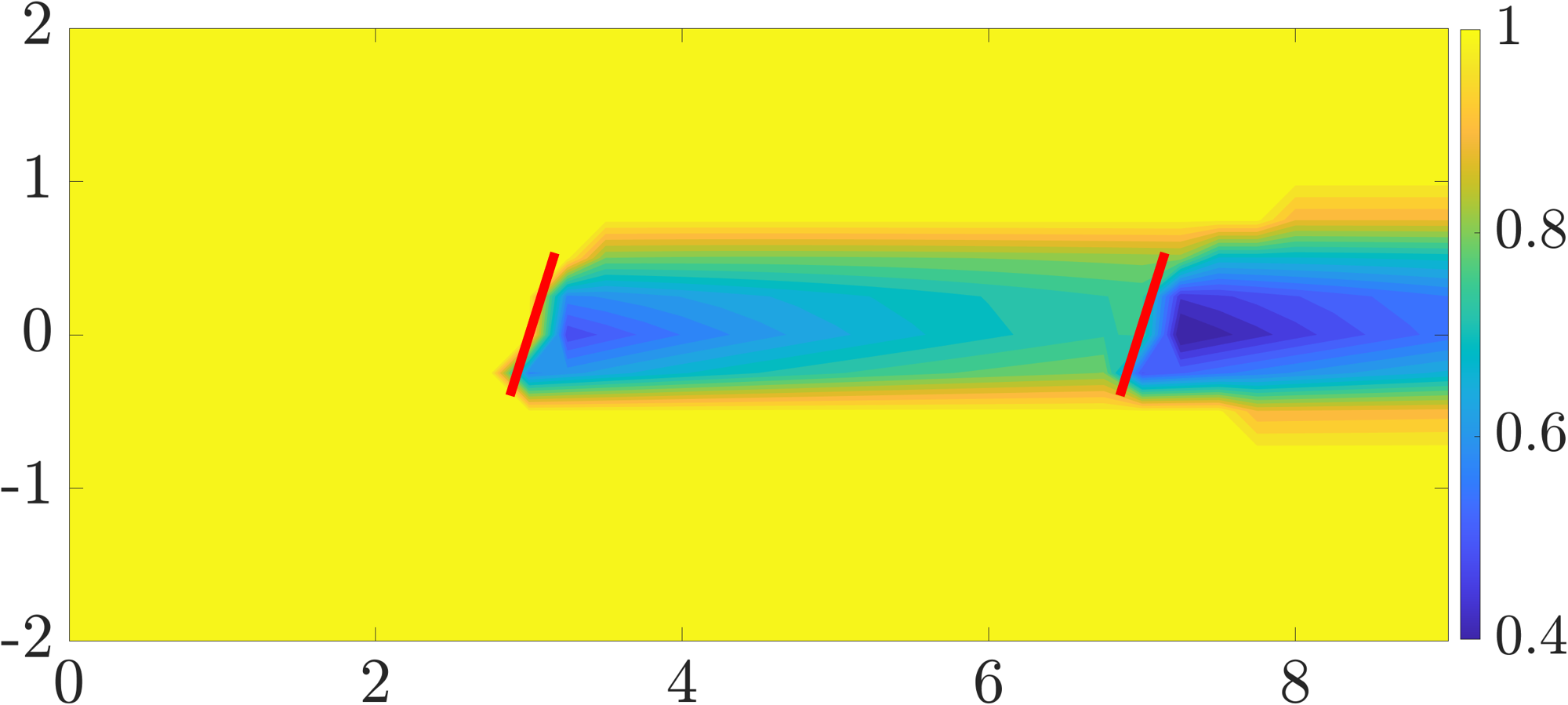}
    \end{tabular}
    \\[0cm]
       &
       {$x$}
       \end{tabular}
    \caption{Streamwise mean velocity provided by Eq.~\eqref{eq.base_flow} at the hub height of a wind farm with $\gamma=15^\degree$. The velocity is non-dimensionalized by the free-stream mean velocity. Thick red lines
        mark turbine rotors and the wind direction is from left to right.}
    \label{fig.MeanVelocity}
\end{figure}

\begin{figure}
\centering
\includegraphics[width=.43\textwidth]{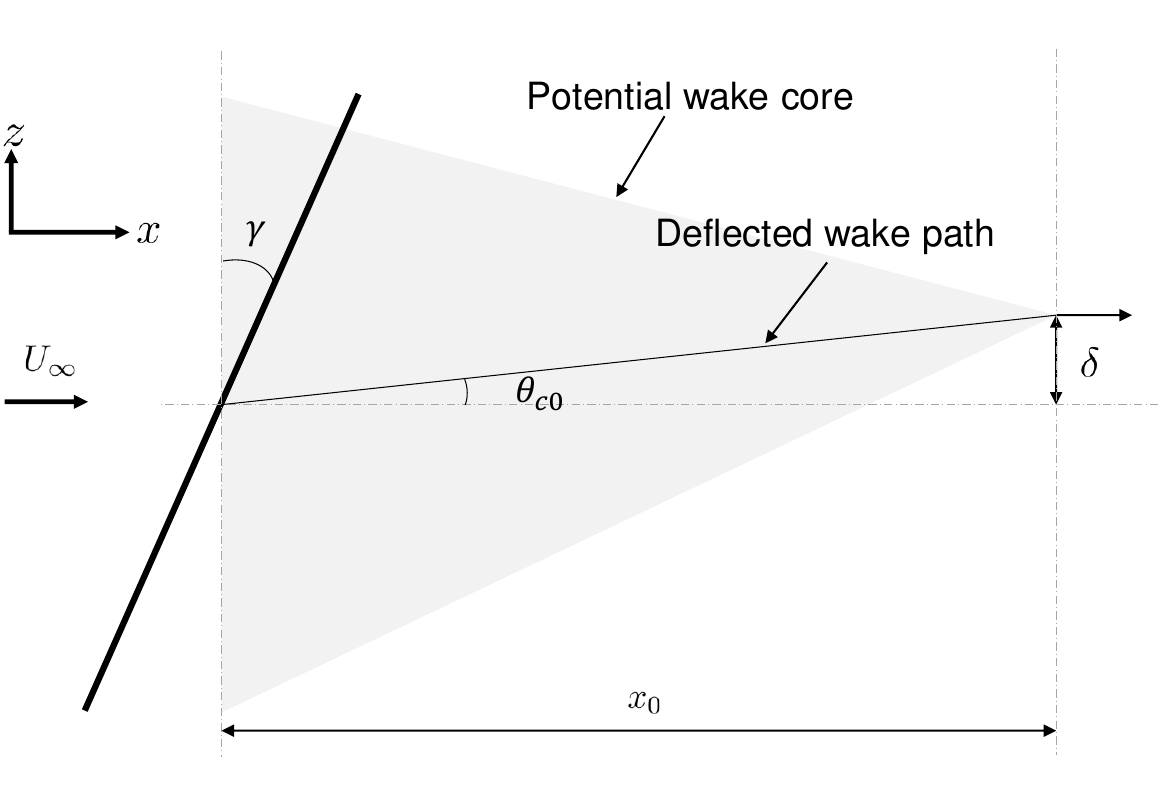}
\vspace{-.2cm}
\caption{A schematic of the potential wake core within the shaded triangular region for a yawed wind turbine.}
\label{fig.wake_core}
\end{figure}

Let the total wind velocity $\bu$ in the 2D horizontal plane at hub height be composed of a static base flow vector $\bar{\bu}$ and zero-mean fluctuations $\bv$, i.e., $\bu = \bar{\bu} \,+\, \bv$, where $\bv = [\,u\,~w\,]^T$ consist of the streamwise $u$ and spanwise $w$ velocity components. We consider the base flow to only consist of a streamwise component, i.e., $\bar{\bu}=[\,U\,~0\,]^T$, given by the Gaussian model of Bastankhah and Port\'e-Agel~\cite{baspor16},
\begin{align}
\non
    \hspace{-.18cm}
    U(x,z) ~=~\;
    &
    U_\infty \;-\; U_\infty\left(1 \,-\, \sqrt{1 \,-\, \dfrac{C_T \cos(\gamma)}{8 \left( \sigma_y\, \sigma_z \right)}}\, \right)
    \\[.15cm]
    \label{eq.base_flow}
    &
    \quad~~\times \exp\! \left( -0.5 \left[ ( \dfrac{y \,-\,y_h}{\sigma_y})^2  + ( \dfrac{z \,-\,\delta}{\sigma_z})^2 \right] \right)\!,\!
\end{align}
which describes the spatial variation of the waked velocity behind a yawed wind turbine. Here, all length scales have been non-dimensionalized by the rotor diameter $d_0$, $U_\infty$ is the free-stream velocity at hub-height, $C_T$ is the thrust coefficient, $\gamma$ is the yaw angle with respect to the free-stream, $y_h$ is the hub height, $\delta$ is the spanwise deflection of the wake centerline, and wake parameters $\sigma_y$ and $\sigma_z$ are given by
\begin{align*}
    \sigma_y\;=\;k\,(x\,-\,x_0) \,+\, \dfrac{1}{\sqrt{8}},
    \quad
    \sigma_z\;=\; k\,(x\,-\,x_0) \,+\,\dfrac{\cos(\gamma)}{\sqrt{8}},
\end{align*}
where $k$ is the wake growth rate and
\begin{align*}
    x_0
    \;\DefinedAs\;
    \dfrac{\cos(\gamma)\left(1\,+\,\sqrt{1-C_T}\right)}
    {\sqrt{2}\left(4\,\alpha\, I_{\mathrm{turb}} \,+\, \beta^\star \left(1-\sqrt{1-C_T}\right)\right)}
\end{align*}
is the wake core length (Fig.~\ref{fig.wake_core}). 
The core length is determined by constant parameters $\alpha$ and $\beta^\star$, which are chosen to be $0.58$ and $0.077$~\cite{baspor16}, and the ambient turbulence variance $I_{\mathrm{turb}}$, which we assume to be constant. The spanwise deflection $\delta$ is determined based on the distance from the rotor: in the near-wake region, i.e., $x\leq x_0$, $\delta=\theta_{c0}\,x$,  whereas in the far-wake region, i.e., $x>x_0$,
\begin{align*}
\delta \;=~&
    \theta_{c0}\,x_0
  \;+\;
  \dfrac{\theta_{c0}}{14.7}\sqrt{\frac{\cos(\gamma)}{k\, C_T}}\left(2.9+1.3\sqrt{1-C_T}-C_T\right)
  \\[.15cm]
  &\quad\quad\times\,
  \ln\left[\dfrac{\left(1.6 \,+\, \sqrt{C_T}\right)\left(1.6\,\sqrt{\dfrac{8\,\sigma_y\,\sigma_z}{\,\cos(\gamma)}} \,-\,\sqrt{C_T}\right)}{\left(1.6 \,-\, \sqrt{C_T}\right)\left(1.6\,\sqrt{\dfrac{8\,\sigma_y\,\sigma_z}{\,\cos(\gamma)}}\,+\,\sqrt{C_T}\right)}\right].
\end{align*}
Here, $\theta_{c0}$ is the deflection angle between the wake centerline and the free-stream, which is given by $\theta_{c0} \DefinedAs \dfrac{0.3\gamma}{\cos(\gamma)}\left(1-\sqrt{1-C_T\cos(\gamma)}\right)$. Finally, we use linear
superposition to capture velocity deficits in overlapping regions where the wakes of different turbines interact. Figure~\ref{fig.MeanVelocity} shows the colormap of the hub-height velocity field generated by this model for a cascade of two turbines that are uniformly yawed against the wind. While the predominant features of the waked velocity field are captured, the time-varying nature of the flow due to atmospheric turbulence is not. We next augment this static velocity profile by modeling the fluctuating velocity field that evolves around it.

\subsection{Stochastically forced linearized Navier Stokes equations}
\label{sec.Noisemodeling}

We assume the dynamics of velocity fluctuations $\bv$ around the base
flow profile $\bar{\bu}$ to be given by the stochastically forced linearized NS equations,
\begin{align}
\non
    \partial_t \bv
    &\;=\;
    -\left(\nabla \cdot \bv \right)\bar{\bu} \,-\, \left(\nabla \cdot \bar{\bu} \right)\bv \,-\, \nabla p \,+\, \dfrac{1}{Re} \Delta \bv \,-\, K^{-1}\bv \,+\, \bd
    \\
    \label{eq.lnse}
    0
    &\;=\;
    \nabla \cdot \bv, 
\end{align}
where $t$ is time, $p$ is the fluctuating component of the pressure field, $\nabla$ is the gradient operator, $\Delta=\nabla\cdot\nabla$ is the Laplacian, and $K(x,z)$ is a 2D permeability function, and $Re= U_\infty d_0/\nu$ is the Reynolds number, which follows the respective non-dimensionalization of length, velocity, time, and pressure by the rotor diameter $d_0$, free-stream velocity $U_\infty$, $d_0/U_\infty$, $\rho\, U_\infty^2$, where $\rho$ is the air density and $\nu$ is the kinematic viscosity. The permeability function $K(x,z)$ captures the effect of turbine rotors and nacelles (and even turbine towers in 3D models) by penalizing the velocity field at or around the coordinates of such solid structures; see Bhatt et al.\cite{bharodberleozarEnergies23} for details. In Eqs.~\eqref{eq.lnse}, $\bd$ is a zero-mean stationary stochastic input that triggers a statistical response from the linear dynamics.
A standard conversion for the elimination of pressure~\cite{schhen01} $p$ and finite-dimensional approximation of the differential operators brings the state-space representation for the linearized dynamics~\eqref{eq.lnse} into the evolution form
\begin{align}
\label{eq.lnse1}
    \dot{\bv}(t)
    \;=\;
    A\,\bv(t) \,+\, B\,{\bd}(t),
\end{align}
where  $\bv$ and $\bd$ are real-valued vectors of the 2D velocity fluctuations in the streamwise and spanwise directions and the stochastic input, respectively. Details on time-invariant matrices $A$ and $B$ can be found in Appendix A of Bhatt et al.\cite{bharodberleozarEnergies23}.

In model~\eqref{eq.lnse1}, the stochastic forcing term $\bd$ can be used to capture the effects of exogenous disturbances~\cite{jovbamJFM05,ranzarhacjovPRF19b,jovARFM20} or the nonlinear terms in the full NS equations~\cite{moajovJFM12,zarjovgeoJFM17,zargeojovARC20,ranzarjovJFM21,abozarJFM23}. Importantly, it has been shown that the performances of Kalman filters, including the ones we employ in the next section, are influenced by the statistics of the disturbance models we use to account for such sources of uncertainty~\cite{hoechebewhen05,chehoebewhen06}. We next offer two approaches for shaping the power spectrum of the additive stochastic forcing $\bd$ as either a white- or colored-in-time process. 
While these approaches vary in their levels of complexity and sophistication, they represent different viewpoints in turbulence modeling. Scaled white-in-time process noise, which results from simple scaling of simulation-based energy spectrum, allows one to match the steady-state kinetic energy as a scalar quantity resulting from integration over all degrees of freedom. On the other hand, colored-in-time process noise, which results from a more involved optimization procedure, allows one to match the spatial variation of steady-state velocity correlations, and thus, encompasses the capability of the white-noise.

\subsubsection{White-in-time process noise}
\label{sec.white-process-noise}
When the stochastic forcing $\bd$ in Eq.~\eqref{eq.lnse1} is zero-mean and white-in-time with covariance matrix $Q = Q^T \succ 0$, i.e., $\left<\bd(t)\bd^T(\tau)\right> = Q\,\delta(t - \tau)$, the steady-state covariance $\bV \DefinedAs \lim_{t \to \infty}\left<\bv(t)\bv^T(t)\right>$ can be found by solving the standard algebraic Lyapunov equation~\cite{kwasiv72}
\begin{align}
\label{eq.lyap}
    A \, \bV \, + \, \bV \, A^T \, =\, - \bar{Q}
\end{align}
where $\bar{Q} \DefinedAs B\,Q\,B^T$ and $\left<\cdot\right>$ is the temporal expectation operator. Following Ref.~[\onlinecite{moajovJFM12}], the covariance of white noise can be scaled to match the turbulent kinetic energy of the flow at hub height in accordance with high-fidelity LES via
\begin{align}
\label{eq.white_cov}
    \bar{Q} \; = \; \dfrac{\bar{E}}{\bar{E}_0} \, Q_0
\end{align}
where the scalar quantity $\bar{E}$ is obtained by integrating the LES-generated energy $E = \left<u^2\right> + \left<w^2\right>$ at hub height over the horizontal dimensions, i.e., $\bar{E}=\int E(x,z) \mrd x\, \mrd z$, and $\bar{E}_0$ is the integrated energy resulting from~\eqref{eq.lnse1} subject to a white-in-time forcing with covariance
\begin{align}
    Q_0 \; = \; \tbt{E(x,z) I}{0}{0}{E(x,z) I}.
\end{align}
Specifically, $\bar{E}_0 = \trace (\bV)$, where $\bV$ solves $A \bV + \bV  A^T = - Q_0$.

\begin{figure}
    \centering
    \begin{tabular}{cc}
    \begin{tabular}{c}
        \rotatebox{90}{$\lambda_i$}
    \end{tabular}
    &
    \hspace{0.15cm}
    \begin{tabular}{c}
         \includegraphics[width=0.3\textwidth]{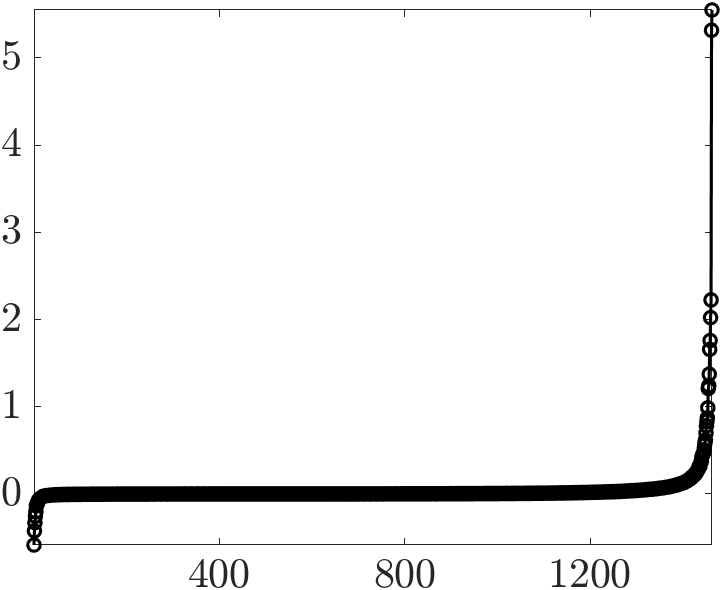}
    \end{tabular}
    \\[0cm]
       &
       {$x$}
       \end{tabular}
    \caption{Eigenvalues of the matrix $-(A\bV_{LES}+\bV_{LES}\, A^T)$, corresponding to a cascade of 2 non-yawed turbines that are separated by 4 diameters. The negative eigenvalues indicate that the turbulent velocity covariance cannot be reproduced by the linearized NS equations~\eqref{eq.lnse1} with white-in-time stochastic input $\bd$.}
    \label{fig.eigenvalue}
\end{figure}

\subsubsection{Colored-in-time process noise}
\label{sec.colored-process-noise}

The scaled white-in-time forcing can be used to match the turbulent kinetic energy using the linearized NS equations. However, this is not adequate in matching the statistical signature of the atmospheric boundary layer, namely, the spatial profile of $\left<u^2\right>$ and $\left<w^2\right>$.
Moreover, the sign indefiniteness of the matrix $-(A\bV_{\mathrm{LES}} + \bV_{\mathrm{LES}}\, A^T)$, where $A$ denotes the generator of the linearized dynamics~\eqref{eq.lnse1} and $\bV_{\mathrm{LES}}$ is the steady-state covariance matrix resulting from LES, further indicates that the second-order statistics of the turbulent flow at hub height cannot be reproduced by the linearized NS equations with white-in-time stochastic input. For a cascade of 2 non-yawed turbines, Fig.~\ref{fig.eigenvalue} demonstrates the coexistence of $694$ negative and $768$ positive eigenvalues in the eigenvalue spectrum of matrix $-(A\bV_{LES} + \bV_{LES}\, A^T)$. In fact, projection of this matrix onto the positive-definite cone results in a non-negligible $9.7\%$ relative error. Thus, we conclude that white-in-time process noise falls short in accurately reproducing the statistical signature of the wind velocity. We follow Bhatt et al.~\cite{bharodberleozarEnergies23} in utilizing the optimization-based framework of Refs.~[\onlinecite{zarchejovgeoTAC17,zarjovgeoJFM17,zargeojovARC20}] to identify the statistics of the colored-in-time input that ensure certain, more dominant, second-order statistics of the velocity field are matched. 

Access to wind speed data in the wind energy industry has grown remarkably in the past decades and such statistics could be computed from the result of high-fidelity simulations or field measurements. Herein, we specifically assume knowledge of a subset of second-order statistics of hub-height velocity upwind of the wind farm and in the wake region up to 4 diameters behind the turbines (Fig.~\ref{fig.hubheight_config}). These velocity correlations correspond to entries on the diagonal of the steady-state covariance matrix ${\bV} \DefinedAs \lim_{t \to \infty}\left<\bv(t)\bv^T(t)\right>$, which solves the Lyapunov-like equation~\cite{geo02a,zarchejovgeoTAC17},
\begin{align}
\label{eq.lyap-like}
	A\,\bV \,+\, \bV A^* ~=~ -B\, H^* \,-\, H\,B^*
\end{align}
where the matrix $H$ quantifies the cross-correlation between the input and the state (see Appendix B in Ref.~[\onlinecite{zarjovgeoJFM17}]), i.e., $H \DefinedAs \lim\limits_{t \to \infty} \left< \bv(t) \bd^*(t)\right> + B \Omega/2$. Coincidentally, it is this matrix that contains information on the coloring filter for generating the stochastic input $\bd$~\cite{zarchejovgeoTAC17}. 

Matrix $H$ and the input matrix $B$ can be obtained from the solution of the covariance completion problem
\begin{align}
	\hspace{-.05cm}
	\ba{cl}
	\minimize\limits_{\bV, \, Z}
	& 
	-\logdet\left(\bV \right) \,+\, \alpha\,\norm{Z}_*
	\\[.25cm]
	\subject 
	&
	~A \, \bV \,+\, \bV A^* \,+\, Z  \;=\; 0
	\\[.1cm]
	&
	\,\,\bV_{i,j} \;=\; G_{i,j}, \quad \forall~\{i,j\} \in \cI.
	 \ea
	\label{eq.CC}
\end{align}
This convex optimization problem solves for Hermitian matrices $\bV$ and $Z$ subject to two linear constraints that ensure consistency with the assumed linear model via satisfaction of the Lyapunov-like equation~\eqref{eq.lyap-like} and the partially known velocity correlations; entries of $G$ corresponding to the set of indices $\cI$ represent partially available second-order statistics of the output $\bv$. The objective function provides a weighted trade-off between the solution to a maximum-entropy problem and a nuclear norm regularizer through determination of $\alpha>0$. The logarithmic barrier function in the first part of the objective function ensures positive definiteness of matrix $\bV$ and the nuclear norm is used as a convex proxy for the rank function~\cite{faz02,recfazpar10}. It is desirable to regulate the rank of matrix $Z$ as it bounds the number of independent input channels or columns in matrix $B$. Without this regularization, a full-rank matrix $B$ permits colored-in-time input $\bd$ to excite all degrees of freedom and completely overwrite the linearized dynamics $A$~\cite{zarchejovgeoTAC17}. 

\begin{figure}
\centering
\includegraphics[width=.47\textwidth]{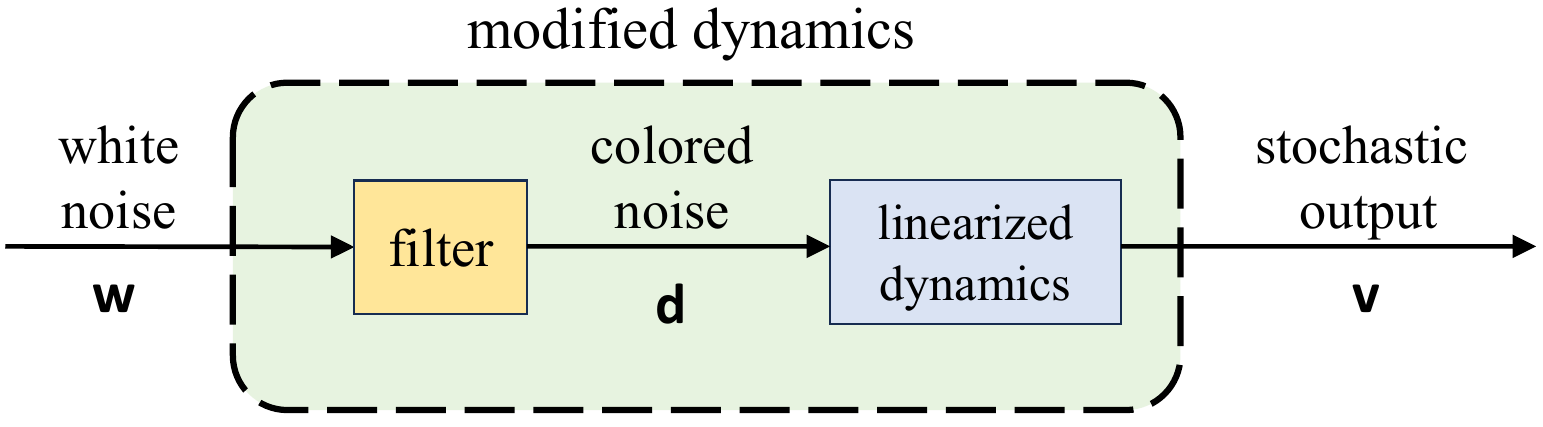}
\caption{Cascade connection of the linearized flow dynamics and a linear coloring filter that is used to match partially available second-order statistics of the output velocity $\bv$.}
\label{fig.modified-dynamics}
\end{figure}

\begin{figure*}
\begin{center}
        \begin{tabular}{cccc}
        \hspace{-1.3cm}
        \subfigure[]{\label{fig.uuLESmap}}
        &&
        \hspace{0.45cm}
        \subfigure[]{\label{fig.uuLNSmap}}
        &
        \\[-.45cm]
        \hspace{-0.9cm}
	\begin{tabular}{c}
        \vspace{.5cm}
        \rotatebox{90}{$z$}
       \end{tabular}
       &
       \hspace{-0.2cm}
	    \begin{tabular}{c}
\includegraphics[width=0.35\textwidth]{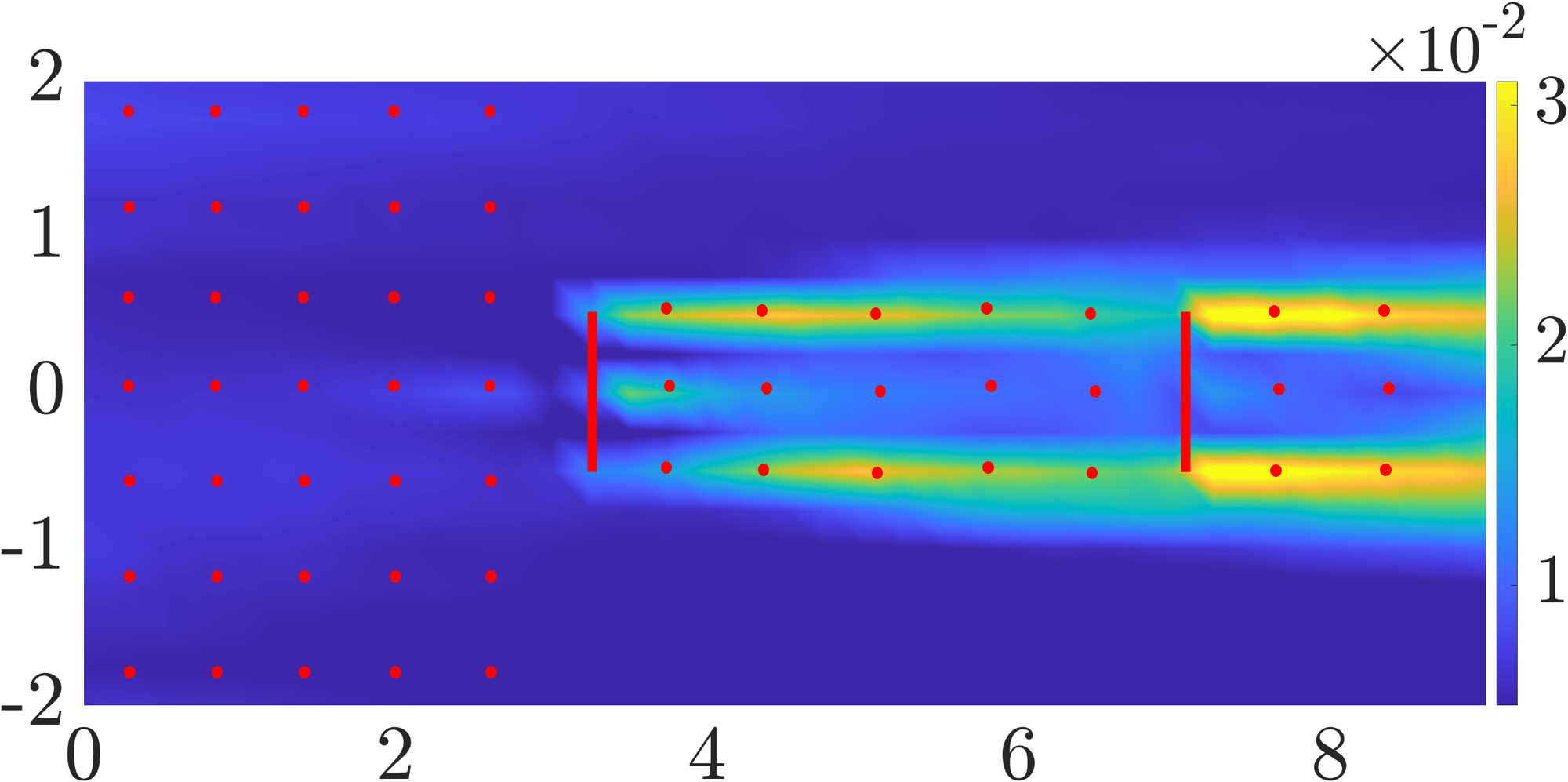}
       \end{tabular}
       &&
       \hspace{.2cm}
        \begin{tabular}{c}
       \includegraphics[width=0.35\textwidth]{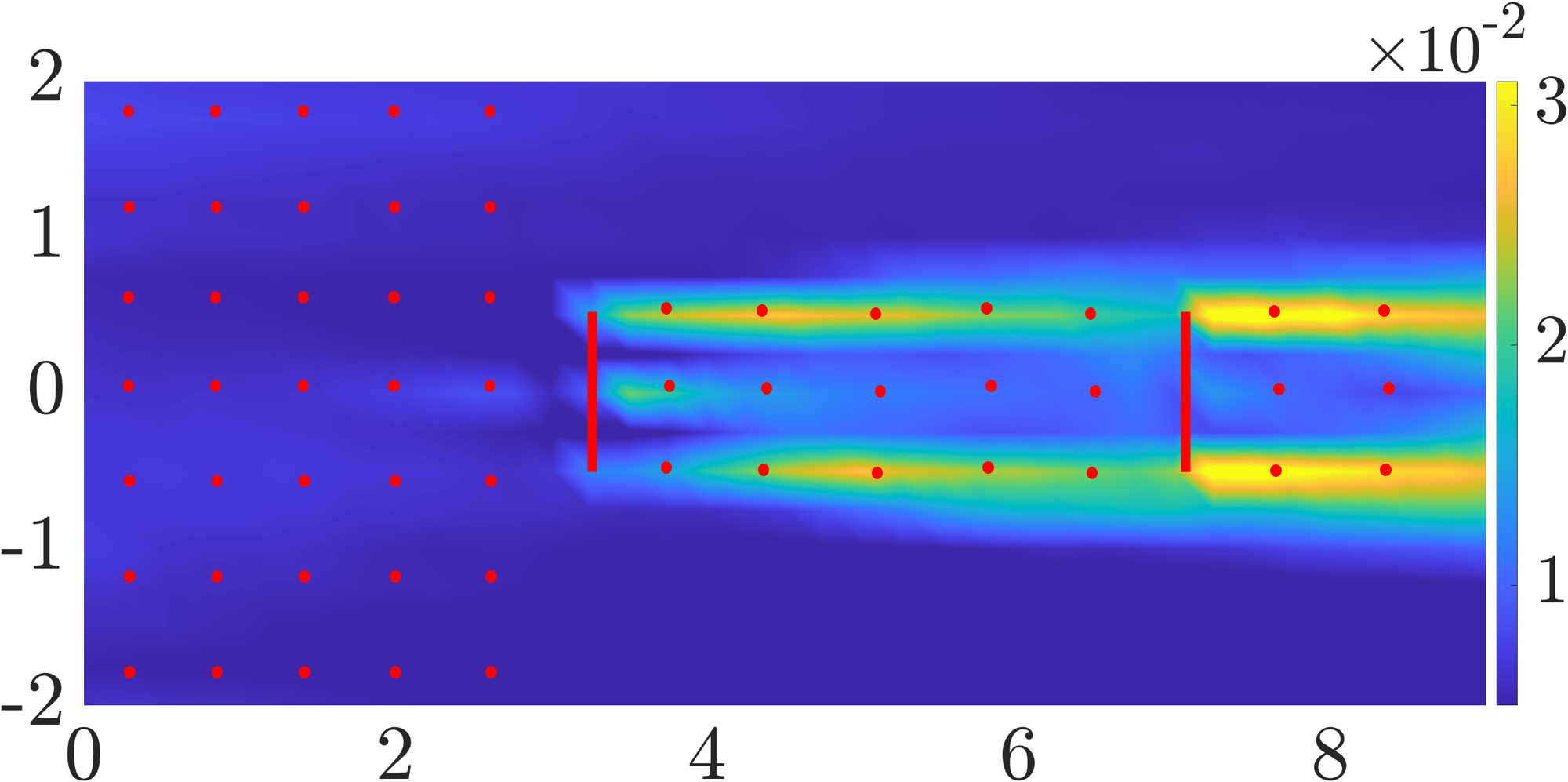}
       \end{tabular}
       \end{tabular}
       \\[-0.1cm]
       \begin{tabular}{cccc}
        \hspace{-1.3cm}
        \subfigure[]{\label{fig.wwLESmap}}
        &&
        \hspace{0.45cm}
        \subfigure[]{\label{fig.wwLNSmap}}
        &
        \\[-.45cm]
        \hspace{-0.9cm}
	\begin{tabular}{c}
        \vspace{.5cm}
        \rotatebox{90}{$z$}
       \end{tabular}
       &
       \hspace{-0.2cm}
	    \begin{tabular}{c}
       \includegraphics[width=0.35\textwidth]{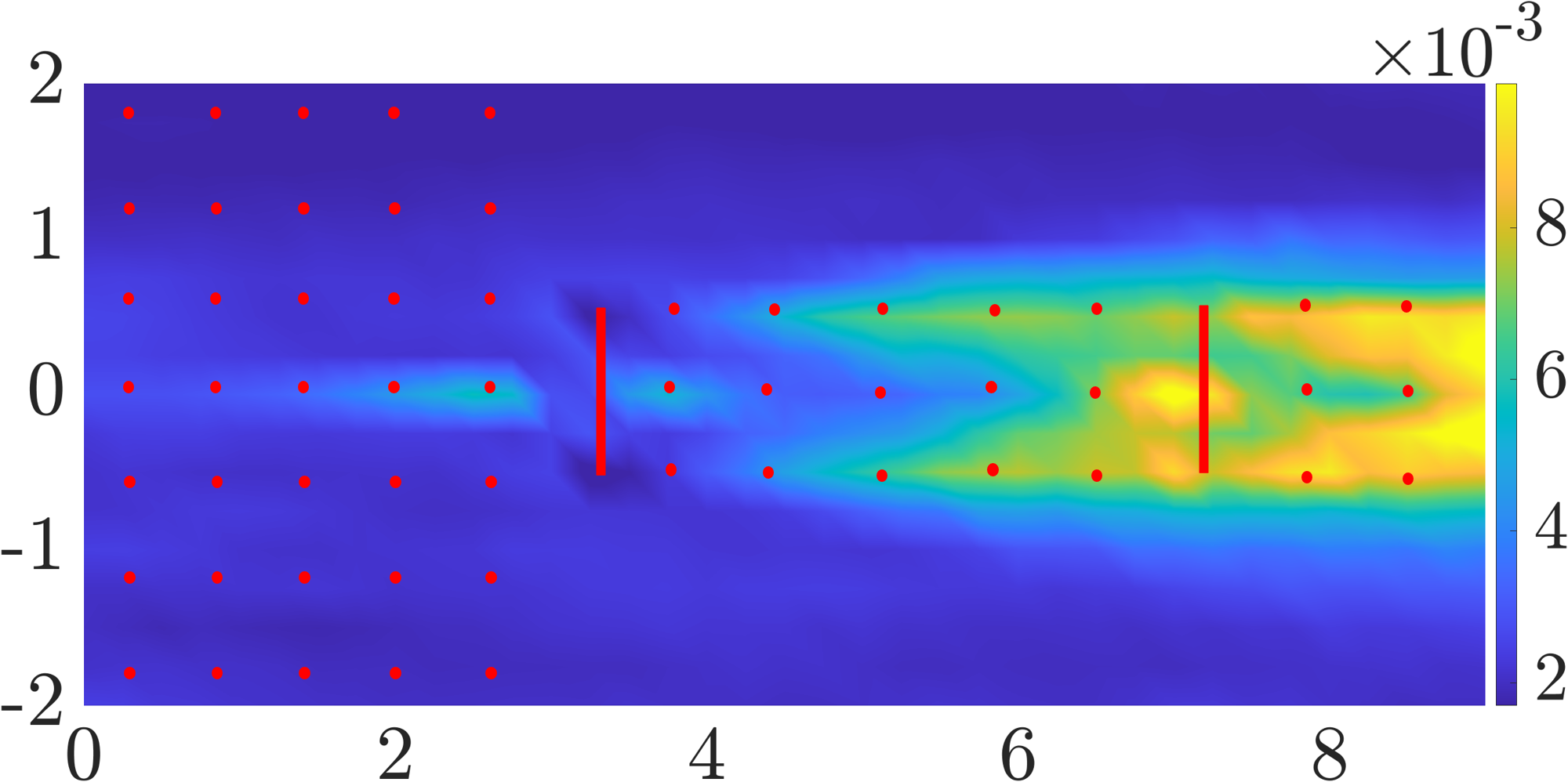}
       \\[-.1cm]
       \hspace{0.4cm}
       {$x$}
       \end{tabular}
       &&
       \hspace{0.2cm}
        \begin{tabular}{c}
       \includegraphics[width=0.35\textwidth]{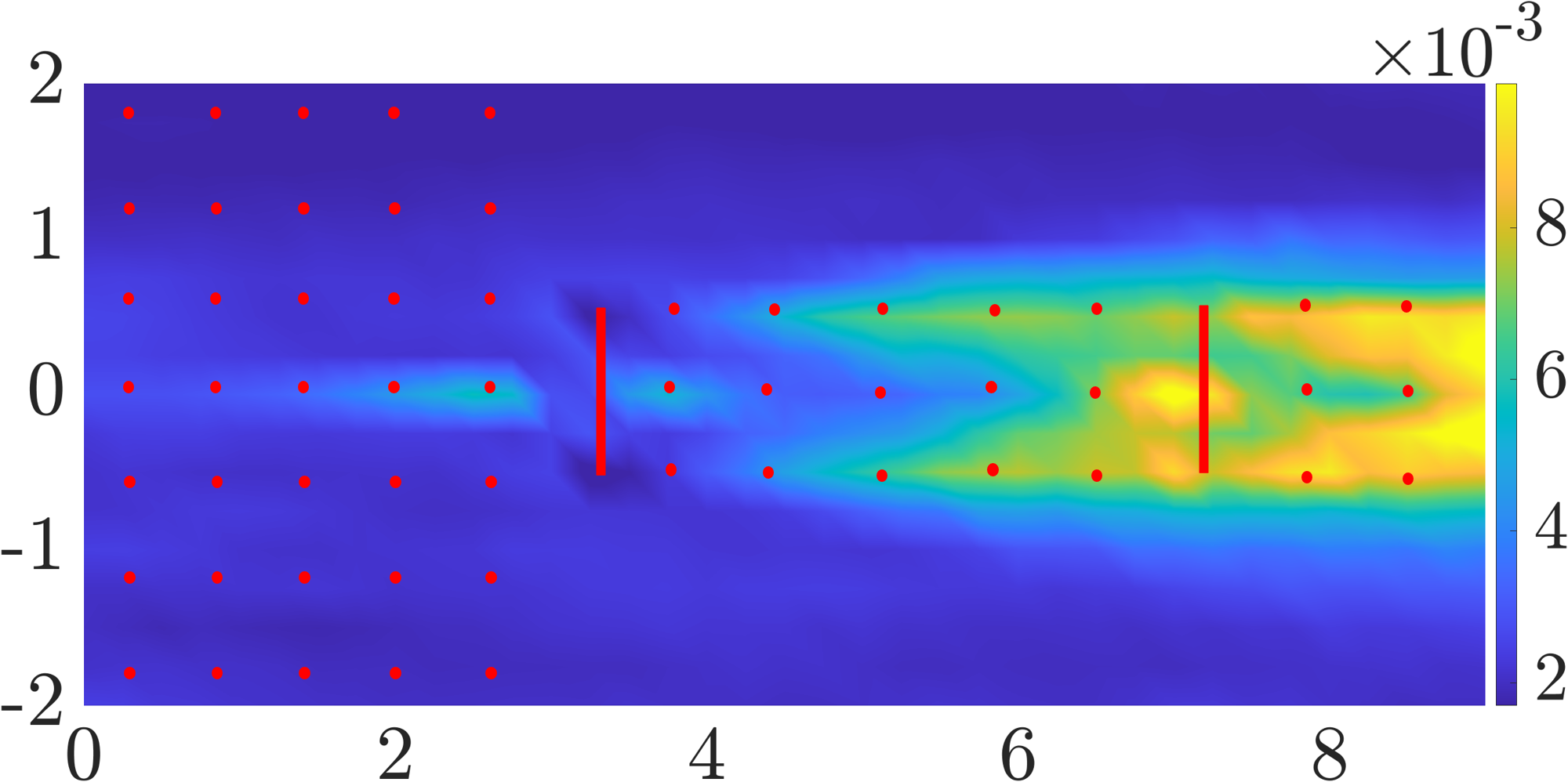}
       \\[-.1cm]
       \hspace{0.4cm}
       {$x$}
       \end{tabular}
       \end{tabular}
        \caption{Streamwise (top) and spanwise (bottom) velocity variances computed from (a,c) LES and (b,d) model~\eqref{eq.modified-dyn}. Red dots denote the spatial location of velocity correlations used in training the stochastic model and thick red lines mark turbine rotors. The wind direction is from left to right.}
        \label{fig.LES_LNS_comp}
\end{center}
\end{figure*}

The solution $Z$ to problem~\eqref{eq.CC} can be decomposed into matrices $B$ and $H$ (cf.~Eq.~\eqref{eq.lyap-like}) via spectral decomposition techniques (Appendix~\ref{sec.appendix1}), which in turn can be used to construct linear coloring filters that realize the input signal $\bd$ (Fig.~\ref{fig.modified-dynamics}). Alternatively, the coloring filter can be absorbed in the LTI dynamics~\eqref{eq.lnse1} in a standard manner yielding the dynamically modified state-space representation
\begin{align}
\label{eq.modified-dyn}
    \dot{\bv}(t)
    \;=\;
    A_f\,\bv(t) \,+\, B\,{\bw}(t).
\end{align}
Here, $A_f \DefinedAs  A - B K_f$, $\bw$ is white noise, and $K_f$ is a parameter of the coloring filter with parameterization offered in Sec.~II.B of Ref.~[\onlinecite{zarchejovgeoTAC17}]. Additional details on the choice of velocity correlations for best recovery, and the robustness of predictions to turbine yawing effects can be found in Refs.~[\onlinecite{bharodberleozarEnergies23,rodburbhaleozarACC23}]. 
 
Figure~\ref{fig.LES_LNS_comp} demonstrates the performance of model~\eqref{eq.modified-dyn} in recovering the streamwise and spanwise velocity variances at the hub-height of a turbine cascade. In this figure, the red dots indicate the locations of the available velocity correlations in the training dataset. In the absence of innovations, the identified colored process noise model ensures statistical consistency between the linear dynamics~\eqref{eq.modified-dyn} and a high-fidelity LES with a rotating actuator disk model~\cite{sancirrotleo15}. The close agreement of the predictions of our model and the result of LES in regions beyond the training dataset warrant its use for wind estimation.

In addition to their differences in complexity and sophistication, as we demonstrate in the next section, the white and colored process noise models also yield quite different estimates of the hub-height wind, where filtering results obtained using the latter are notably more accurate. While prior studies have modeled turbulence using both white-~\cite{moajovJFM12,ranzarjovJFM21} and colored-in-time~\cite{zarchejovgeoTAC17,zarjovgeoJFM17,zargeojovARC20,abozarJFM23} processes, the choice of process noise for any specific application should contemplate a trade-off between accuracy and computational complexity.

\section{Flow estimation via ground pressure and nacelle velocity measurements}
\label{KF_algorithm}

\begin{figure}
\centering
\includegraphics[width=.47\textwidth]{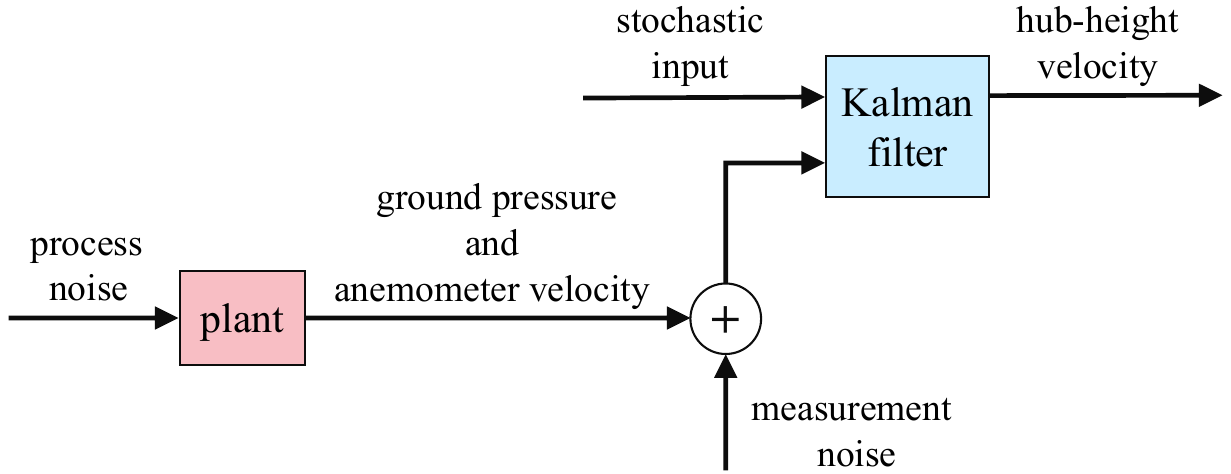}
\vspace{-.1cm}
\caption{Block diagram illustrating our data-assimilation strategy in updating model-based predictions using real-time measurements.}
\label{fig.model_KF_blockdiagram}
\end{figure}

We use Kalman filters to update our model-based predictions of the hub-height velocity field in accordance with real-time readings of ground-level pressure sensors and nacelle-mounted anemometers that are affected by atmospheric variations and can be contaminated by noise (Fig.~\ref{fig.model_KF_blockdiagram}). In addition to noisy measurements that inform innovations in the Kalman filtering algorithm, a stochastic input is used to induce a statistical response from our prior model, i.e., the linearized NS dynamics~\eqref{eq.lnse1} or the modified dynamics~\eqref{eq.modified-dyn}. As mentioned in Sec.~\ref{sec.Noisemodeling}, the spectral content of the stochastic inputs are
shaped to achieve consistency with high-fidelity simulations in matching statistical features of hub-height turbulence. While turbulence modeling is conducted in continuous time, we implement discrete-time analogues of conventional linear and nonlinear Kalman filters, namely, the LKF, the EKF, the EnKF, and the UKF, in our numerical experiments. These filters differ in handling the nonlinear measurement equation
\begin{align}
\label{eq.state-output-approx-filter}
    \varphi(t)
    \;\DefinedAs\;
    \tbo{p_\mathrm{g}(t)}{\bv_\mathrm{nacelle}(t)}
    \;=\;
    \tbo{\cH_\mathrm{gh}\, {\cal P} (\bv(t))}{E\, \bv(t)} 
    \;=:\;
    {\cal C} (\bv (t)),
\end{align}
which relates the hub-height velocity fluctuations $\bv$ from model~\eqref{eq.modified-dyn} to ground pressure $p_\mathrm{g}$. This is done by first solving for the hub-height pressure $p_\mathrm{h}$ using the pressure Poisson equation,
\begin{align}
    \non
    p 
    \;=&\, 
    {\cal P}(\bv) \;=\;
    -\Delta^{-1} \!\left[ \left(\dfrac{\partial u}{\partial x}\right)^2\! +\; 2\, \dfrac{\partial u}{\partial z}\, \dfrac{\partial w}{\partial x} \,+\, \left(\dfrac{\partial w}{\partial z}\right)^2 \!
    \right.
    \\
    \label{eq.Poisson}
    & \left. \hspace{2.6cm}
    +\; 2\, \dfrac{\partial U}{\partial x}\, \dfrac{\partial u}{\partial x}
    \;+\; 2\, \dfrac{\partial U}{\partial z}\, \dfrac{\partial w}{\partial x}
    \right]
\end{align}
followed by a projection to the ground using the transfer kernel $\cH_{\mathrm{gh}}$, i.e., $p_\mathrm{g} = \cH_{\mathrm{gh}} p_\mathrm{h}$.  
In Eq.~\eqref{eq.state-output-approx-filter}, $E$ is a matrix with the same number of rows as the number of nacelle-mounted anemometers with unit rows that each have $1$ in the entry corresponding to the spatial location of an anemometer and zeros in all other entries. In what follows, we provide details of the Kalman filtering algorithms we use to estimate hub-height velocity as well as the transfer kernel $\cH_{\mathrm{gh}}$.

\subsection{Kalman filtering algorithms}
\label{sec.KF-slgorithms}

The discretized LKF uses the linearization of the pressure Poisson equation
such that at each iteration $k$ an estimate of the pressure on the ground $\hat{p}_{\mathrm{g},k}$ is computed from the estimated velocity at hub-height $\hat{\bv}_k = [\,\hat{u}_k\,~ \hat{w}_k\,]^T$ via
\begin{align*}
    \hat{p}_{\mathrm{g},k} 
        \;=\;
        -2\,\cH_\mathrm{gh}\, \Delta^{-1} \left( \dfrac{\partial U}{\partial x}\dfrac{\partial \hat{u}_k}{\partial x} \,+\, \dfrac{\partial U}{\partial z}\dfrac{\partial \hat{w}_k}{\partial x}\right)
\end{align*}
and follows the standard form of the optimal state estimator; see Sec.~5.1 of Ref.~[\onlinecite{sim06}].
The estimate of the output for the LKF is, therefore, given by $\hat{\varphi}_k = C\, \hat{\bv}_k$, with
\begin{align}
\label{eq.C}
    C\, 
    &=
     \begin{bmatrix} 
     -2\,\cH_\mathrm{gh}\,\Delta^{-1}
        \ds{\begin{bmatrix}
        \dfrac{\partial U}{\partial x}\dfrac{\partial}{\partial x}
        \\[.35cm]
        \dfrac{\partial U}{\partial z}\dfrac{\partial}{\partial x}
        \end{bmatrix}^T} 
    \\[.95cm] 
    E
    \end{bmatrix}.
\end{align}
For uncorrelated, white, zero-mean measurement noise $\eta_k$ with covariance $R \succ 0$, Algorithm~\ref{alg.LKF} summarizes the steps of the LKF in obtaining the estimate for the hub-height velocity fluctuation field $\hat{\bv}_k$. Here, $\Psi_{\bv,k} \DefinedAs \left<(\bv_k - \hat{\bv}_k)(\bv_k - \hat{\bv}_k)^T\right>$ provides an approximation for the covariance of the estimation error, and $L_k$ is the Kalman gain. When considering white-in-time process noise the discrete-time dynamics for a time step of $\Delta t$ are given by $F = \mre^{A \Delta t}$ and the covariance of the process noise follows the construction prescribed in Sec.~\ref{sec.white-process-noise}, i.e., $M=\bar{Q}$ (cf.~Eq.~\eqref{eq.white_cov}). On the other hand, when considering colored-in-time process noise (Sec.~\ref{sec.colored-process-noise}), $F = \mre^{A_f \Delta t}$, where $A_f$ is the generator of the modified dynamics~\eqref{eq.modified-dyn} and $M=G\,\Omega\, G^T$. Here, $G = \left(\mre^{A_f \Delta t} - I\right) A_f^{-1} B$ with $\Omega = \Omega^T \succ 0$ denoting the covariance of white noise $\bw_k$ and $I$ denoting the identity operator of appropriate size. Finally, the negative and positive superscripts in Algorithm~\ref{alg.LKF} respectively denote a priori and a posteriori states, which correspond to estimated quantities before and after incorporating information from new measurements.

\begin{algorithm}[H]
\caption{\enskip Linearized Kalman Filter}\label{alg.LKF}
\begin{algorithmic}
\State \textbf{input:} $F$, $C$, $M$, $R$, and measurements $\varphi_k$
\\ \vspace{-0.3cm}
\State \textbf{initialize:} $\hat{\bv}^+_0$ and $\Psi^+_{\bv,0}$
\\ \vspace{-0.3cm}
\State \textbf{for} $k=1,2, \dots$ 
\\
\vspace{.15cm}
\quad\begin{tabular}{rcl}
     $\Psi_{\bv,k}^-$ & $=$ & $F\,\Psi_{\bv,k-1}^+F^T \,+\, M$
     \\[.2cm]
     $\hat{\bv}_k^-$ & $=$ & $F\, \hat{\bv}_{k-1}^+$
     \\[.1cm]
     $L_k$ & $=$ & $\Psi_{\bv,k}^- \, C^T \left(C\,\Psi_{\bv,k}^- C^T \,+\,R \right)^{-1}$
     \\[.2cm]
     $\hat{\bv}_k^+$ & $=$ & $\hat{\bv}_k^- \,+\, L_k\left(\varphi_k \,-\, C\,\hat{\bv}_k^-\right)$
     \\[.2cm]
     $\Psi_{\bv,k}^+$ & $=$ & $\left(I \, - \, L_k C\right)\Psi_{\bv,k}^- \left(I\,-\,L_k C\right)^T +  L_k R L_k^T$
\end{tabular}
\vspace{.15cm}
\State \textbf{endfor}
\\  \vspace{-0.3cm}
\State \textbf{output:} $\hat{\bv}_k^+$  and $\Psi^+_{\bv,k}$ as the posterior estimates of the velocity field and the covariance of the error
\end{algorithmic}
\end{algorithm}

Assuming a time-invariant output relation can result in errors when using the LKF. The EKF addresses this problem by accounting for the evolution of the state via iterative linearization of the measurement equation. In our problem, this amounts to an update in the output relation per iteration, i.e.,  $\hat{\varphi}_k = C_k \hat{\bv}_k$, where
\begin{align}
\label{eq.Ck}
    C_k\, 
    &= 
    \begin{bmatrix}
    \cH_\mathrm{gh} \left(\dfrac{\partial {\cal P}}{\partial \bv} \Big|_{\hat{\bv}_k^-} \right)^T
    \\[.25cm] 
    E 
    \end{bmatrix}
    \\[.1cm]
    &=
    \non
     \begin{bmatrix} 
     -2\,\cH_\mathrm{gh}\,\Delta^{-1}
        \ds{\begin{bmatrix}
        \dfrac{\partial \hat{u}^-_{k}}{\partial x}\dfrac{\partial}{\partial x} 
        \,+\, 
        \dfrac{\partial \hat{w}^-_{k}}{\partial x}\dfrac{\partial }{\partial z} 
        \,+\, 
        \dfrac{\partial U}{\partial x}\dfrac{\partial}{\partial x}
        \\[.35cm]
        \dfrac{\partial \hat{u}^-_{k}}{\partial z}\dfrac{\partial }{\partial x} 
        \,+\, 
        \dfrac{\partial U}{\partial z}\dfrac{\partial}{\partial x}
        \,+\, 
        \dfrac{\partial \hat{w}^-_k}{\partial z}\dfrac{\partial}{\partial z}
        \end{bmatrix}^T} 
    \\[.95cm] 
    E
    \end{bmatrix}
\end{align}
where $\hat{\bv}_k^- = [\,\hat{u}_k^- \,~ \hat{w}_k^-\,]^T$ is the a priori state estimate at the $k$th iteration.
The EKF also uses the nonlinear output relation~\eqref{eq.state-output-approx-filter} to obtain the a posteriori state estimate $\hat{\bv}^+_k$; see Algorithm~\ref{alg.EKF}.

\begin{algorithm}[H]
\caption{\enskip Extended Kalman Filter}\label{alg.EKF}
\begin{algorithmic}
\State \textbf{input:} $F$, $M$, $R$, and measurements $\varphi_k$
\\ \vspace{-0.3cm}
\State \textbf{initialize:} $\hat{\bv}^+_0$ and $\Psi^+_{\bv,0}$
\\ \vspace{-0.3cm}
\State \textbf{for} $k=1,2, \dots$ 
\\ \vspace{.15cm}
\quad\begin{tabular}{rcl}
     $\Psi_{\bv,k}^-$ & $=$ & $F\,\Psi_{\bv,k-1}^+F^T \,+\, M$
     \\ [.15cm]
     $\hat{\bv}_k^-$ & $=$ & $F\, \hat{\bv}_{k-1}^+$
     \\ [.1cm]
     $L_k$ & $=$ & $\Psi_{\bv,k}^- \, C_k^T \left(C_k\,\Psi_{\bv,k}^- C_k^T \,+\,R \right)^{-1}$
     \\ [.2cm]
     $\hat{\bv}_k^+$ & $=$ & $\hat{\bv}_k^- \,+\, L_k\left(\varphi_k \,-\, \mathcal{C}(\hat{\bv}_k^-)\right)$
     \\ [.2cm]
     $\Psi_{\bv,k}^+$ & $=$ & $\left(I\,-\,L_k C_k\right)\Psi_{\bv,k}^- \left(I\,-\,L_k C_k\right)^T\,+\, L_k R L_k^T$
\end{tabular}
\vspace{0.15cm}
\State \textbf{endfor}
\\ \vspace{-0.3cm}
\State \textbf{output:} $\hat{\bv}_k^+$  and $\Psi^+_{\bv,k}$ as the posterior estimates of the velocity field and the covariance of the error
\end{algorithmic}
\end{algorithm}

In spite of its popularity in nonlinear state estimation, the EKF comes with errors due to linearization. To address this issue, we also consider two filtering algorithms that respect the nonlinearity in the measurement equation~\eqref{eq.state-output-approx-filter}, the UKF (Algorithm~\ref{alg.UKF}) and the EnKF (Algorithm~\ref{alg.EnKF}), which compute a priori updates by averaging over an ensemble of trajectories resulting from a set of deterministic (Sec.~14 of Ref.~[\onlinecite{sim06}]) or non-deterministic~\cite{eve03} initial profiles, also known as ``sigma points", respectively. More specifically, the UKF algorithm begins by propagating a set of $2n+1$ deterministic sigma points for $n$ states in our state dynamics. These points are then used in computing a priori estimates of the measurement vector $\hat{\varphi}_{i}^-$ via the nonlinear output relation~\eqref{eq.state-output-approx-filter}. In Algorithm~\ref{alg.UKF}, $\mathrm{e}_i\in\bbR^{n\times 1}$ is the unit vector with one in its $i$th entry, covariance matrices $\Psi_{\bv}$ and $\Psi_{\varphi}$ quantify deviations of the resulting state $\bv_i$ and output $\varphi_i$ trajectories from their ensemble average over all sigma points, and the matrix $\Psi_{\bv \varphi}$ quantifies the cross-correlation between these estimates. The a posteriori estimate of the state is ultimately computed as the ensemble average of the effect of various sigma points. 

\begin{algorithm}[H]
\caption{\enskip Unscented Kalman Filter}\label{alg.UKF}
\begin{algorithmic}
\State \textbf{input:} $F$, $M$, $R$, and measurements $\varphi_k$
\\ \vspace{-0.3cm}
\State \textbf{initialize:} Choose $m=2n+1$ sigma points, with $n$ equal to the number of states and initialize $\hat{\bv}^+_0$ and $\Psi^+_{\bv,0}$.
\\ \vspace{-0.3cm}
\State \textbf{for} $k=1,2, \dots$ 
\\ \vspace{-0.3cm}
\State \quad \quad \textbf{if} $i=1$
\\
\quad \quad \quad \quad $\hat{\bv}^+_{k-1,i} \;=\; \hat{\bv}^+_{k-1}$
\\ \vspace{-0.2cm}
\State \quad \quad \quad \textbf{elseif} $i<n+2$
\\ \vspace{.1cm}
\quad \quad \quad \quad $\hat{\bv}^+_{k-1,i} \;=\; \hat{\bv}^+_{k-1} \,+\, \sqrt{n}\,\mathrm{e}^T_i\,(\Psi^+_{\bv,k-1})^{1/2}$
\\ \vspace{-0.3cm}
\State \quad \quad \quad \textbf{else}
\\ \vspace{.1cm}
\quad \quad \quad \quad $\hat{\bv}^+_{k-1,i} \;=\; \hat{\bv}^+_{k-1} \,-\,\sqrt{n}\,\mathrm{e}^T_i \, (\Psi^+_{\bv,k-1})^{1/2}$
\\ \vspace{-0.2cm}
\State \quad \quad \textbf{endif} $i=1$
\\ 
\quad \quad \quad $\hat{\bv}^-_{k,i}\;=\;F\hat{\bv}^+_{k-1,i}$
\\ \vspace{0.1cm}
\quad \quad \quad $\hat{\varphi}_{k,i}\;=\;\mathcal{C}(\hat{\bv}^-_{k,i})$
\\ \vspace{-.3cm}
\State \quad \textbf{endfor}
\\ 
    \quad\begin{tabular}{rcl}
    $\hat{\bv}^-_{k}$ & $=$ & $\dfrac{1}{m}\sum_{i=1}^{m} \hat{\bv}^-_{k,i}$
    \\ [0.25cm]
     $\hat{\varphi}_{k}$ & $=$ & $\dfrac{1}{m}\sum_{i=1}^{m} \hat{\varphi}_{k,i}$
     \\ [0.25cm]
     $\Psi_{\bv,k}^-$ & $=$ & $\dfrac{1}{m}\sum_{i=1}^{m} (\hat{\bv}^-_{k,i} - \hat{\bv}^-_k)(\hat{\bv}^-_{k,i} - \hat{\bv}^-_k)^T + M$ 
    \\ [0.25cm]
     $\Psi_{\varphi,k}^-$ & $=$ & $\dfrac{1}{m}\sum_{i=1}^{m} (\hat{\varphi}_{k,i} - \hat{\varphi}_{k})(\hat{\varphi}_{k,i} - \hat{\varphi}_{k})^T +R$
    \\  [0.25cm]
     $\Psi_{\bv \varphi,k}^-$ & $=$ & $\dfrac{1}{m}\sum_{i=1}^{m} (\hat{\bv}^-_{k,i} - \hat{\bv}^-_{k})(\hat{\varphi}_{k,i} - \hat{\varphi}_{k})^T$
    \\ [0.25cm]
    $L_k$ & $=$ & $\Psi^-_{\bv \varphi,k} \, (\Psi^-_{\varphi,k})^{-1}$
    \\  [0.25cm]
     $\hat{\bv}^+_k$ & $=$ & $\hat{\bv}^-_k \, + \, L_k \, (\varphi_{k} \, - \, \hat{\varphi}_{k})$
    \\ [0.25cm]
     $\Psi^+_{\bv,k}$ & $=$ & $\Psi^-_{\bv,k} \, - \, L_k \, (\Psi^-_{\varphi,k}) \, L_k^T$
    \end{tabular}
\\ \vspace{-0.1cm}
\State \textbf{endfor}
\\ \vspace{-0.3cm}
\State \textbf{output:} $\hat{\bv}_k^+$  and $\Psi^+_{\bv,k}$ as the posterior estimates of the velocity field and the covariance of the error
\end{algorithmic}
\end{algorithm}

The EnKF algorithm follows a similar sequence of steps to the UKF but with mean and covariances computations conducted over a significantly smaller number of randomly generated sigma points. In practice, the number of sigma points considered in the EnKF is determined empirically as the number beyond which the estimated fluctuation field does not meaningfully change. 
While this approach reduces the number of trajectories, and thus, computations, it can come at the cost of persistent underestimation of the estimation error covariance matrices (i.e., inbreeding) and the appearance of spurious two-point correlations within these matrices~\cite{pet08}. A common remedy to the former issue is to inflate the a priori estimates of the covariance $\Psi_\varphi^-$ and cross-correlation $\Psi_{\bv \varphi}^-$ matrices at each iteration $k$ by correcting the a priori state estimate of each trajectory $\hat{\bv}^-_i$ in the direction of its deviation from the ensemble mean $\hat{\bv}^-$, i.e., 
$
    \hat{\bv}^-_{i} \,\leftarrow\; \hat{\bv}^- +\, l\left(\hat{\bv}^-_{i} -\, \hat{\bv}^-\right)
$,
where $l \in [1.01,1.25]$ is an inflation parameter~\cite{pet08}. On the other hand, potential long-range spurious correlations are typically eliminated via covariance localization, which involves the truncation of weighted variants of the error covariance matrices~\cite{houmit01,gascoh99}. As such long-range correlations are already negligible (compared to one-point correlations on the main diagonal of the covariance matrices) in our study, we do not make use of the latter technique. We note that such remedies, work in favor of maintaining the computational edge of EnKF over UKF, and that such issues can generally be overcome by increasing the number of sigma points.

\begin{algorithm}[H]
\caption{\enskip Ensemble Kalman Filter}\label{alg.EnKF}
\begin{algorithmic}
\State \textbf{input:} $F$, $G$, $\bw_{k-1}$, $\eta_k$, and measurements $\varphi_k$
\\ \vspace{-0.3cm}
\State \textbf{initialize:} Choose $m$ random sigma points and initialize $\hat{\bv}^+_0$ and $\Psi^+_{\bv,0}$.
\\ \vspace{-0.2cm}
\State \textbf{for} $k=1,2, \dots$
\vspace{.1cm}
\State \quad \textbf{for} $i=1,\dots,m$
\\ \vspace{.1cm}
\quad \quad $\hat{\bv}^-_{k,i}\;=\; F\,\hat{\bv}^+_{k-1,i} \,+\, G\,\bw_{k-1}$
\\  \vspace{-0.2cm}
\State \quad \textbf{endfor}
\\
\quad $\hat{\bv}^-_{k} \;=\; \dfrac{1}{m}\sum_{i=1}^{m}\hat{\bv}_{k,i}^-$
\\ \vspace{-0.2cm}
\State \quad \textbf{for} $i=1,\dots,m$
\\ \vspace{.1cm}
\qquad $\hat{\bv}^-_{k,i} \;=\; \hat{\bv}^-_k \,+\, l\,(\hat{\bv}^-_{k,i}\,-\, \hat{\bv}^-_k)$
\\ \vspace{.15cm}
\quad \quad $\hat{\varphi}_{k,i}\;=\;\mathcal{C}(\hat{\bv}_{k,i}^-) \,+\, \eta_k$
\\ \vspace{-0.2cm}
\State \quad \textbf{endfor}
\\
\quad \begin{tabular}{rcl} 
$\hat{\bv}^-_{k}$ & $=$ & $\dfrac{1}{m}\sum_{i=1}^{m}\hat{\bv}_{k,i}^-$
\\ [.2cm]
$\hat{\varphi}_{k}$ & $=$ & $\dfrac{1}{m}\sum_{i=1}^{m} \hat{\varphi}_{k,i}$
\\ [.2cm]
$\Psi_{\varphi,k}^-$ & $=$ & $\dfrac{1}{m}\sum_{i=1}^{m} (\hat{\varphi}_{k,i} \, - \, \hat{\varphi}_{k}) \, (\hat{\varphi}_{k,i} \, - \, \hat{\varphi}_{k})^T$
\\ [.2cm]
$\Psi_{\bv \varphi,k}^-$ & $=$ & $\dfrac{1}{m}\sum_{i=1}^{m} (\hat{\bv}^-_{k,i} \, - \, \hat{\bv}^-_{k}) \, (\hat{\varphi}_{k,i} \, - \, \hat{\varphi}_{k})^T$
\\ [.2cm]
$L_k$ & $=$ & $\Psi^-_{\bv \varphi,k} \, (\Psi^-_{\varphi,k})^{-1}$
\end{tabular}
\\  \vspace{0cm}
\State \quad \quad \textbf{for} $i=1,\dots,m$
\\ \vspace{.1cm}
\quad \quad \quad $\hat{\bv}^+_{k,i} \;=\; \hat{\bv}^-_{k,i}\;+\;L_k(\varphi_{k} \, - \, \hat{\varphi}_{k,i})$
\\ \vspace{-0.2cm}
\State \quad \quad \textbf{endfor}
\\ \vspace{0.15cm}
\quad \quad \;\;\;$\hat{\bv}^+_{k}=\dfrac{1}{m}\sum_{i=1}^{m}\hat{\bv}^+_{k,i}$
\vspace{.1cm}
\State \textbf{endfor}
\\ \vspace{-0.3cm}
\State \textbf{output:} $\hat{\bv}_k^+$ as the posterior estimate of the velocity field 
\end{algorithmic}
\end{algorithm}

\subsection{Pressure projection via linear stochastic estimation}
\label{sec.LSE}
\begin{figure}
\vspace{.3cm}
    \centering
    \begin{tabular}{cc}
    \begin{tabular}{c}
        \rotatebox{90}{$z$}
    \end{tabular}
    &
    \hspace{0cm}
    \begin{tabular}{c}
         \includegraphics[width=0.4\textwidth]{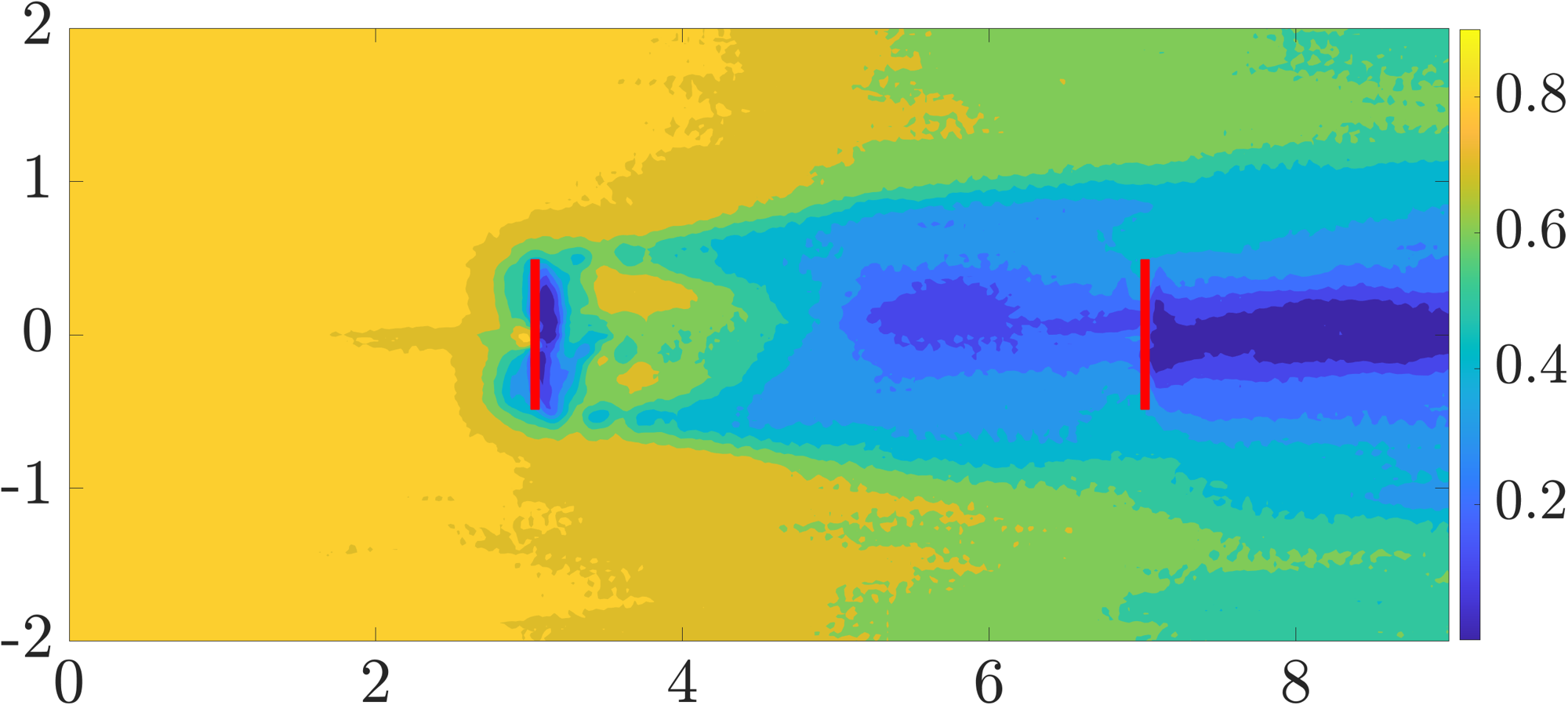}
    \end{tabular}
    \\[0cm]
       &
       {$x$}
       \end{tabular}
    \caption{The coherence measure $\gamma_\mathrm{gh}^2$ (Eq.~\eqref{eq.LCM}) between the hub-height and the ground pressure. Thick red lines mark turbine rotors and the wind direction is from left to right.}
    \label{fig.LCM}
\end{figure}

The existence of coherent motions in wall-bounded flows can result in strong two-point correlations of flow quantities, e.g., pressure and velocity, between points that are near the wall and points that are away from it~\cite{marmon19}. Recently,  pressure fluctuations near the wall have been shown to maintain high levels of correlation with those in wall-separated planes~\cite{baalee22}. This is in contrast to the velocity field which does not demonstrate significant vertical correlation. 
This motivates the use of linear stochastic estimators for projecting wall-separated measurements to the near-wall region (and vice versa) using normalized variants of such two-point correlations~\cite{baadaclee24}. 
The normalized two-point correlation between the pressure on the ground and at hub height 
\begin{align}
\label{eq.LCM}
	\gamma_{\mathrm{gh}}^2 (x,z)
	\;\DefinedAs\;
	\dfrac{\langle p_\mathrm{g}(x,z) \, p_\mathrm{h}(x,z) \rangle ^2}{\langle p_\mathrm{g}^2(x,z) \rangle \langle p_\mathrm{h}^2(x,z) \rangle}
\end{align}
provides a metric for evaluating the strength of such two-point correlations over the 2D computational domain. 
Ideally, the temporal expectation 
need to be computed over a sufficiently long time window to ensure statistical convergence.
Figure~\ref{fig.LCM} shows this measure of vertical coherence over the computational domain considered in the numerical experiments conducted in Sec.~\ref{sec.KF_comparison}.
A high level of coherence is observed over most of the domain except the wake region.

Assuming a nonzero correlation between pressure at hub height and on the ground, an estimate of ground-level pressure can be provided through the linear transformation
\begin{align}
\label{eq.projection}
	p_\mathrm{g} (x,z)
	\;=\;
	{\cal H}_\mathrm{gh}(x,z) \, p_\mathrm{h}(x,z).
\end{align}
Here, $\mathcal{H}_\mathrm{gh}$ is a kernel transfer function
\begin{align*}
	\cH_\mathrm{gh} (x,z)
	\;\DefinedAs\;
	\dfrac{\langle p_\mathrm{h}(x,z)\, p_\mathrm{g}(x,z) \rangle}{\langle p_\mathrm{h}(x,z)\, p_\mathrm{h}(x,z) \rangle}
\end{align*}
that is trained offline using a time-resolved pressure field resulting from high-fidelity simulations or field measurements and is related to the coherence measure $\gamma_\mathrm{gh}^2$ via
\begin{align*}
	\cH_\mathrm{gh} (x,z)
	\;=\; 
    \sqrt{\gamma_{\mathrm{gh}}^2(x,z) \dfrac{\langle p^2_\mathrm{g} (x,z) \rangle}{\langle p^2_\mathrm{h} (x,z) \rangle}}.
\end{align*}
We note that $\cH_\mathrm{gh}$ is a statistical quantity that has been shown to provide reasonable estimates under statistically-steady conditions (see, e.g., Ref.~[\onlinecite{baadaclee24}]), but cannot ensure an accurate representation at each time instance. 

\begin{figure*}
\begin{center}
        \begin{tabular}{cccc}
        \hspace{-1.2cm}
        \subfigure[]{\label{fig.LES_pressure}}
        &&
        \hspace{0.45cm}
        \subfigure[]{\label{fig.LSE_pressure}}
        &
        \\[-.45cm]
        \hspace{-0.9cm}
	\begin{tabular}{c}
        \vspace{.5cm}
        \rotatebox{90}{$z$}
       \end{tabular}
       &
       \hspace{-0.2cm}
	    \begin{tabular}{c}
\includegraphics[width=0.35\textwidth]{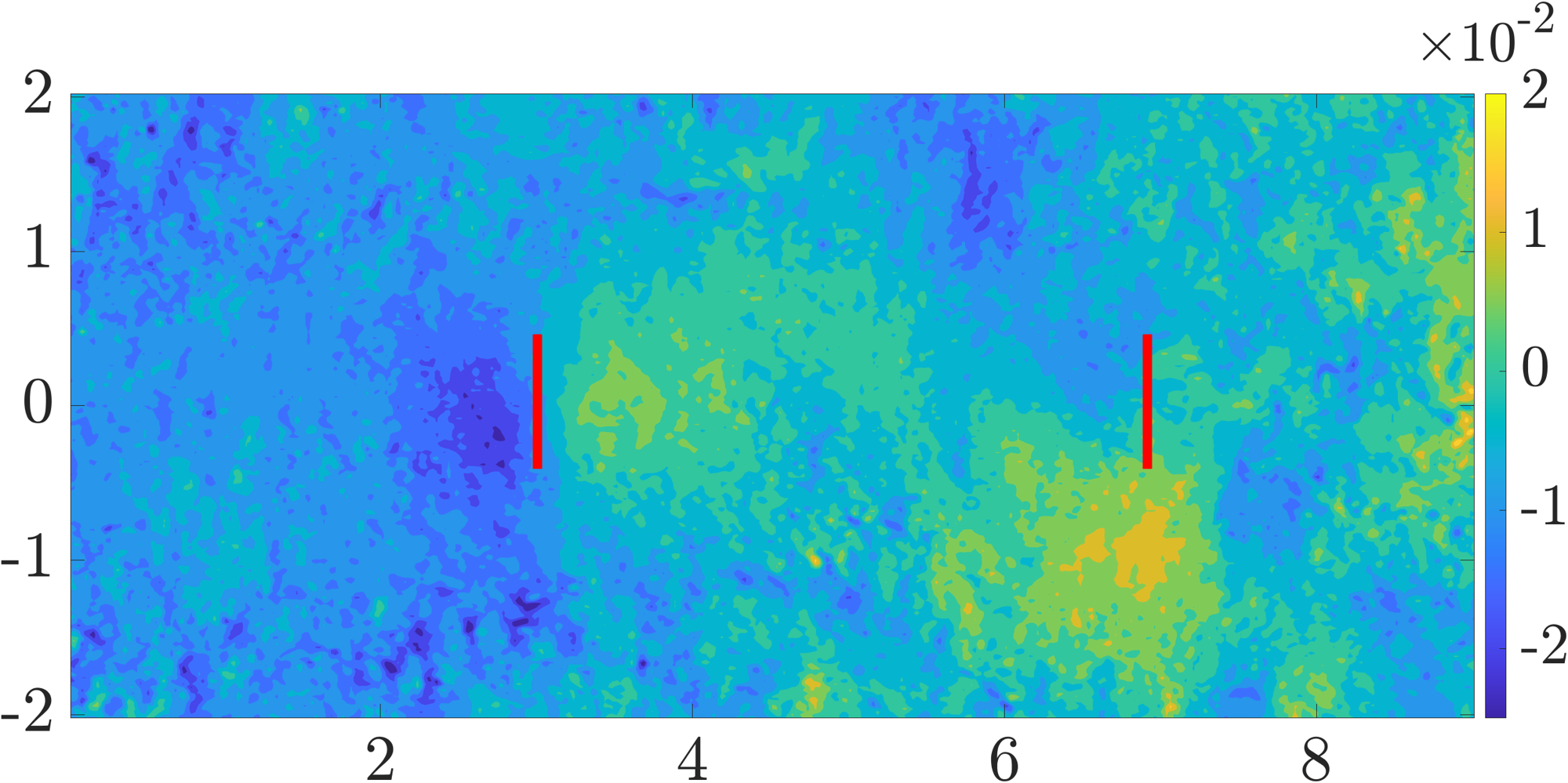}
       \end{tabular}
       &&
       \hspace{.2cm}
        \begin{tabular}{c}
       \includegraphics[width=0.35\textwidth]{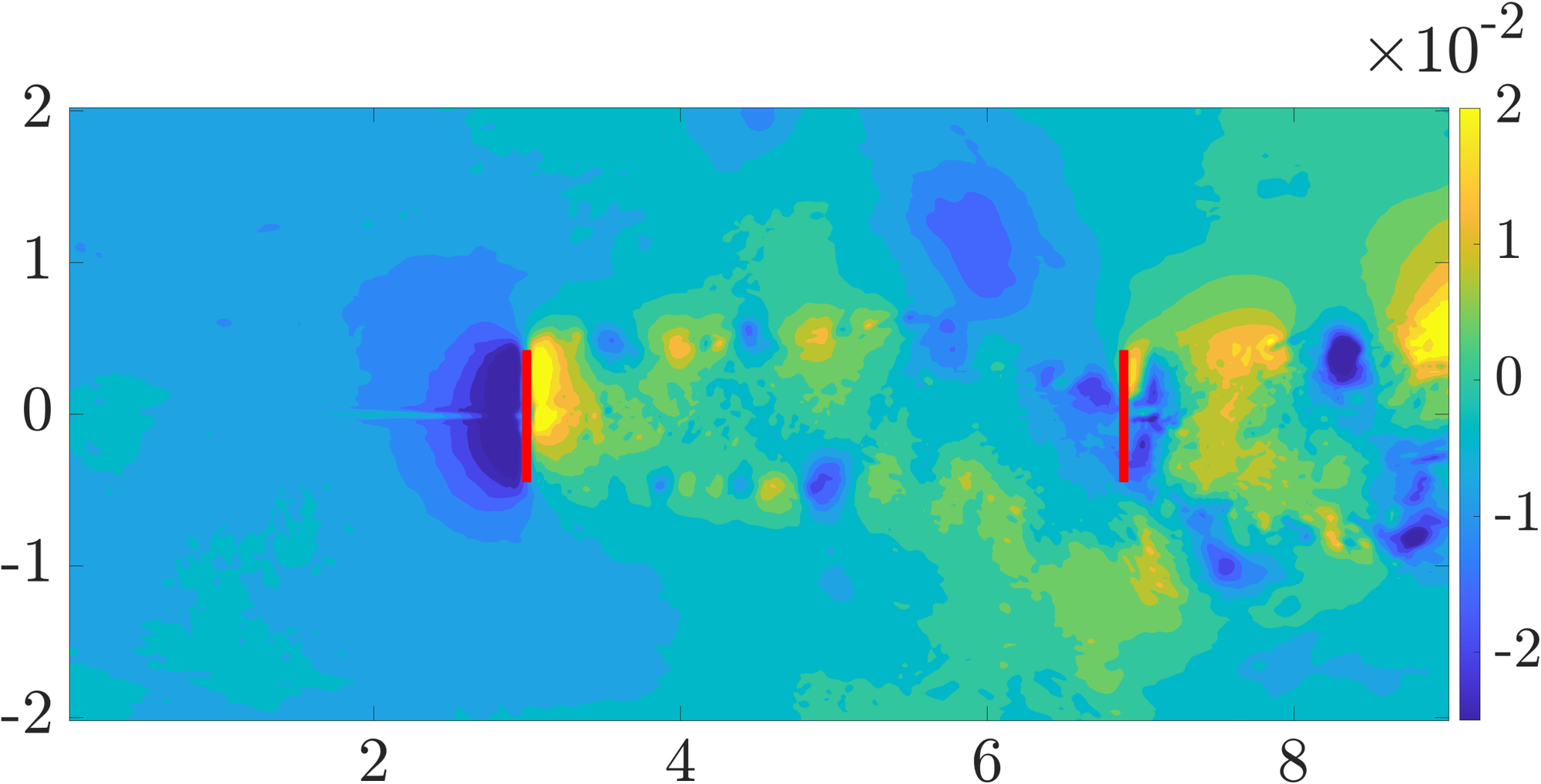}
       \end{tabular}
       \end{tabular}
       \\[-0.1cm]
       \begin{tabular}{cccc}
        \hspace{-1.3cm}
        \subfigure[]{\label{fig.LES_pressure_yaw15}}
        &&
        \hspace{0.45cm}
        \subfigure[]{\label{fig.LSE_pressure_yaw15}}
        &
        \\[-.45cm]
        \hspace{-0.9cm}
	\begin{tabular}{c}
        \vspace{.5cm}
        \rotatebox{90}{$z$}
       \end{tabular}
       &
       \hspace{-0.2cm}
	    \begin{tabular}{c}
       \includegraphics[width=0.35\textwidth]{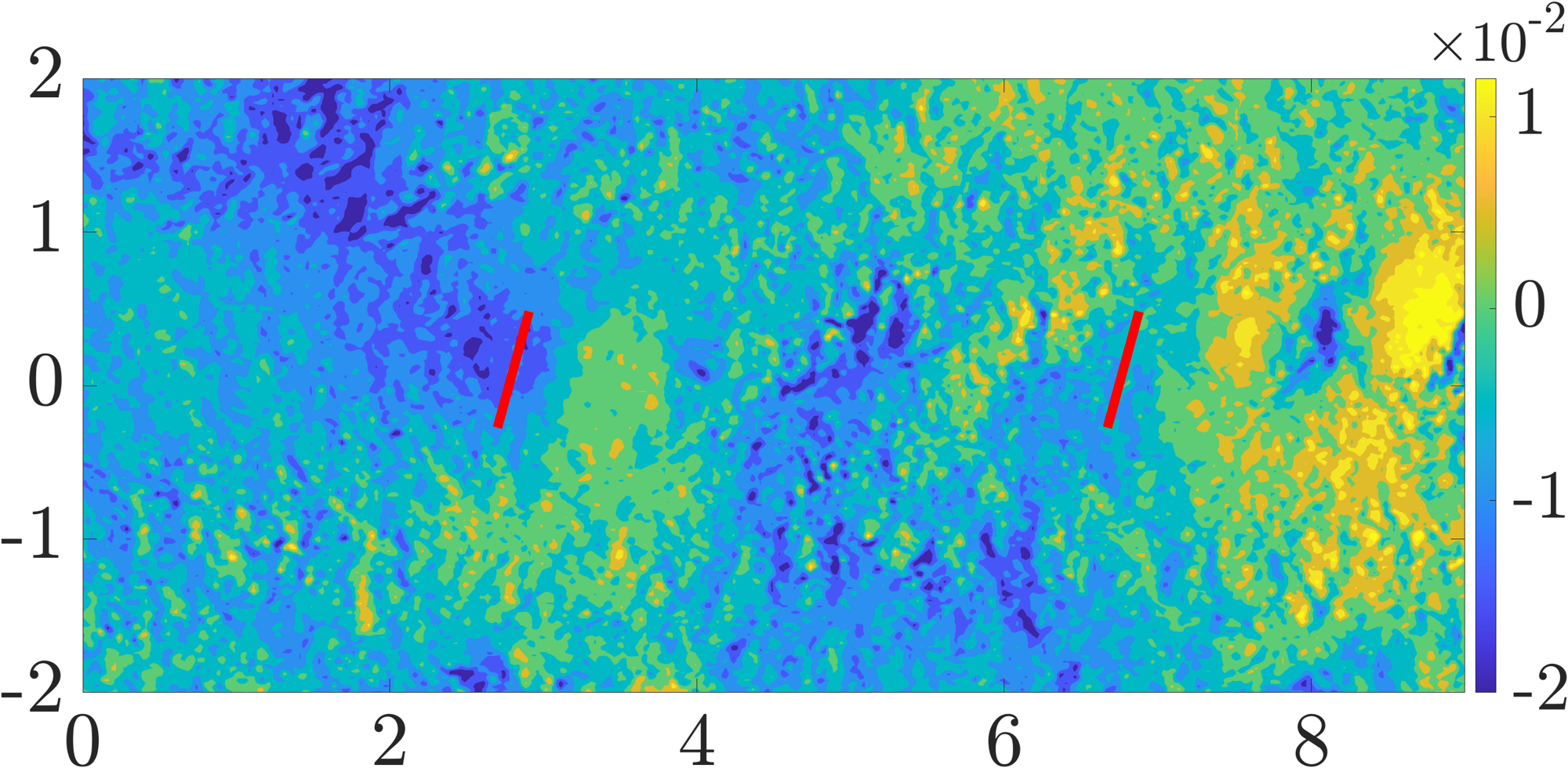}
       \\[-.1cm]
       \hspace{0.4cm}
       {$x$}
       \end{tabular}
       &&
       \hspace{0.2cm}
        \begin{tabular}{c}
       \includegraphics[width=0.35\textwidth]{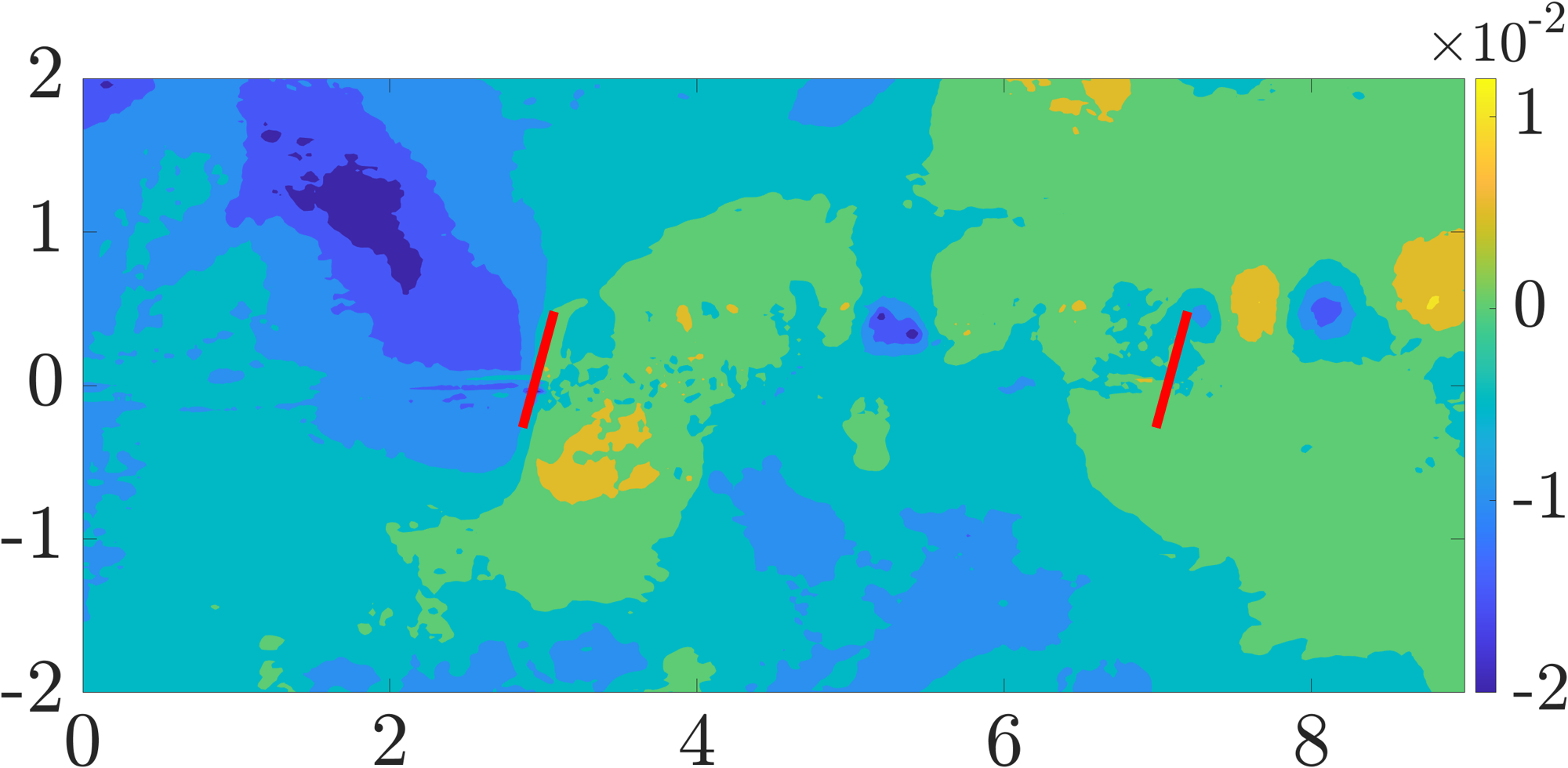}
       \\[-.1cm]
       \hspace{0.4cm}
       {$x$}
       \end{tabular}
       \end{tabular}
       \vspace{-.1cm}
        \caption{Snapshots of ground-level pressure field obtained by LES (left) and the projection of LES-generated hub-height pressure at the same time instance using Eq.~\eqref{eq.projection} (right) in the absence (a,b) and presence of a $\gamma=15^\degree$ yaw (c,d) with respect to the free-stream wind direction. Turbine rotors are marked by thick red lines and the wind direction is from left to right.}
        \label{fig.LSE_comparison}
\end{center}
\end{figure*}

Figure~\ref{fig.LSE_comparison} shows the ground-level pressure across a two-turbine farm resulting from LES and the projection of hub-height pressure using Eq.~\eqref{eq.projection}. It is evident that the projected pressure field captures the dominant pressure variations on the ground, which justifies the use of the linear stochastic estimation 
method for our purposes. As evident from Figs.~\ref{fig.LSE_comparison}(c,d), the pressure variations are well-captured by the proposed projection scheme even when the turbine rotors are not aligned with the direction of the wind. Note that, while the state of our prior model excludes the pressure field at hub height, we obtain this quantity using Eq.~\eqref{eq.Poisson} and the velocity field predicted by the linear stochastic model~\eqref{eq.modified-dyn}.

\section{Comparison of filtering algorithms}
\label{sec.KF_comparison}

We present results for velocity fluctuation estimation in a wind farm with two turbines using the proposed estimation framework. The Reynolds number is set to $10^8$ in accordance with the LES that generated the training dataset; see Appendix~\ref{sec.appendix2} for details on the high-fidelity LES. When considering a colored noise process the modification to the linearized dynamic generator (Eq.~\eqref{eq.modified-dyn}) is obtained via the optimization-based modeling framework summarized in Sec.~\ref{sec.Noisemodeling} to achieve consistency with the LES in matching velocity correlations at the hub-height locations highlighted in Fig.~\ref{fig.LES_LNS_comp}. 

For our numerical experiments, we consider a computational domain with $x \in [0 ~ 9]$ and $z \in [-2, 2]$ that includes two turbines of unit diameter located at $(x,z) = (3,0)$ and $(7,0)$ (Fig.~\ref{fig.hubheight_config}). Similar to the considerations taken in determining the base flow (Sec.~\ref{sec.baseflow}), all length scales are non-dimensionalized by the rotor diameter $d_0$. We use a uniform grid with $\Delta x=\Delta z=0.25$ (i.e., $N_x=43$ and $N_z=17$ grid points in the streamwise and spanwise directions, respectively), which leads to $\bv\in\bbR^{1462\times 1}$ in the linearized model (Eq.~\eqref{eq.lnse1}). A second-order central difference scheme is used to discretize the computational grid and differential operators. Simulations of all discrete-time Kalman filtering algorithms are conducted using a time step of $\Delta t= 0.18$ over the same span of time provided by the LES. Our numerical experiments did not show a significant change in quality of estimation when smaller time steps were used. Given the chaotic nature of hub-height turbulence and to provide a fair comparison with LES, we initialize all simulations with the initial flow conditions of the LES. While this choice yields a sensible starting condition (and level of error) at $t=0$, our numerical experiments show that the asymptotic convergence of the algorithms are not impacted by the choice of the initial condition. The covariance of measurement noise is set to $R=10^{-4}\times I_{n/2+2}$, where $n$ is the number of states in the prior model. Moreover, the covariance of white noise $\Omega$ is taken as a scalar multiple of identity, which can be tuned to achieve the best performance; in our numerical experiments, $\Omega=[{281}, 221,3.5,5.3]\times I_{n}$ for LKF, EKF, EnKF, and UKF, respectively. As expected, the significance of the process noise (with covariance $M = G\,\Omega\, G^T$), which accounts for modeling uncertainties due to linearization, is higher in the case of the LKF and EKF~\cite{wanmer00}. 

In this section, we perform estimation assuming that pressure measurements are uniformly available at all grid points on the ground.
To quantify performance, we propose two error metrics: (i) a running relative error in matching the variance of streamwise velocity at hub height, i.e.,
\begin{align}
\label{eq.RelativeError}
    \mathrm{err}_\mathrm{rel}(\bx,t)
    \;=\;
     \dfrac{\left|\int_0^t \left(u_\mathrm{LES}^2(\bx,\tau) \,-\, \hat{u}^2(\bx,\tau) \right) \mrd \tau\right|}{\int_0^t u_\mathrm{LES}^2(\bx,\tau) \, \mrd \tau}
\end{align}
and (ii) the error in matching the evolution of streamwise velocity normalized by the base flow $U$, i.e.,
\begin{align}
\label{eq.NormalizedError}
    \mathrm{err}_\mathrm{norm}(\bx,t)
    \;=\;
    \dfrac{|u_\mathrm{LES}(\bx,t)-\hat{u}(\bx,t)|}{U(\bx)}.
\end{align}
Normalization by the local base flow value gives more weight to dynamically significant regions of the flow (e.g., within the turbine wakes) that can be more difficult to estimate.

\subsection{Non-yawed wind farm}
\label{sec.non-yawed}
\begin{figure}
    \centering
    \begin{tabular}{cc}
    \begin{tabular}{c}
        \rotatebox{90}{$\mathrm{err_{rel}}$}
    \end{tabular}
    &
    \hspace{0.15cm}
    \begin{tabular}{c}
         \includegraphics[width=0.3\textwidth]{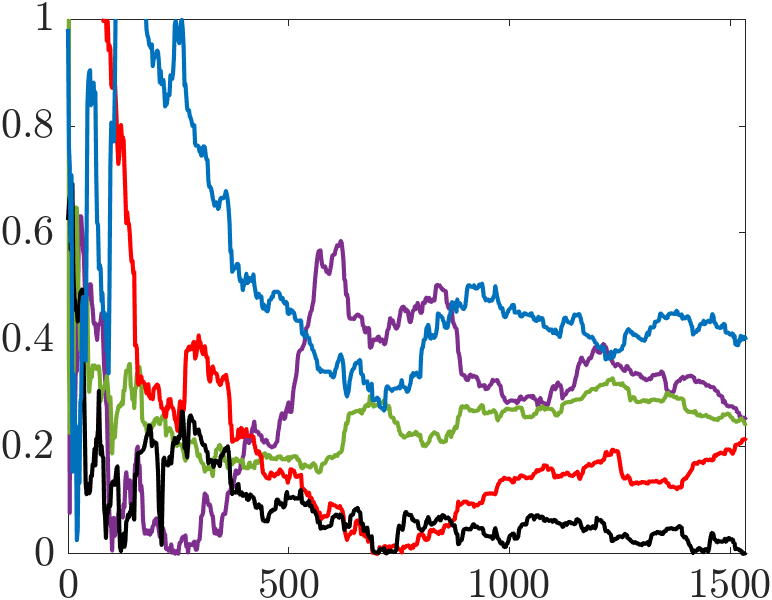}
    \end{tabular}
    \\[0cm]
       &
       {time (sec)}
       \end{tabular}
    \caption{Relative error in estimating the velocity variance, $\mathrm{err}_\mathrm{rel}$ (Eq.~\eqref{eq.RelativeError}), at $(x,z)=(4,0)$ using the LKF (blue), EKF with scaled white process noise (purple), EKF using colored process noise (green), EnKF (red), and UKF (black).}
    \label{fig.variance_error_comparison}
\end{figure}

We first evaluate the performance of our estimation framework for ideal conditions in which both turbines are perfectly aligned with the direction of the free-stream wind. Besides the EKF, we consider colored-in-time process noise models for all Kalman filters discussed in Sec.~\ref{sec.KF-slgorithms}. Figure~\ref{fig.variance_error_comparison} shows the relative statistical error $\mathrm{err}_\mathrm{rel}$ achieved by all Kalman filters at $(x,z) = (4,0)$ (i.e., one diameter behind the leading turbine) as a function of time. Evaluating the quality of estimation at this point is of particular interest as it is located in the dynamically complex near-wake region, where wake recovery is typically underestimated by conventional engineering wake models. A reasonably accurate estimation at this point can also be significant for preview control. Due to its distance from the second row of turbines (3 diameters for our experimental farm configuration), a sufficiently precise estimate of wind variations at this location would result in a significant lead time that is critical to the success of turbine control in mitigating the effects of atmospheric variability on structural loads and power production. In fact, by relying on the frozen turbulence hypothesis, which assumes that turbulent flow structures are transported as frozen entities by the mean wind~\cite{sch11}, approximately $54$ seconds of preview time could be provided if the maximum speed of wind in the wake region was around $7$ m/s.

\begin{figure}
\hspace{-.4cm}
    \begin{tabular}{cccc}
    \subfigure[]{\label{fig.rel_err}}
    &&
    \hspace{-0.4cm}
    \subfigure[]{\label{fig.avg_err}}
    \\
    \hspace{.2cm}
        \begin{tabular}{c}
             \rotatebox{90}{$\mathrm{err_{rel}}$}
        \end{tabular}
         &
         \hspace{-.2cm}
        \begin{tabular}{c}
             \includegraphics[width=0.19\textwidth]{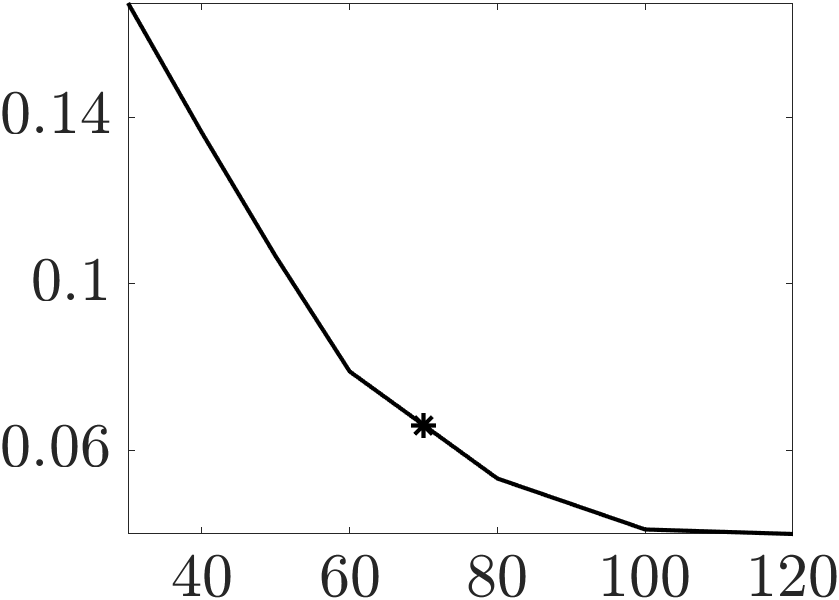}
        \end{tabular}
        &
        \begin{tabular}{c}
             \rotatebox{90}{$\langle \mathrm{err_{norm}}\rangle$}
        \end{tabular}
         &
         \hspace{-0.2cm}
        \begin{tabular}{c}
             \includegraphics[width=0.19\textwidth]{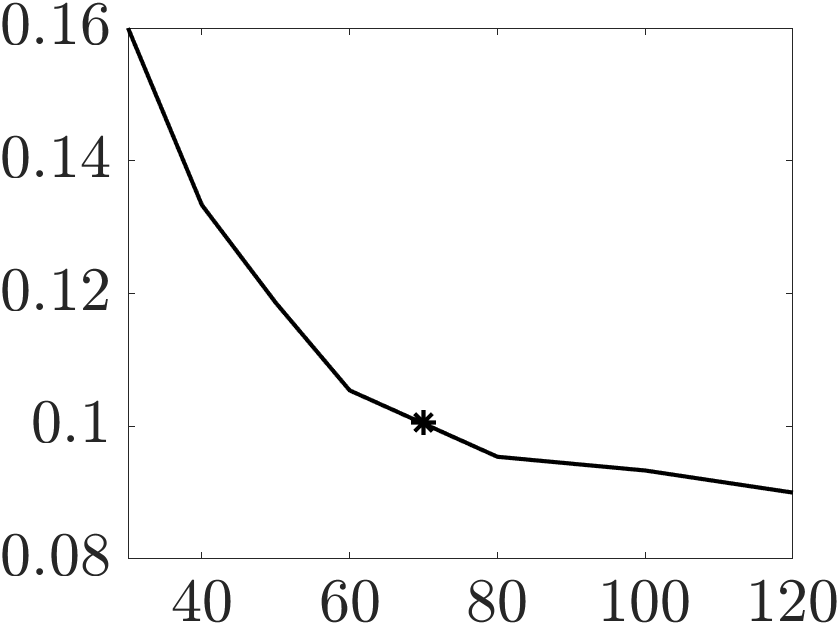}
        \end{tabular}
        \\
        &
        {\small number of sigma points}
        &&
        {\small number of sigma points}
    \end{tabular}
    \caption{{(a) Relative error in estimating the velocity variance  $\mathrm{err}_\mathrm{rel}$ (Eq.~\eqref{eq.RelativeError}); and (b) averaged normalized error $\langle\mathrm{err}_\mathrm{norm}\rangle$ at $(x,z)=(4,0)$ as a function of the number of sigma points used by the EnKF.}}
    \label{fig.sigma_points_number}
\end{figure}

While the relative error shown in Fig.~\ref{fig.variance_error_comparison} remains above $60\%$ for the LKF, it settles at a much lower value (despite a transient phase) for the EKF. Due to this high error, we exclude the LKF from the results presented in the remainder of this section. It can also be seen that the use of scaled white process noise (Sec.~\ref{sec.white-process-noise}), lowers the quality of estimation based on $\mathrm{err}_\mathrm{rel}$ for the EKF. As expected, both EnKF and UKF outperform the EKF models as they do not involve linearization errors. Overall, the UKF estimator manifests a better performance compared to the EnKF although it uses significantly more sigma points (2925 vs 70). It is important to note that while the UKF framework relies on the evolution of $2n + 1$ sigma points ($n$ being the number of degrees-of-freedom in our prior model, i.e., $1462$ in this numerical example), the EnKF framework allows for flexibility in exploring the trade-off between the number of randomly chosen sigma points and the resulting estimation error. {Both of the error metrics proposed in Eqs.~\eqref{eq.RelativeError} and~\eqref{eq.NormalizedError} can be used to choose the number of sigma points for the EnKF. Figures~\ref{fig.rel_err} and~\ref{fig.avg_err} show the dependence of statistical error $\mathrm{err_{rel}}$ and the time-averaged normalized error $\langle\mathrm{err_{norm}}\rangle$, respectively, on the number of sigma points used by the EnKF in estimating the velocity fluctuations at $(x,z)=(4,0)$.} As evident from this figure, the performance of the EnKF rapidly declines if we use less than 70 sigma points {(marked by $*$ in Fig.~\ref{fig.rel_err})} but does not significantly improve if we use more than 100. In what follows, we achieve a balance between computational speed and estimation quality by using 70 sigma points in constructing the EnKF. {As shown in Fig.~\ref{fig.avg_err}, a similar trend can be observed in the time-averaged normalized error, which also rapidly increases if less than 70 sigma points are used.}

\begin{figure*}
\begin{center}
        \begin{tabular}{cccccccc}
        \hspace{-0.2cm}
        \subfigure[]{\label{fig.white_EKF_estimation}}
        &&
        \hspace{0.3cm}
        \subfigure[]{\label{fig.EKF_estimation}}
        &&
        \hspace{-4.3cm}
        \subfigure[]{\label{fig.EnKF_estimation}}
        &&
        \hspace{-0.2cm}
        \subfigure[]{\label{fig.UKF_estimation}}
        &
        \\[-.4cm]
        \hspace{0.15cm}
	\begin{tabular}{c}
        \vspace{.5cm}
        \rotatebox{90}{$\hat{u}$}
       \end{tabular}
        &
               \hspace{-.1cm}
	    \begin{tabular}{c}
       \includegraphics[width=0.2\textwidth]{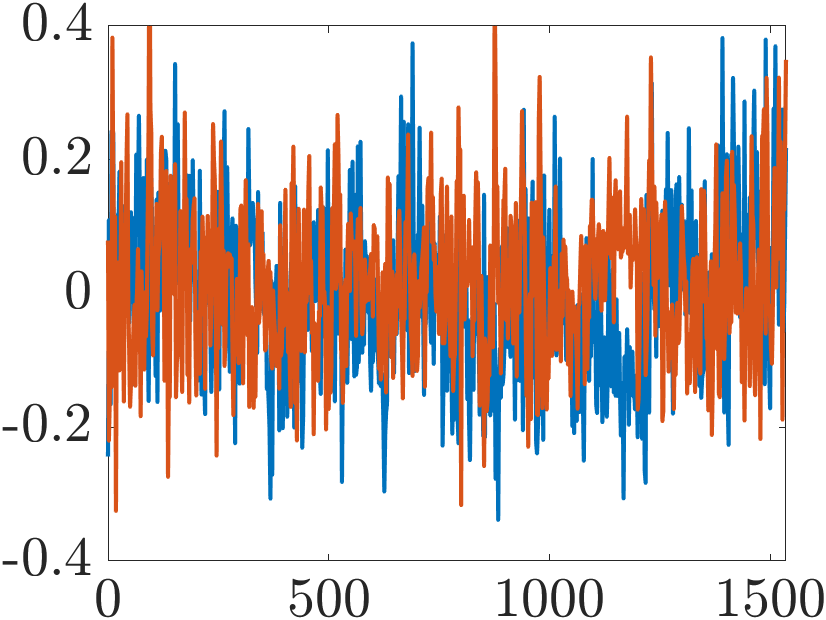}
       \hspace{-.5cm}
       \end{tabular}
       &&
       \hspace{.1cm}
        \begin{tabular}{c}
       \includegraphics[width=0.2\textwidth]{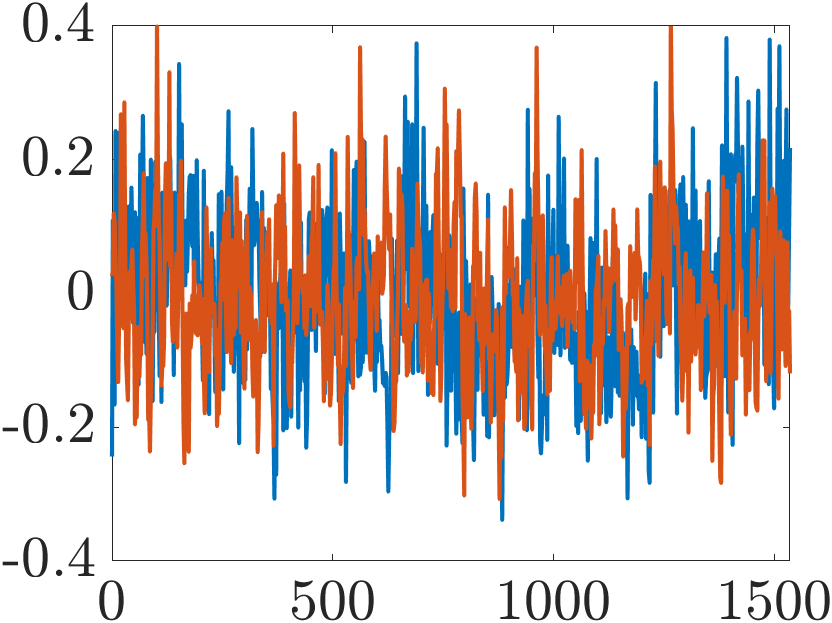}
       \hspace{2cm}
       \end{tabular}
       &&
        \hspace{-2cm}
        \begin{tabular}{c}
       \includegraphics[width=0.2\textwidth]{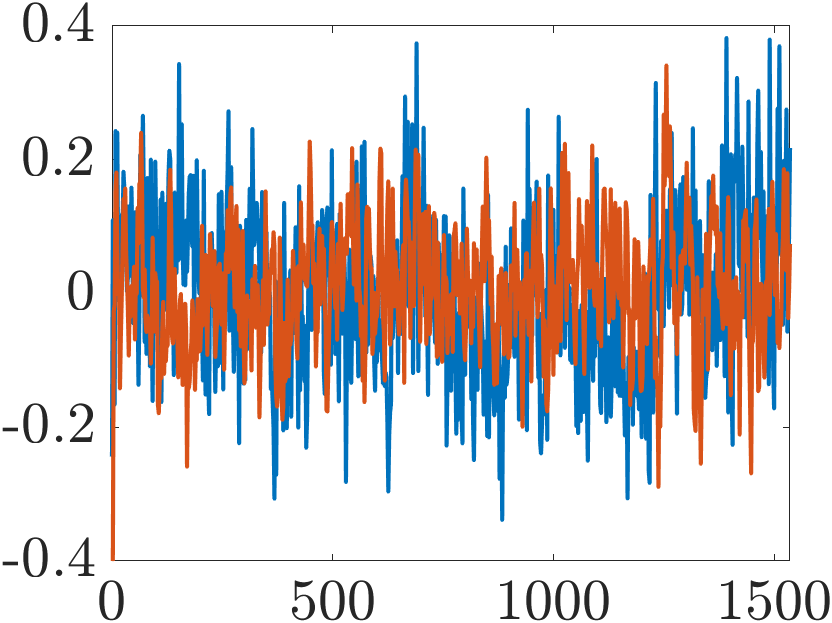}
       \end{tabular}
       &&
        \hspace{-0cm}
        \begin{tabular}{c}
       \includegraphics[width=0.2\textwidth]{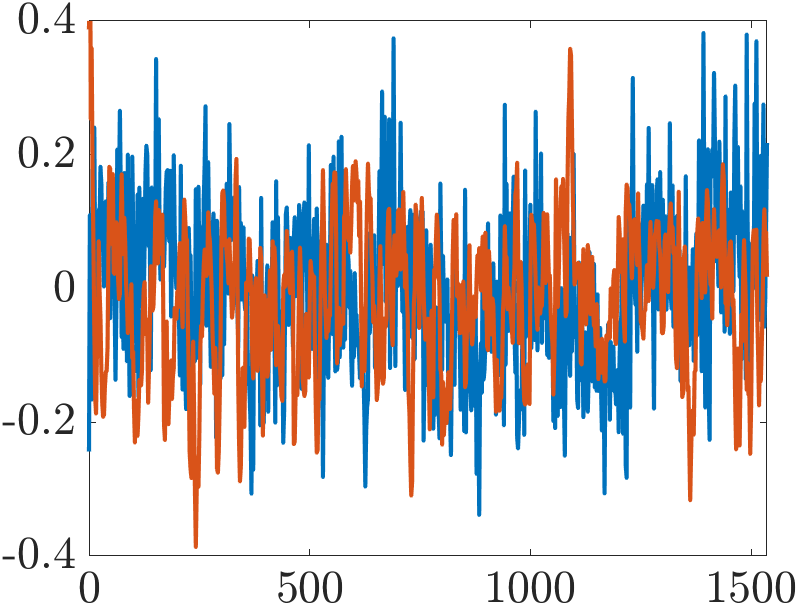}
       \end{tabular}
       \\[-.2cm]
       \hspace{-0.2cm}
       \subfigure[]{\label{fig.white_EKF_error_stat}}
        &&
        \hspace{0.3cm}
        \subfigure[]{\label{fig.EKF_error_stat}}
        &&
        \hspace{-4.3cm}
        \subfigure[]{\label{fig.EnKFK_error_stat}}
        &&
        \hspace{-0.2cm}
        \subfigure[]{\label{fig.UKF_error_stat}}
        &
        \\[-.4cm]
        \hspace{.15cm}
	\begin{tabular}{c}
        \vspace{.5cm}
        \rotatebox{90}{$\mathrm{err}_\mathrm{norm}$}
        \hspace{0cm}
       \end{tabular}
        &
        \hspace{-.1cm}
	    \begin{tabular}{c}
       \includegraphics[width=0.2\textwidth]{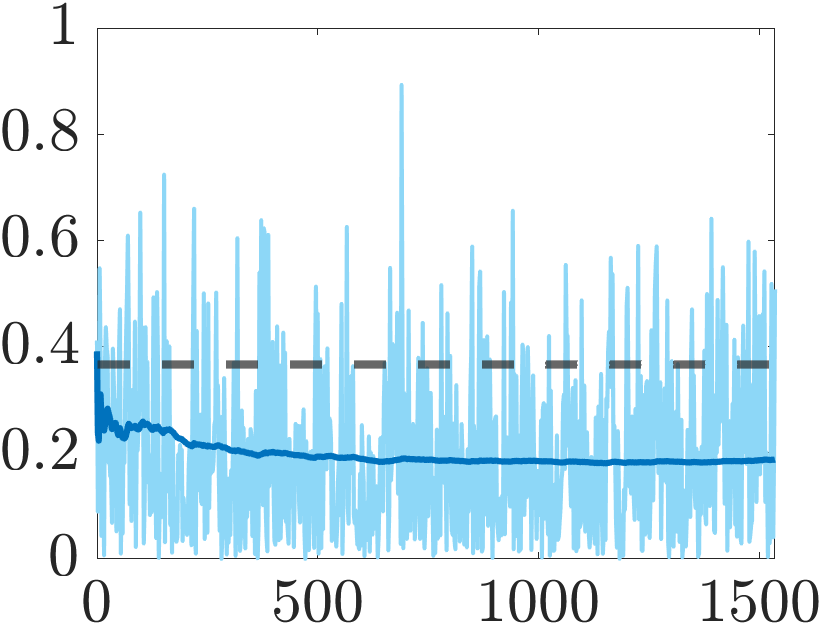}
       \hspace{-.5cm}
       \\[-.1cm]
       \hspace{.5cm}
       {time(s)}
       \end{tabular}
       &&
       \hspace{.1cm}
        \begin{tabular}{c}
       \includegraphics[width=0.2\textwidth]{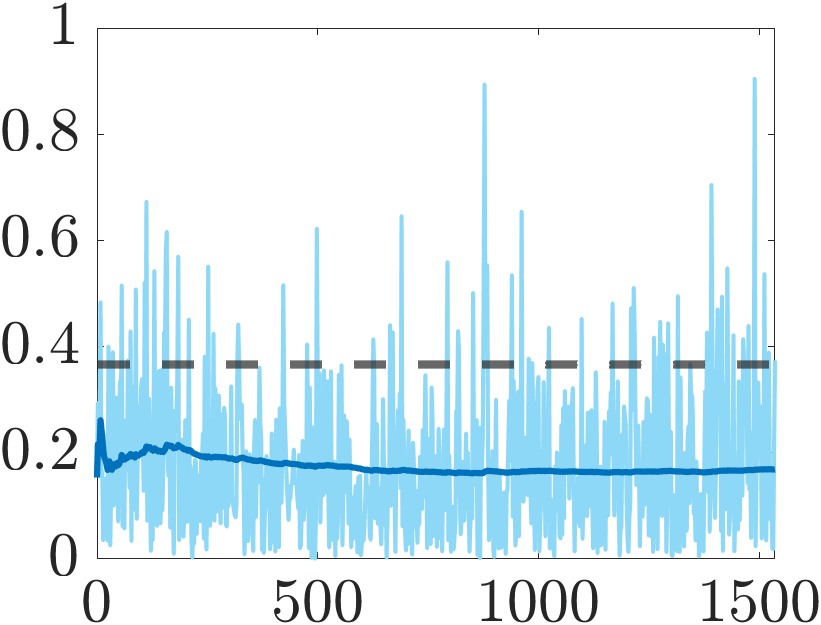}
              \hspace{2cm}
       \\[-.1cm]
       \hspace{-2cm}
       {time (s)}
       \end{tabular}
       &&
              \hspace{-2cm}
        \begin{tabular}{c}
       \includegraphics[width=0.2\textwidth]{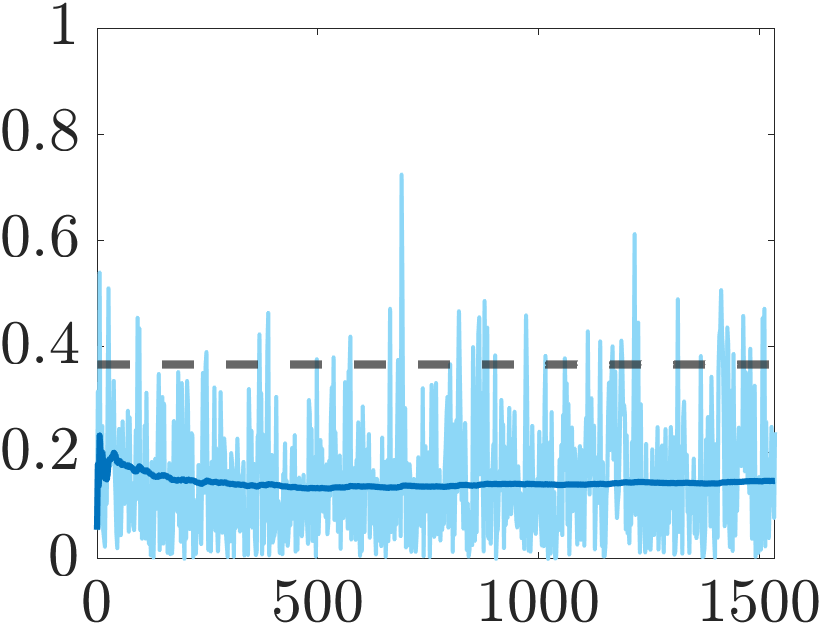}
       \\[-.1cm]
       \hspace{.1cm}
       {time (s)}
       \end{tabular}
       &&
              \hspace{-0cm}
        \begin{tabular}{c}
       \includegraphics[width=0.2\textwidth]{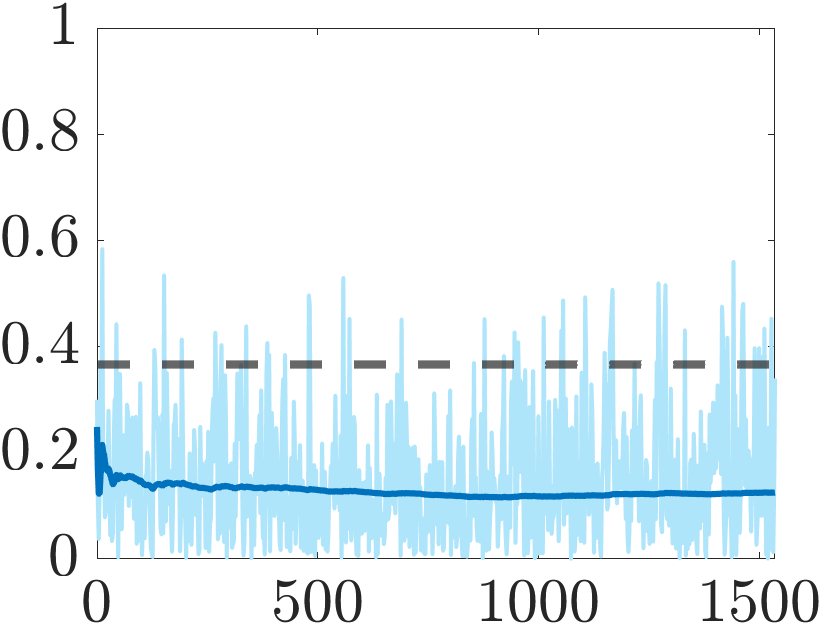}
       \\[-.1cm]
       \hspace{.3cm}
       {time (s)}
       \end{tabular}
       \end{tabular}
       \vspace{-.1cm}
        \caption{{Top row: Temporal variations in the streamwise velocity fluctuation at $(x,z)=(4,0)$ resulting from the reference LES (blue) and Kalman filtering (orange). Bottom row: The normalized error $\mathrm{err}_\mathrm{norm}$ (light blue) and its time-averaged trajectory (dark blue) in estimating streamwise velocity fluctuation at $(x,z)=(4,0)$ as a function of time. (a,e) EKF with scaled white process noise; (b,f) EKF with colored process noise; (c,g) EnKF; and (d,h) UKF. Black dashed lines in the normalized error plots denote the $3\sigma$ error bounds.}}
        \label{fig.EKF-EnKF-UKF-normalizederror}
        \end{center}
\end{figure*}

\begin{figure}
    \begin{tabular}{cc}
    \begin{tabular}{c}
        \rotatebox{90}{$\hat{u}$}
    \end{tabular}
    &
    \hspace{-0.15cm}
    \begin{tabular}{c}
         \includegraphics[width=0.42\textwidth]{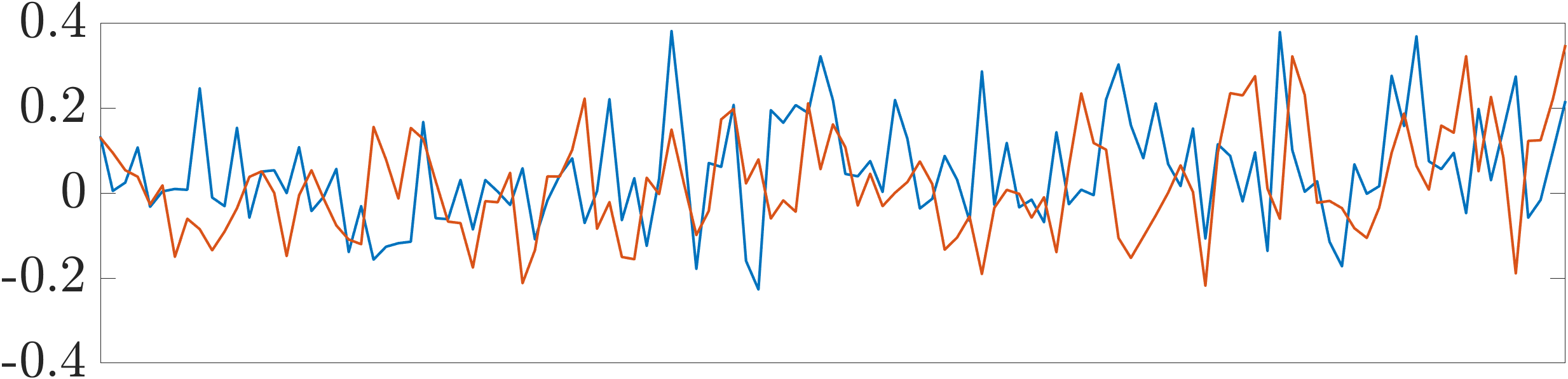}
    \end{tabular}
    \\
    \begin{tabular}{c}
        \rotatebox{90}{$\hat{u}$}
    \end{tabular}
    &
    \hspace{-0.15cm}
    \begin{tabular}{c}
         \includegraphics[width=0.42\textwidth]{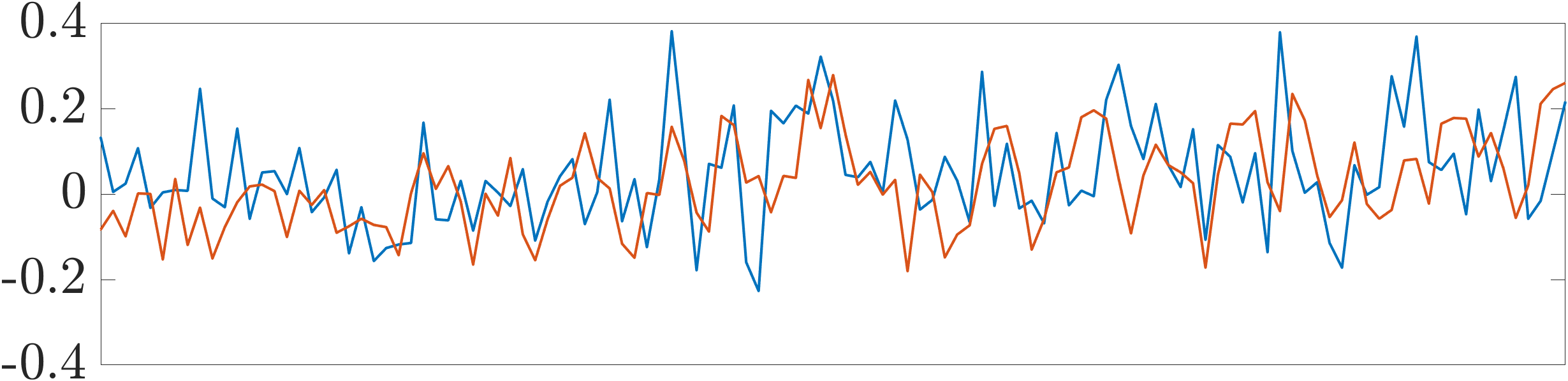}
    \end{tabular}
    \\
    \begin{tabular}{c}
        \rotatebox{90}{$\hat{u}$}
    \end{tabular}
    &
    \hspace{-0.15cm}
    \begin{tabular}{c}
         \includegraphics[width=0.42\textwidth]{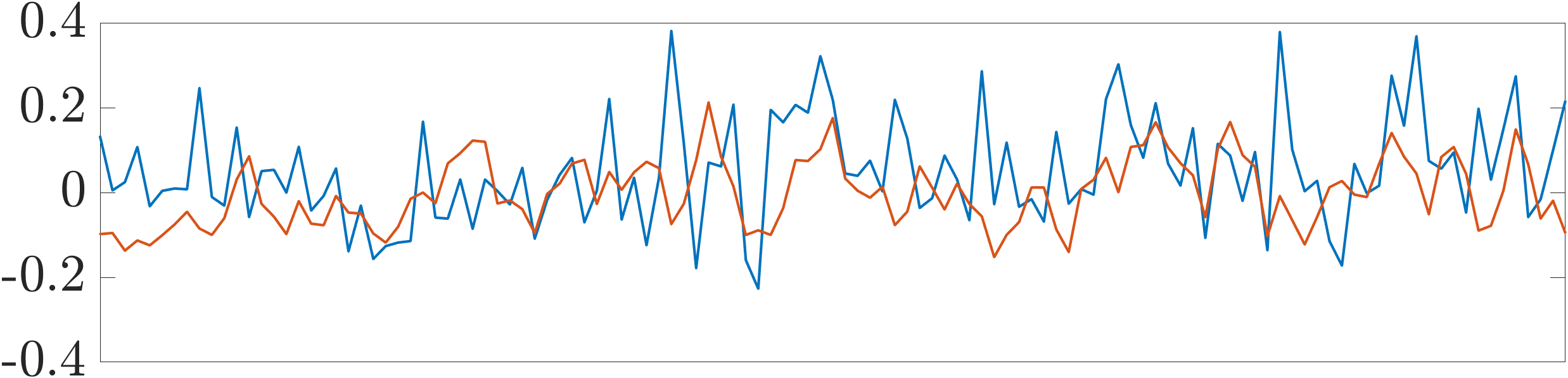}
    \end{tabular}
    \\[.8cm]
    \begin{tabular}{c}
        \rotatebox{90}{$\hat{u}$}
    \end{tabular}
    &
    \hspace{-0.15cm}
    \begin{tabular}{c}
         \includegraphics[width=0.42\textwidth]{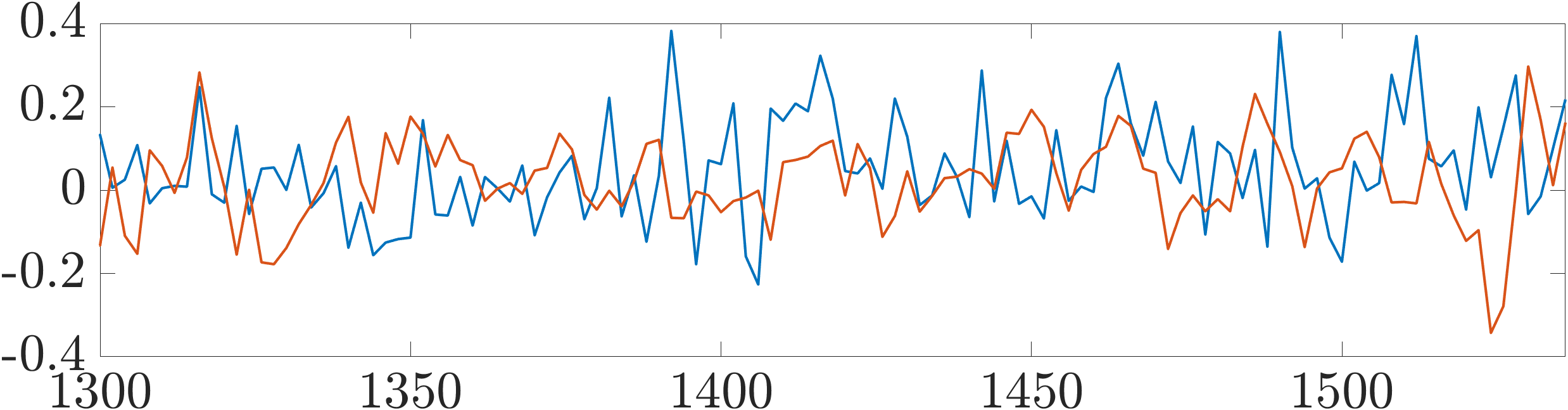}
    \end{tabular}
    \\
       &
       {\small time (sec)}
       \end{tabular}
    \caption{Final 250 seconds of the estimation results presented in Fig.~\ref{fig.EKF-EnKF-UKF-normalizederror}(a-d), with the reference LES streamwise velocity in blue and estimation results in orange.  (a) EKF with scaled white process noise; (b) EKF with colored process noise; (c) EnKF; and (d) UKF.}
    \label{fig.KF_zoomed_inst_vel}
\end{figure}

Focusing on the same point in the wake of the first turbine, i.e., $(x,z) = (4,0)$, Figs.~\ref{fig.EKF-EnKF-UKF-normalizederror}(a-d) compare the time evolution of fluctuation estimates (blue lines) generated by the various filtering algorithms against the result of LES (orange lines). Zoomed-in versions of these plots, which show the estimated velocity over the final 250 seconds of simulation time, are provided in Fig.~\ref{fig.KF_zoomed_inst_vel}. Figures~\ref{fig.EKF-EnKF-UKF-normalizederror}(e-h) show the normalized errors $\mathrm{err}_\mathrm{norm}$ (Eq.~\eqref{eq.NormalizedError}) together with time-averages $\langle \mathrm{err}_\mathrm{norm}\rangle$ and their corresponding  $3\sigma$ error bounds. While these plots demonstrate a clear advantage in using the EnKF and the UKF, they are also indicative of reasonably good estimation levels (with $\langle \mathrm{err}_\mathrm{norm}\rangle \lesssim 0.2$) and quick recovery from sudden changes when using the EKF with both white and colored noise processes.

\begin{figure*}
\begin{center}
\hspace{1cm}
        \begin{tabular}{cccccccc}
        \hspace{-1.7cm}
        \subfigure[]{\label{fig.EKF_relerr_wn}}
        &&
        \hspace{0.5cm}
        \subfigure[]{\label{fig.EKF_relerr}}
        &&
        \hspace{-3.8cm}
        \subfigure[]{\label{fig.EnKF_relerr}}
        &&
        \hspace{0.1cm}
        \subfigure[]{\label{fig.UKF_relerr}}
        &
        \\[-.4cm]
        \hspace{-1.2cm}
	\begin{tabular}{c}
        \vspace{.5cm}
        \rotatebox{90}{$x$}
       \end{tabular}
        &
               \hspace{0.15cm}
	    \begin{tabular}{c}
       \includegraphics[width=0.15\textwidth]{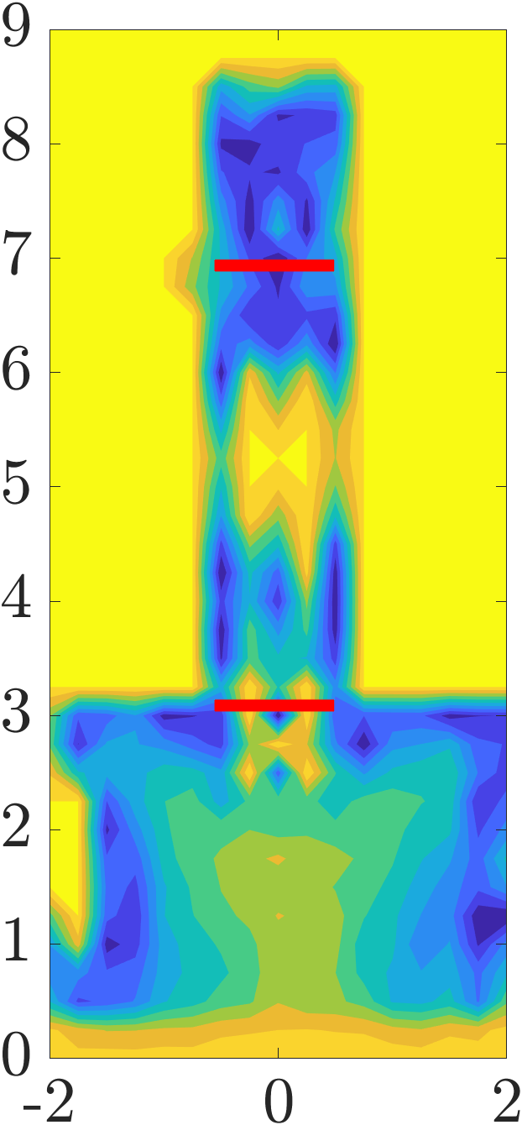}
       \hspace{-.5cm}
       \end{tabular}
       &&
       \hspace{.3cm}
        \begin{tabular}{c}
       \includegraphics[width=0.15\textwidth]{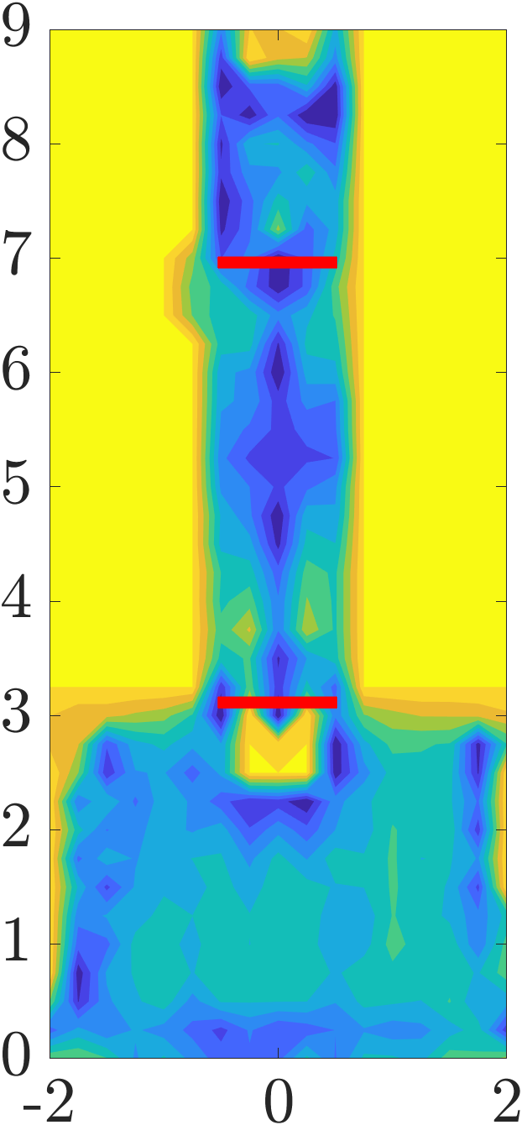}
       \hspace{2cm}
       \end{tabular}
       &&
        \hspace{-1.6cm}
        \begin{tabular}{c}
       \includegraphics[width=0.15\textwidth]{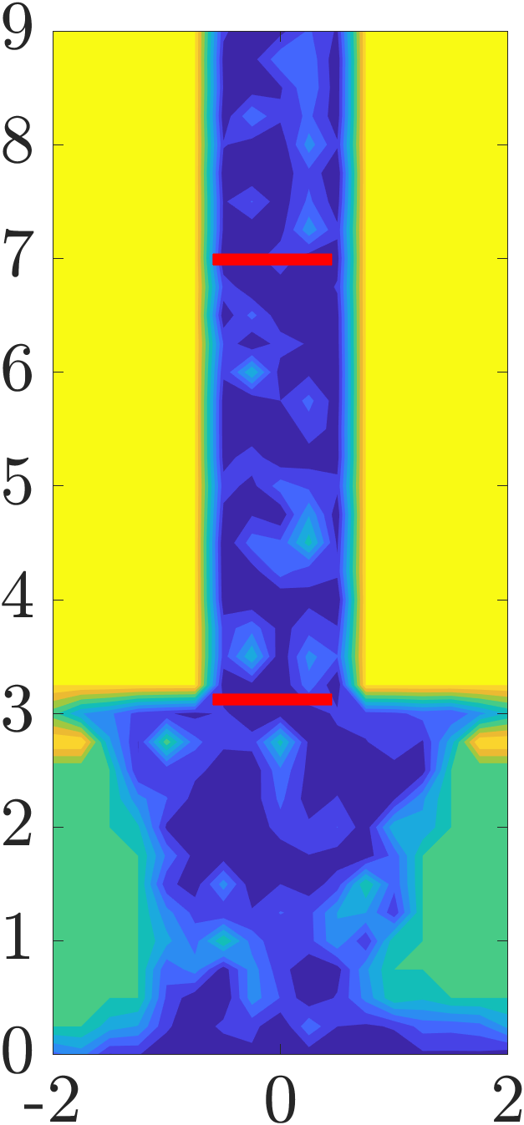}
       \end{tabular}
       &&
        \hspace{.3cm}
        \begin{tabular}{c}
       \includegraphics[width=0.195\textwidth]{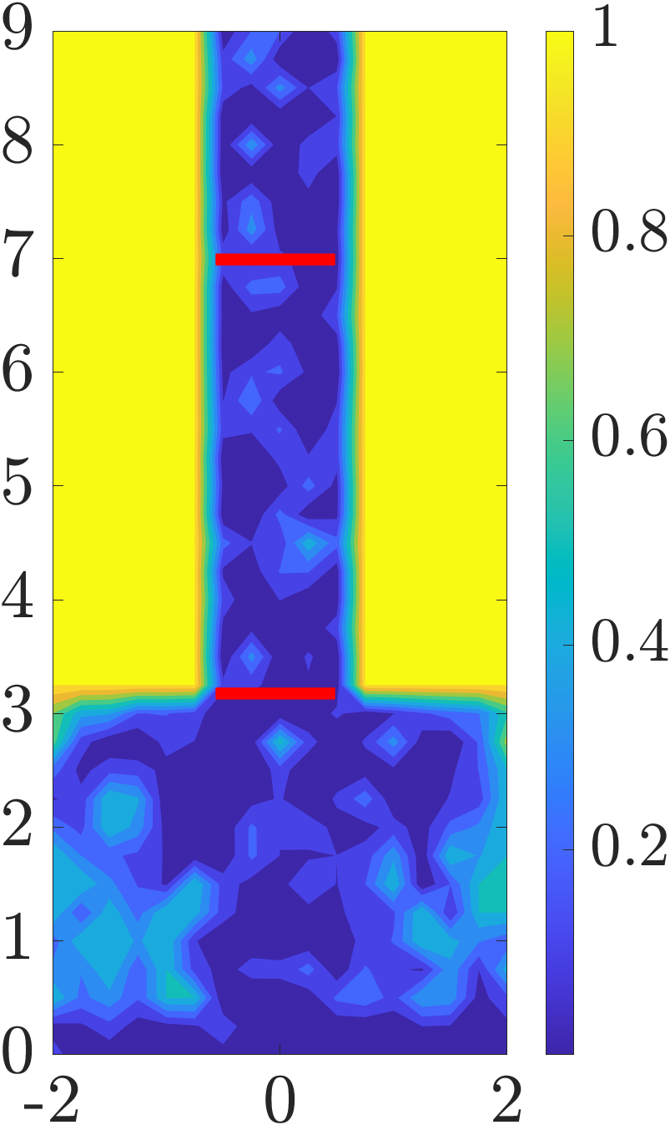}
       \end{tabular}
       \\[-.2cm]
       \hspace{-1.7cm}
       \subfigure[]{\label{fig.EKF_normerr_wn}}
        &&
        \hspace{0.5cm}
        \subfigure[]{\label{fig.EKF_normerr}}
        &&
        \hspace{-3.8cm}
        \subfigure[]{\label{fig.EnKF_normerr}}
        &&
        \hspace{.1cm}
        \subfigure[]{\label{fig.UKF_normerr}}
        &
        \\[-.4cm]
        \hspace{-1.2cm}
	\begin{tabular}{c}
        \vspace{.5cm}
        \rotatebox{90}{$x$}
        \hspace{0cm}
       \end{tabular}
        &
        \hspace{0.15cm}
	    \begin{tabular}{c}
       \includegraphics[width=0.15\textwidth]{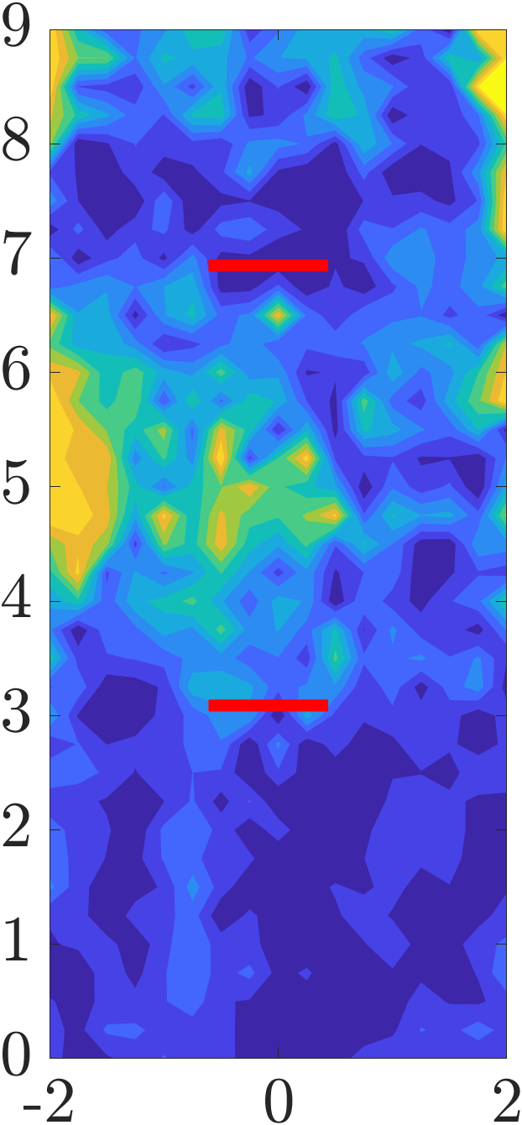}
       \hspace{-.5cm}
       \\[-.1cm]
       \hspace{.6cm}
       {$z$}
       \end{tabular}
       &&
       \hspace{.3cm}
        \begin{tabular}{c}
       \includegraphics[width=0.15\textwidth]{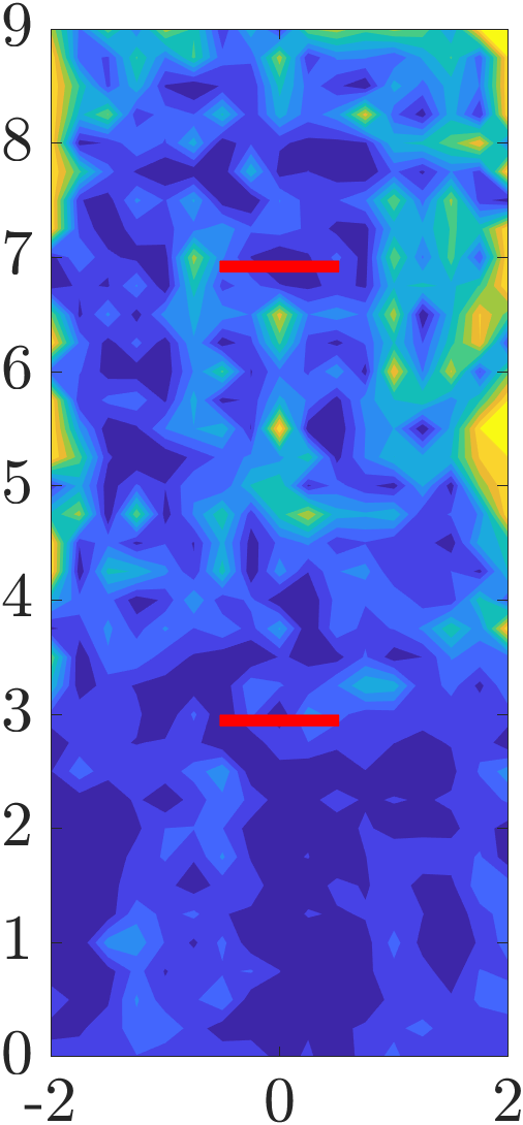}
              \hspace{2cm}
       \\[-.1cm]
       \hspace{-1.9cm}
       {$z$}
       \end{tabular}
       &&
              \hspace{-1.6cm}
        \begin{tabular}{c}
       \includegraphics[width=0.15\textwidth]{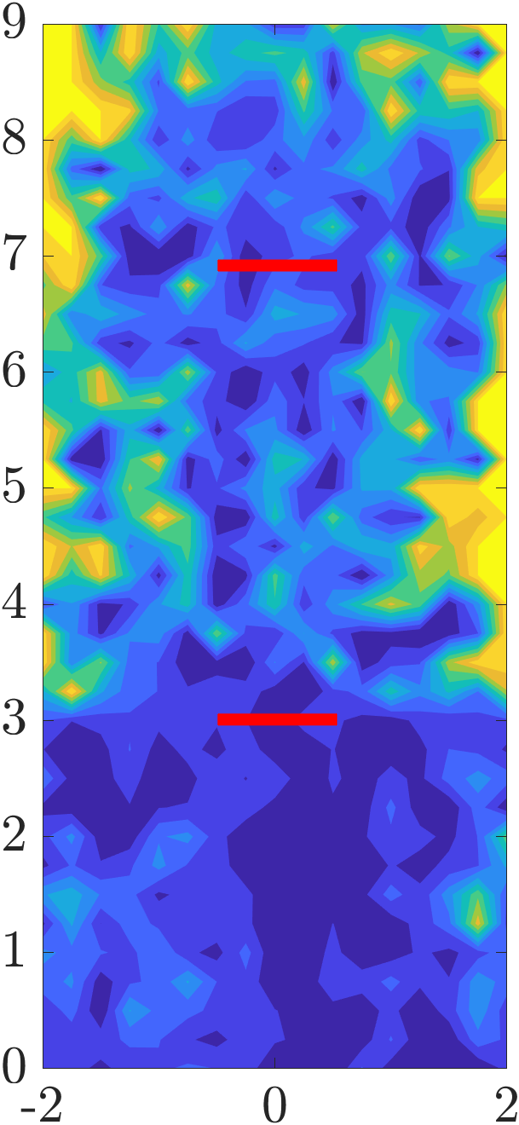}
       \\[-.1cm]
       \hspace{.1cm}
       {$z$}
       \end{tabular}
       &&
              \hspace{.3cm}
        \begin{tabular}{c}
       \includegraphics[width=0.195\textwidth]{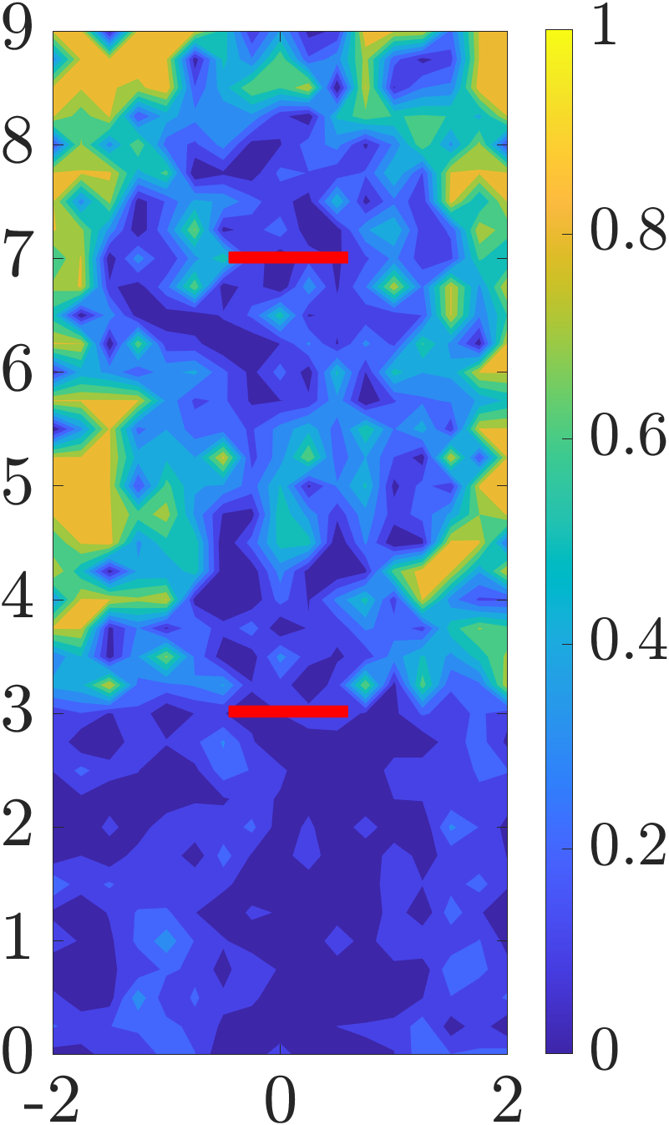}
       \\[-.1cm]
       \hspace{-.8cm}
       {$z$}
       \end{tabular}
       \end{tabular}
       \vspace{-.1cm}
        \caption{Colormaps of relative $\mathrm{err}_\mathrm{rel}$ (top row) and normalized $\mathrm{err}_\mathrm{norm}$ (bottom row) errors achieved by the (a,e) EKF with scaled white process noise, (b,f) EKF with colored process noise, (c,g) EnKF, and (d,h) UKF over the hub-height plane at $t=1536$ sec. Turbine rotors are marked by thick red lines and the wind direction is from bottom to top.}
        \label{fig.comparisons}
        \end{center}
\end{figure*}

It is also important to compare the performance of the Kalman filters in estimating the hub-height velocity field over the entire wind farm domain. Figures~\ref{fig.comparisons}(a-d) compare the statistical error, $\mathrm{err}_\mathrm{rel}$ (Eq.~\eqref{eq.RelativeError}), obtained using the various Kalman filters
for $t=1536$ sec over the 2D domain. While the nonlinear filters (UKF and EnKF) generally outperform the EKF in capturing the velocity variance, the EKF subject to colored process noise performs reasonably well in the wake region behind the first turbine, which is of importance as the velocity field in this region ultimately impacts the second row. It is also noteworthy that the EKF achieves this level of accuracy in significantly less computation time (4 minutes vs approximately 35 minutes and 2 hours for EnKF and UKF, respectively). All the models show higher levels of error downwind of rotor tips and outside of the wake region. This can be attributed to the absence of training points in those regions (cf.\ red dots in Fig.~\ref{fig.LES_LNS_comp}) when generating the prior stochastic model~\eqref{eq.modified-dyn} (Sec.~\ref{sec.Noisemodeling}) and to the less significant contribution of velocity fluctuations beyond the wake region.

Figures~\ref{fig.comparisons}(e-h) show the normalized error, $\mathrm{err}_\mathrm{norm}$ (Eq.~\eqref{eq.NormalizedError}), within the wind field at $t=1536$ sec. Contrary to the relative error, the normalized error is higher in the wake regions. This is because the mean velocity $U$ that normalizes the difference between the estimated streamwise velocity and the result of LES in Eq.~\eqref{eq.NormalizedError} is larger outside of the wake region. While the results are mixed on the sides of the domain, the EnKF and UKF outperform the EKF in the wake region, where fluctuation variance (and turbulence intensity) is relatively more significant. Meanwhile, the EKF with scaled white process noise shows particularly higher levels of error in the wake region behind the first turbine (Fig.~\ref{fig.EKF_normerr_wn}). 

The lack of accuracy in tracking instantaneous velocity fluctuations observed in Figs.~\ref{fig.EKF-EnKF-UKF-normalizederror}, \ref{fig.KF_zoomed_inst_vel}, and \ref{fig.comparisons}(e-h) can be attributed to various factors such as weak observability due to the short length of the time series of the data used for assimilation; see Appendix~\ref{sec.appendixC} for an infinite-time observability analysis. These factors may also be related to linearization of the flow dynamics.
Our results demonstrate the ability of the EKF, EnKF, and UKF in capturing the general trend of variations within regions in which statistical modeling is conducted (i.e., the inflow and wake regions). Because of this, the filters (especially the nonlinear ones, i.e., the EnKF and the UKF), better match velocity statistics (cf.~Figs.~\ref{fig.comparisons}(a-d)) rather than instantaneous changes.

\subsection{{Nacelle-based sign correction}}
\label{sec.signcorr}

{In the Kalman filtering algorithms, the a priori estimate $\hat{v}^-$ is provided by a stochastic dynamical model of the hub-height velocity, which is trained to match second-order statistics of the velocity field, i.e., $\left<u^2\right>$ and $\left<w^2\right>$ (Sec.~\ref{sec.colored-process-noise}). The colored-in-time stochastic process noise is thereby shaped to ensure statistical consistency in steady state. While desirable from a turbulence modeling stand-point, this method does not capture the sign of velocity fluctuations. While this itself motivates the use of Kalman filters for tracking transients of the fluctuation field, it can also result in a posteriori estimates of correct magnitude but incorrect sign (cf.~figure~\ref{fig.KF_zoomed_inst_vel}). Limited to the sensory information provided thus far, we address this issue by leveraging potential correlations of flow quantities in the horizontal direction, which can be justified by the prevalence of coherent motions in wall-bounded flows.
}

\begin{figure*}
        \begin{tabular}{cccc}
        \hspace{-1.2cm}
        \subfigure[]{\label{fig.sign_corr_turb1}}
        &&
        \hspace{0.35cm}
        \subfigure[]{\label{fig.sign_corr_turb2}}
        &
        \\[-.3cm]
        \hspace{-.9cm}
         \vspace{-.6cm}
	\begin{tabular}{c}
            \rotatebox{90}{$z$}
       \end{tabular}
               &
       \hspace{-0.1cm}
       \vspace{0.1cm}
	    \begin{tabular}{c}
    \includegraphics[width=0.35\textwidth]{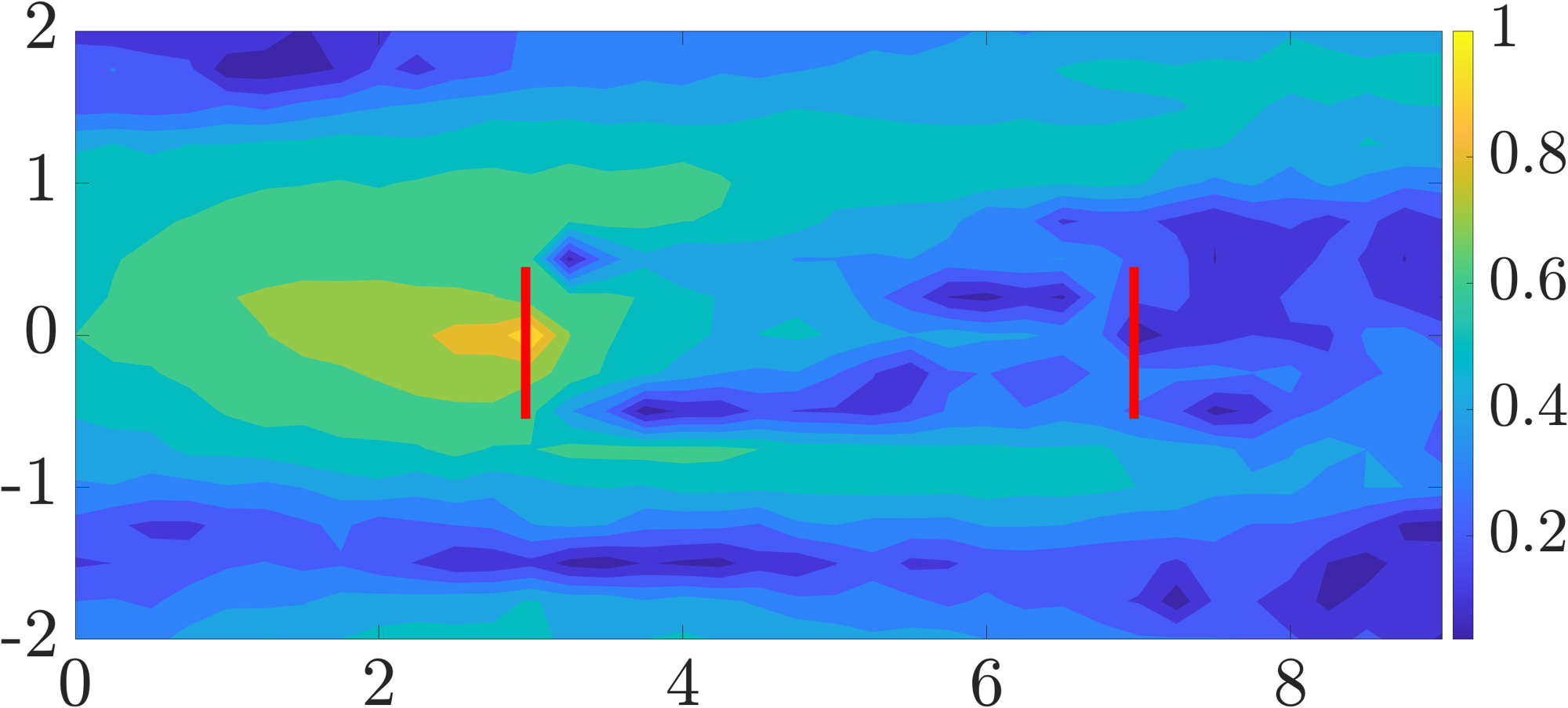}
       \\[0.1cm]
       {\normalsize $x$}
       \end{tabular}
       &
       &
      \hspace{0.6cm}
      \vspace{0.1cm}
        \begin{tabular}{c}
 \includegraphics[width=0.35\textwidth]{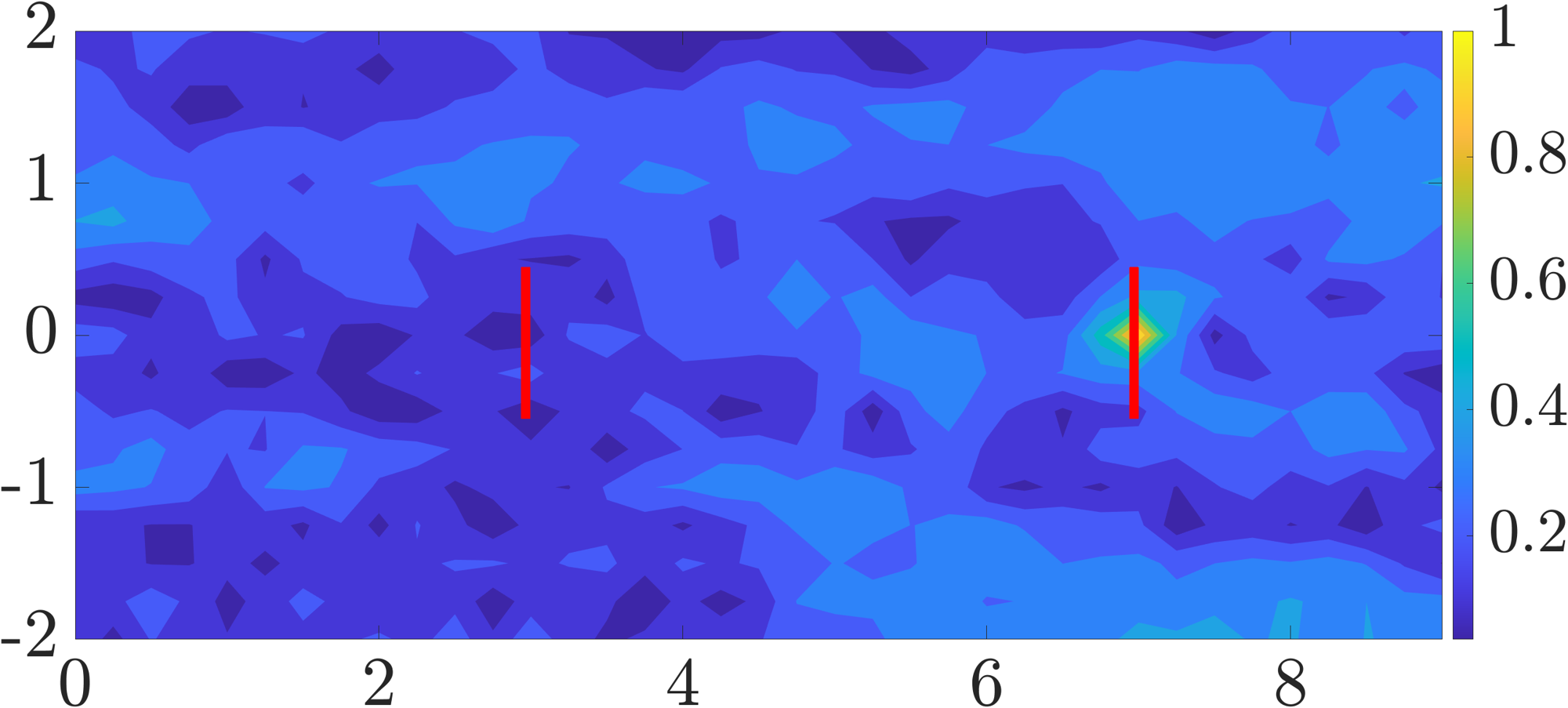}
       \\[0.1cm]
       \hspace{.4cm}
       {$x$}
       \end{tabular}
       \end{tabular}
       \vspace{-.1cm}
        \caption{{Sign correlation $\Phi_\mathrm{sign}$ (Eq.~\eqref{eq.sgncorr}) between the streamwise velocity at various locations across the hub-height plane and the streamwise velocity recorded at the first (a) and second (b) turbines. Thick red lines mark turbine rotors and the wind direction is from left to right.}}
        \label{fig.signcorr}
\end{figure*}

\begin{figure}
\centering
\includegraphics[width=.48\textwidth]{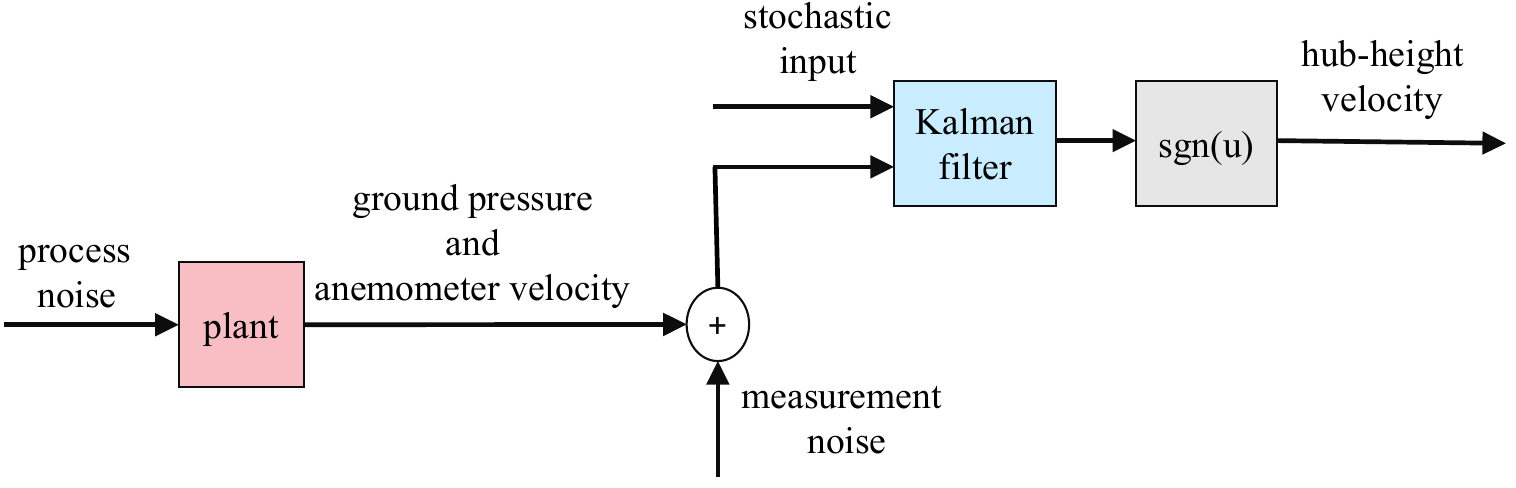}
\vspace{-.1cm}
\caption{{Block diagram illustrating the inclusion of the correlation-based sign synchronization into our proposed data-assimilation strategy.}}
\label{fig.model_KF_blockdiagram_sgn}
\end{figure}

{The correlation between the sign of fluctuations recorded by the anemometer mounted on the nacelle of each turbine and the hub-height velocity at other points can be quantified using the temporal expectation
\begin{align}
	\label{eq.sgncorr}
	\Phi_\mathrm{sign}(x,z) \;\DefinedAs\; 
 \langle \mathrm{sgn}(u_\mathrm{nacelle}(t)) \,\mathrm{sgn}(u(x,z,t)) \rangle,
\end{align}
where $\mathrm{sgn}(\cdot)$ is the sign function and $u_\mathrm{nacelle}$ is the streamwise velocity recorded at the nacelle. We evaluate this potential correlation using time-resolved data from LES. Figure~\ref{fig.sign_corr_turb1} shows that the sign of velocity fluctuations recorded at the nacelle of the leading turbine is correlated with the sign of fluctuations at nearby points, particularly upwind, where the flow is not affected by the wake. In contrast, Fig.~\ref{fig.sign_corr_turb2} shows that due to turbulent mixing in the wake region, the same levels of correlation do not exist for the sign of velocity fluctuations recorded at the second turbine. Based on this observation, we apply a correction to the sign of a posteriori velocity estimates in region where $\Phi_\mathrm{sign} > 0.8$. We note that this correction is performed outside the Kalman filtering algorithm and does not change the evolution of the Kalman filter; see Fig.~\ref{fig.model_KF_blockdiagram_sgn}. Introducing such information (i.e., sign of fluctuations) into the filtering algorithms, which would affect the trajectory of the Kalman filter, requires theoretical developments that are beyond the scope of the current work.}

\begin{figure*}
\begin{center}
        \begin{tabular}{cccc}
        \hspace{-0.3cm}
        \subfigure[]{\label{fig.EKF_vel_nocorr}}
        &&
        \hspace{0.45cm}
        \subfigure[]{\label{fig.EnKF_vel_nocorr}}
        &
        \\[-.45cm]
        \hspace{-0.1cm}
	\begin{tabular}{c}
        \vspace{.5cm}
        \rotatebox{90}{$\hat{u}$}
       \end{tabular}
       &
       \hspace{0cm}
	    \begin{tabular}{c}
\includegraphics[width=0.45\textwidth]{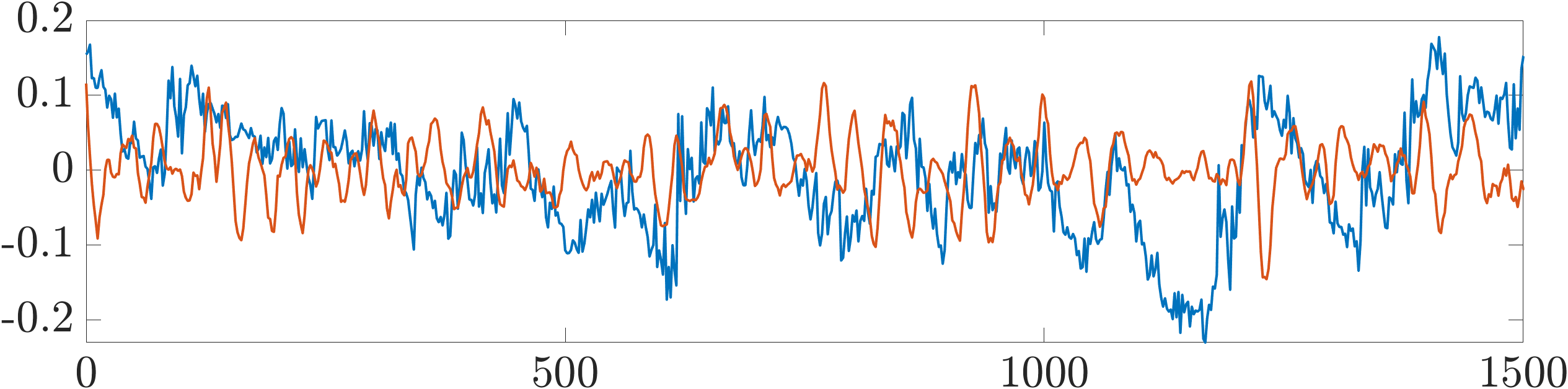}
       \end{tabular}
       &&
       \hspace{.2cm}
        \begin{tabular}{c}
       \includegraphics[width=0.45\textwidth]{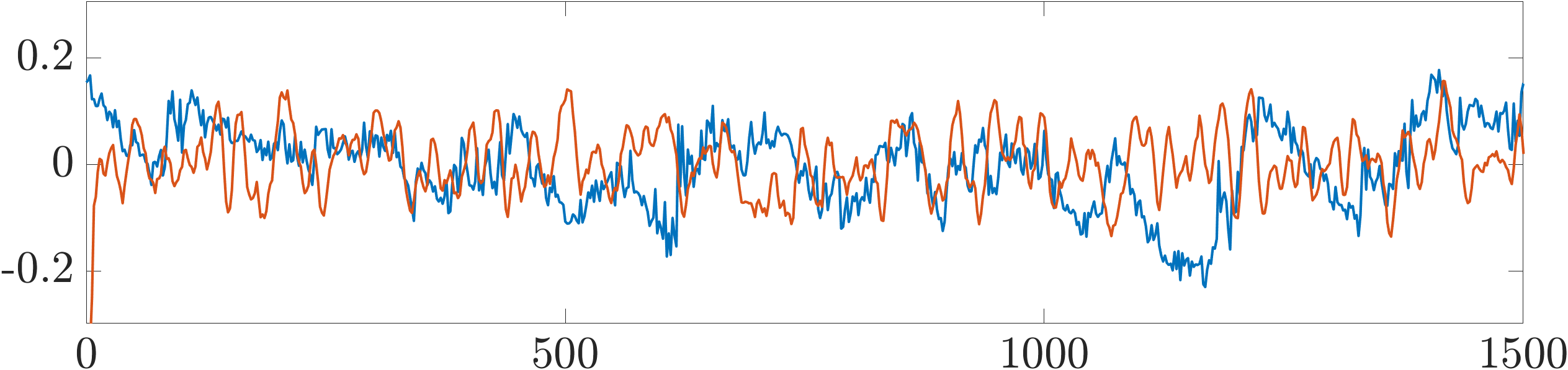}
       \end{tabular}
       \end{tabular}
       \\[-0.1cm]
       \begin{tabular}{cccc}
        \hspace{-0.3cm}
        \subfigure[]{\label{fig.EKF_vel_corr}}
        &&
        \hspace{0.45cm}
        \subfigure[]{\label{fig.EnKF_vel_corr}}
        &
        \\[-.45cm]
        \hspace{-0.1cm}
	\begin{tabular}{c}
        \vspace{.5cm}
        \rotatebox{90}{$\hat{u}$}
       \end{tabular}
       &
       \hspace{0cm}
	    \begin{tabular}{c}
       \includegraphics[width=0.45\textwidth]{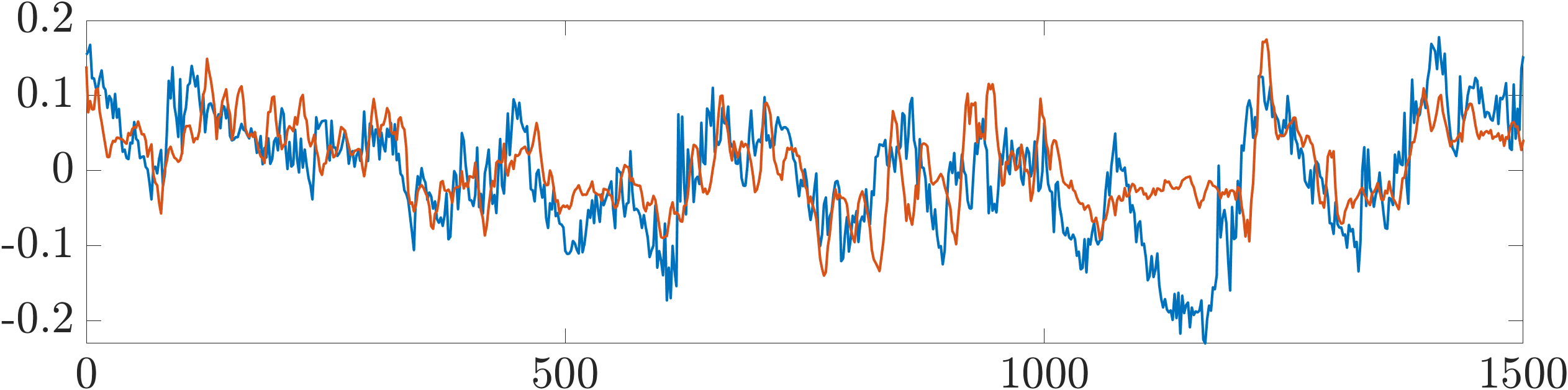}
       \\[-.1cm]
       \hspace{0.4cm}
       {time (sec)}
       \end{tabular}
       &&
       \hspace{0.2cm}
        \begin{tabular}{c}
       \includegraphics[width=0.45\textwidth]{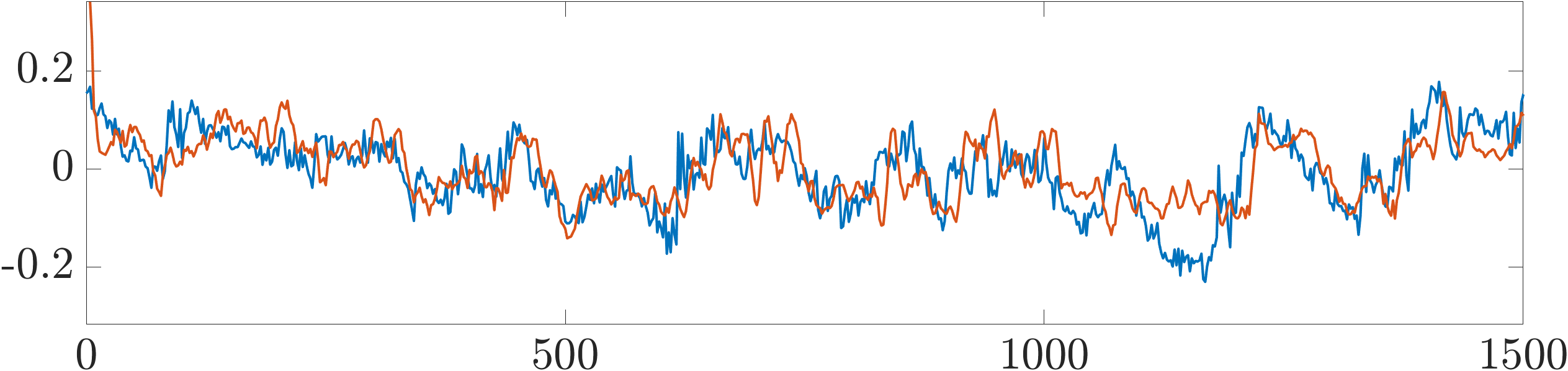}
       \\[-.1cm]
       \hspace{0.4cm}
       {time (sec)}
       \end{tabular}
       \end{tabular}
       \vspace{-.1cm}
        \caption{{Streamwise velocity fluctuations estimated at $(x,z)=(2,0)$ by the (a,c) EKF and the (b,d) EnKF before (first row) and after (second row) nacelle-based sign corrections. The results of the LES and filtering are shown in blue and orange, respectively.}}
        \label{fig.vel_est_upwind}
\end{center}
\end{figure*}

\begin{figure}
    \centering
    \begin{tabular}{cc}
    \begin{tabular}{c}
    \hspace{-0.25cm}
       {$\hat{p}$}
    \end{tabular}
    &
    \hspace{-0.35cm}
    \begin{tabular}{c}
        \includegraphics[width=0.45\textwidth]{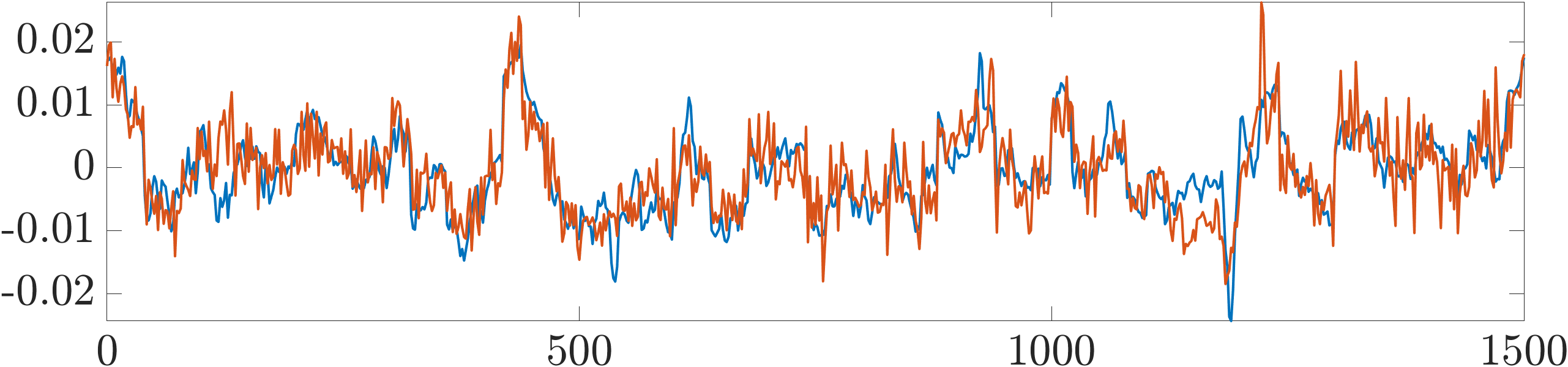}
    \end{tabular}
    \\[0.2cm]
       &
       {time (sec)}
       \end{tabular}
    \caption{{The projected pressure from the hub height to the ground (blue) plotted against the LES-based ground pressure (orange) at $(x,z)=(2,0)$.}}
    \label{fig.projection_comparison}
\end{figure}

{The significant sign correlation observed upwind of the first turbine (Fig.~\ref{fig.sign_corr_turb1}) is of particular interest due to its potential benefit for estimation-based preview control. Figure~\ref{fig.vel_est_upwind} compares the quality of estimation at $(x, z)=(2,0)$, i.e., one diameter upwind of the leading turbine, using the EKF and EnKF with and without correlation-based corrections to the estimated velocity. It is evident that the proposed correction significantly improves the ability of both Kalman filters to track variations in the streamwise velocity. In spite of the improved estimation, a region of high error remains untouched for $1100<t<1200$, which originates from an undesirable inference of ground pressure based on that of the hub-height (Sec.~\ref{sec.LSE}); see fig.~\ref{fig.projection_comparison}.}

{
We next examine the performance of the EKF with colored process noise in estimating the velocity fluctuations at the hub-height of turbines that are yawed against the direction of incoming wind.
In spite of the slight superiority of nonlinear filters, we will focus on the computationally efficient EKF estimator with colored noise process in the remainder of the study. We also refrain from further evaluation of the EKF with white process noise as it fails to provide reasonable estimations of the instantaneous velocity field downwind of the turbine rotors.}

\subsection{Yawed wind farm}
\begin{figure*}
\begin{center}
        \begin{tabular}{cccc}
        \hspace{-1.3cm}
        \subfigure[]{\label{fig.LESmap_yaw15}}
        &&
        \hspace{0.45cm}
        \subfigure[]{\label{fig.LNSmap_yaw15}}
        &
        \\[-.45cm]
        \hspace{-0.9cm}
	\begin{tabular}{c}
        \vspace{.5cm}
        \rotatebox{90}{$z$}
       \end{tabular}
       &
       \hspace{-0.2cm}
	    \begin{tabular}{c}
\includegraphics[width=0.35\textwidth]{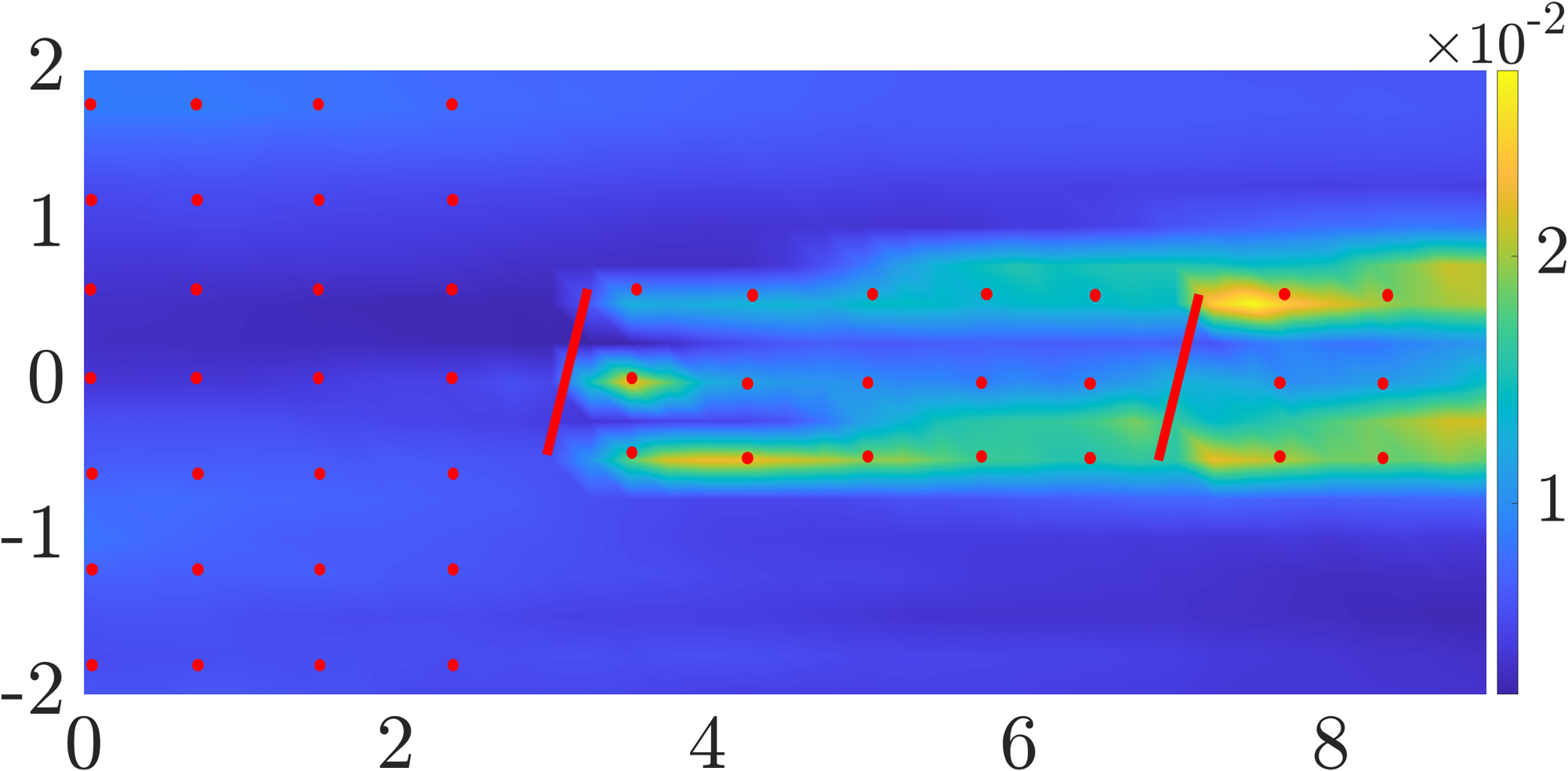}
       \end{tabular}
       &&
       \hspace{.2cm}
        \begin{tabular}{c}
       \includegraphics[width=0.35\textwidth]{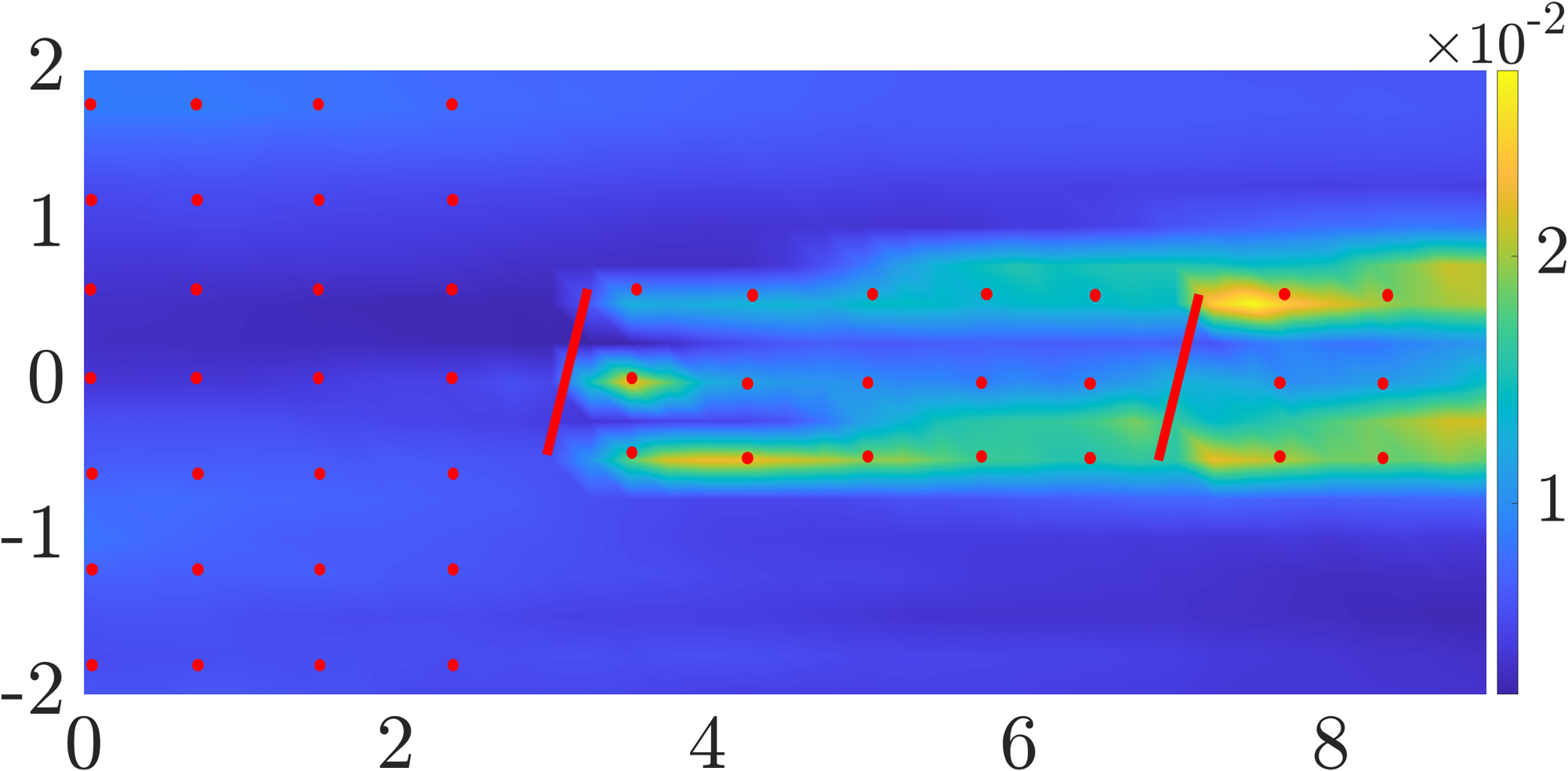}
       \end{tabular}
       \end{tabular}
       \\[-0.1cm]
       \begin{tabular}{cccc}
        \hspace{-1.3cm}
        \subfigure[]{\label{fig.wwLESmap_yaw15}}
        &&
        \hspace{0.45cm}
        \subfigure[]{\label{fig.wwLNSmap_yaw15}}
        &
        \\[-.45cm]
        \hspace{-0.9cm}
	\begin{tabular}{c}
        \vspace{.5cm}
        \rotatebox{90}{$z$}
       \end{tabular}
       &
       \hspace{-0.2cm}
	    \begin{tabular}{c}
       \includegraphics[width=0.35\textwidth]{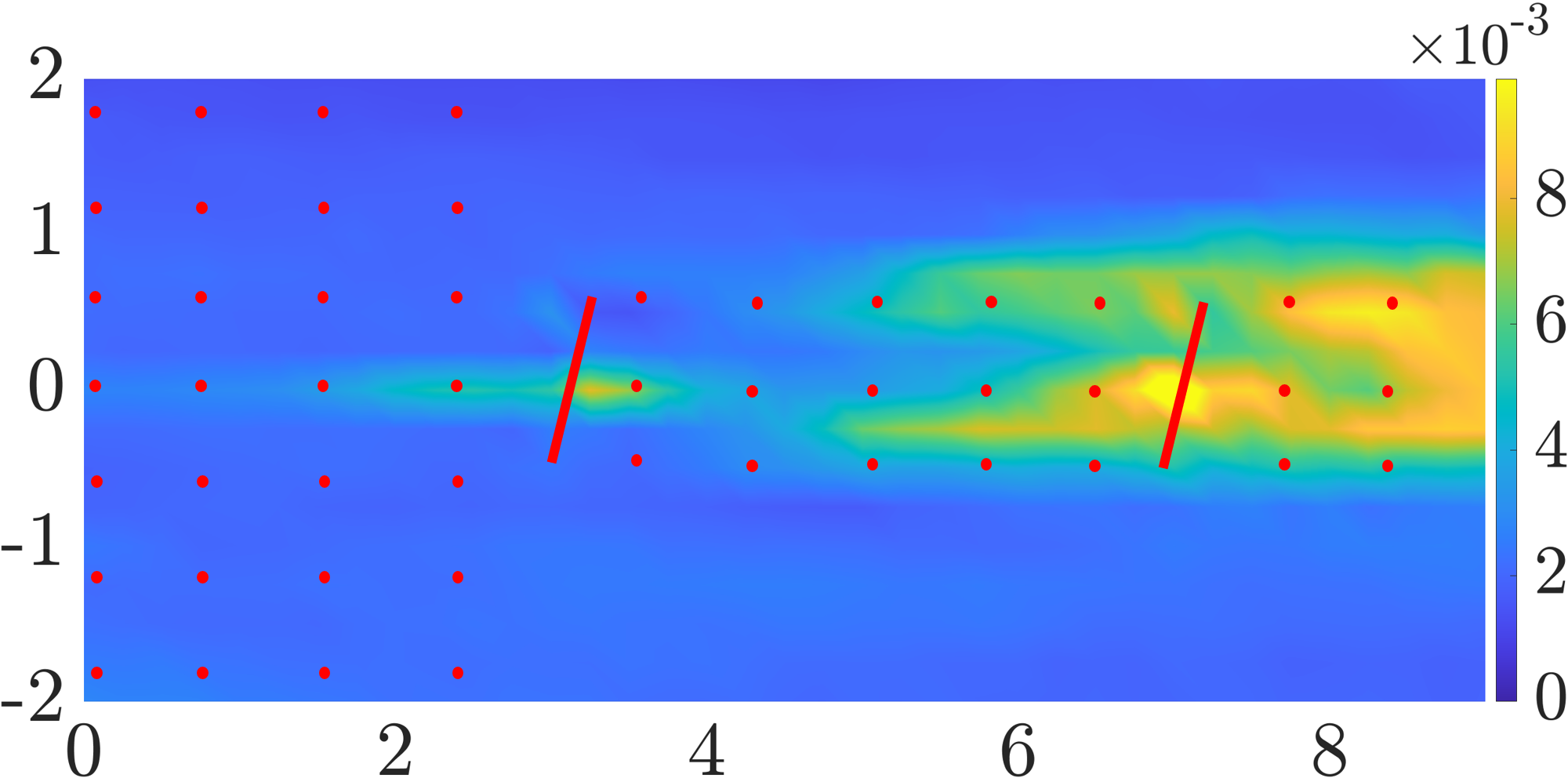}
       \\[-.1cm]
       \hspace{0.4cm}
       {$x$}
       \end{tabular}
       &&
       \hspace{0.2cm}
        \begin{tabular}{c}
       \includegraphics[width=0.35\textwidth]{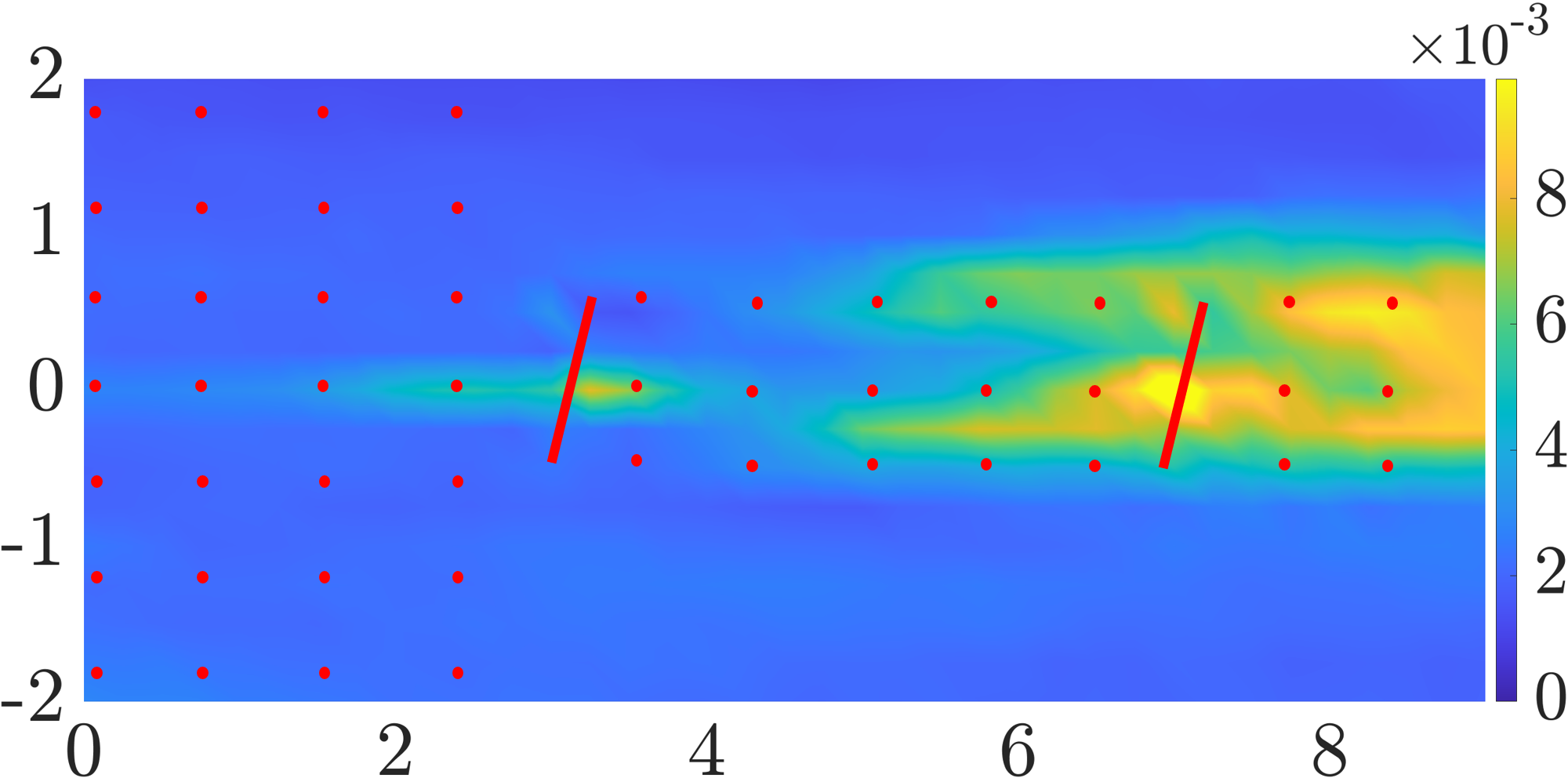}
       \\[-.1cm]
       \hspace{0.4cm}
       {$x$}
       \end{tabular}
       \end{tabular}
        \caption{Streamwise (top) and spanwise (bottom) velocity variances computed from (a,c) LES and (b,d) model~\eqref{eq.modified-dyn} for $15^\degree$ yaw angle. Red dots denote the spatial location of velocity correlations used in training the stochastic model and thick red lines mark turbine rotors. The wind direction is from left to right.}
        \label{fig.LES_LNS_comp_yaw15}
\end{center}
\end{figure*}

Yaw misalignment with respect to the free stream occurs either intentionally, with the aim of maximizing power production, or due to atmospheric variability. It is thus important to estimate variations in the fluctuating velocity field even when the wind is not perfectly perpendicular to the rotor disks. Herein, we examine the performance of the EKF with colored process noise in capturing wind variations under such conditions. To examine the necessity for re-training the stochastic prior model after a yaw misalignment has occurred, we consider two types of models that differ in the dynamic modifier ($-B K_f$) of the linearized NS equations (cf.~construction of $A_f$ in Eq.~\eqref{eq.modified-dyn}):
\bi
    \item[]Type A: the dynamic modifier $-B K_f$ is generated using a statistical dataset that is collected from an LES of non-yawed wind turbines;
    \item[]Type B: the dynamic modifier $-B K_f$ is generated using a statistical dataset that is collected from an LES of uniformly yawed wind turbines.
\ei
Given that the linearization of the NS equations would happen around an appropriately yawed base flow profile ($\gamma\neq 0$), Type B models should result in a more accurate estimation. For a uniformly yawed turbine configuration with $\gamma=15^\degree$, Fig.~\ref{fig.LES_LNS_comp_yaw15} demonstrates the streamwise and spanwise variances resulting from model~\eqref{eq.modified-dyn} after the dynamic modification has been trained to achieve second-order statistical consistency with LES at the spatial locations highlighted by the red dots. Robustness features that are attributed to the physics-based nature of the underlying dynamics~\cite{rodburbhaleozarACC23}, however, provide a plausible argument for dynamically modifying the linearized equations with an a priori computed $-B K_f$ term that corresponds to a non-yawed setup (cf. Fig.~\ref{fig.LES_LNS_comp}) instead of the yawed setup, especially for small yaw angles $\gamma$.

\begin{figure*}
\begin{center}
\hspace{5.4cm}
        \begin{tabular}{cccccc}
        \hspace{-5.5cm}
        \subfigure[]{\label{fig.EKF_relerr_yaw15}}
        &\hspace{0.6cm}{Type A}& \hspace{0cm}{Type B}&
        \hspace{.8cm}
        \subfigure[]{\label{fig.EKF_relerr_yaw30}}
        &\hspace{0.7cm}{Type A}&\hspace{-1cm}{Type B}
        \\[-.1cm]
        \hspace{-5.5cm}
	\begin{tabular}{c}
        \vspace{.5cm}
        \rotatebox{90}{$x$}
                \hspace{-0.9cm}
       \end{tabular}
        &
       \hspace{.3cm}
	    \begin{tabular}{c}
\includegraphics[width=0.15\textwidth]{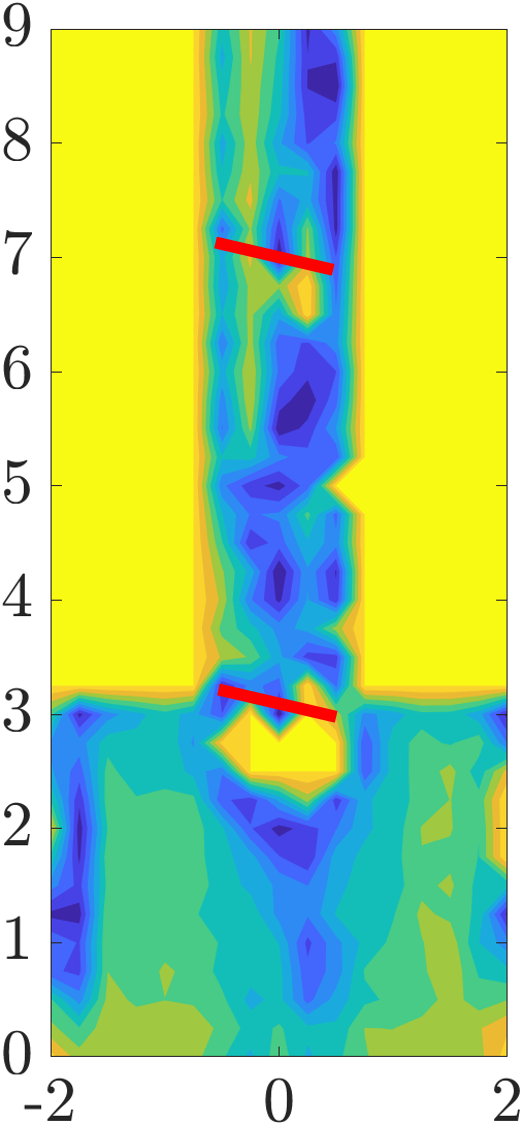}
       \end{tabular}
       &
       \hspace{-0.3cm}
        \begin{tabular}{c}
       \includegraphics[width=0.15\textwidth]{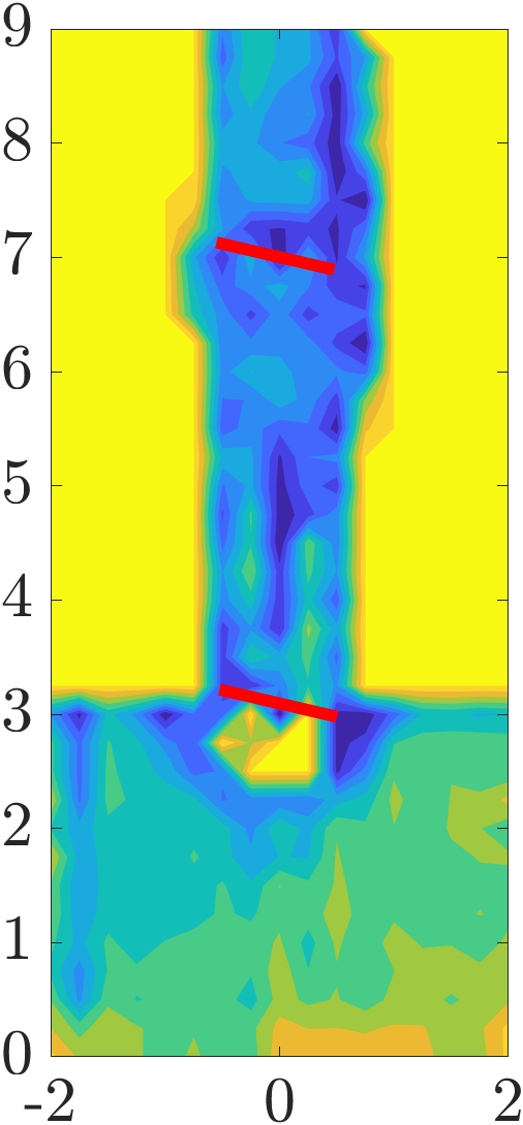}
       \end{tabular}
       &&
        \hspace{0.3cm}
        \begin{tabular}{c}
\includegraphics[width=0.15\textwidth]{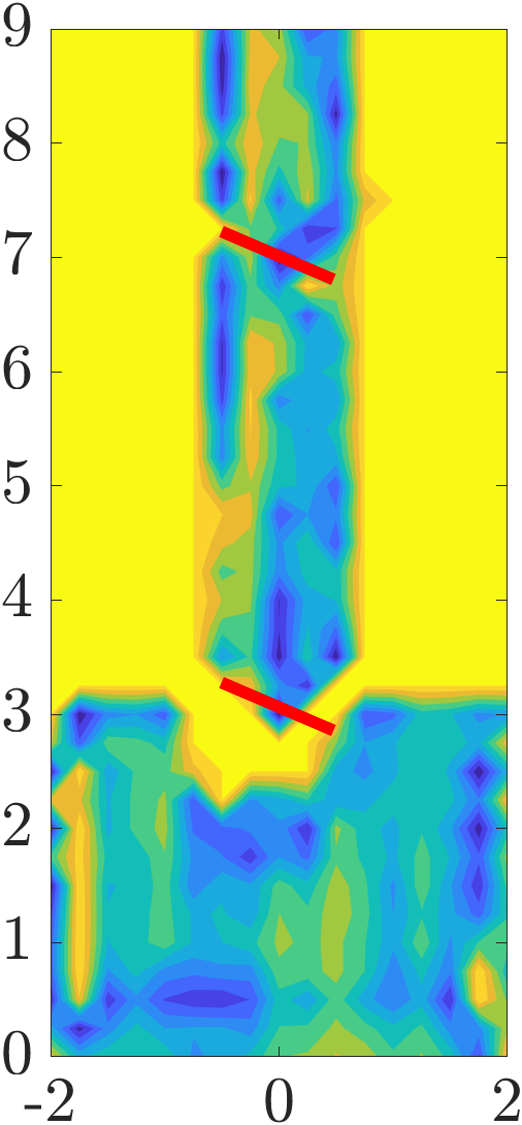}
       \end{tabular}
       &
        \hspace{-0.3cm}
        \begin{tabular}{c}
\includegraphics[width=0.196\textwidth]{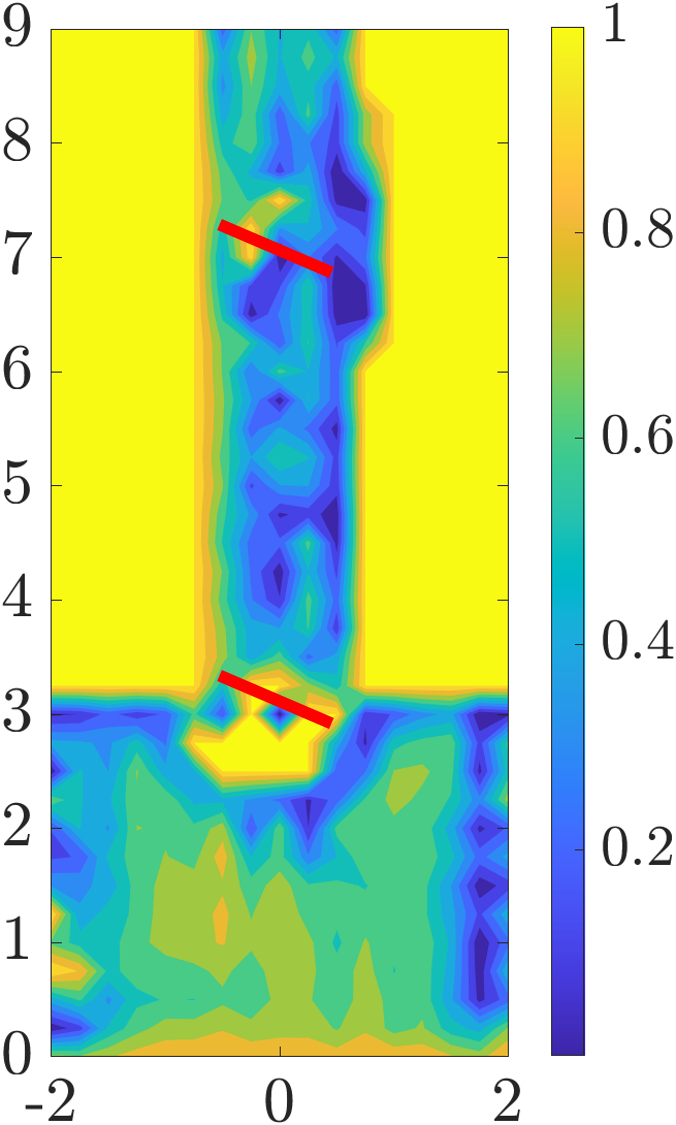}
       \end{tabular}
       \\[-.2cm]
       \hspace{-5.5cm}
        \subfigure[]{\label{fig.EKF_inst_err_yaw15}}
        &&&
        \hspace{.8cm}
        \subfigure[]{\label{fig.EKF_inst_err_yaw30}}
        &&
        \\[-.4cm]
        \hspace{-5.5cm}
	\begin{tabular}{c}
        \vspace{.5cm}
        \rotatebox{90}{$x$}
         \hspace{-0.9cm}
       \end{tabular}
        &
       \hspace{.3cm}
	    \begin{tabular}{c}
       \includegraphics[width=0.15\textwidth]{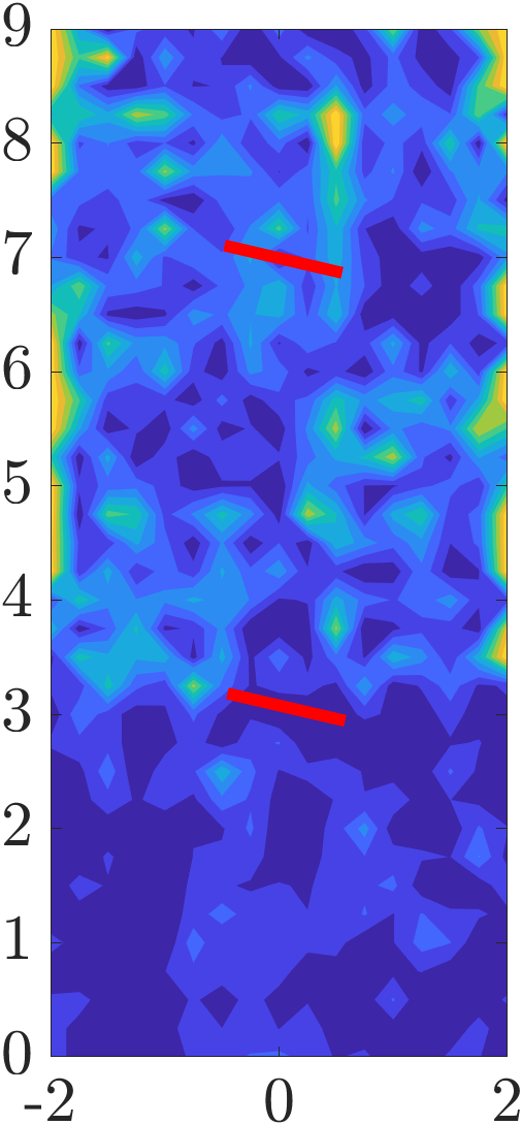}
       \\[-.1cm]
       \hspace{0cm}
       {$z$}
       \end{tabular}
       &
       \hspace{-0.3cm}
        \begin{tabular}{c}
       \includegraphics[width=0.15\textwidth]{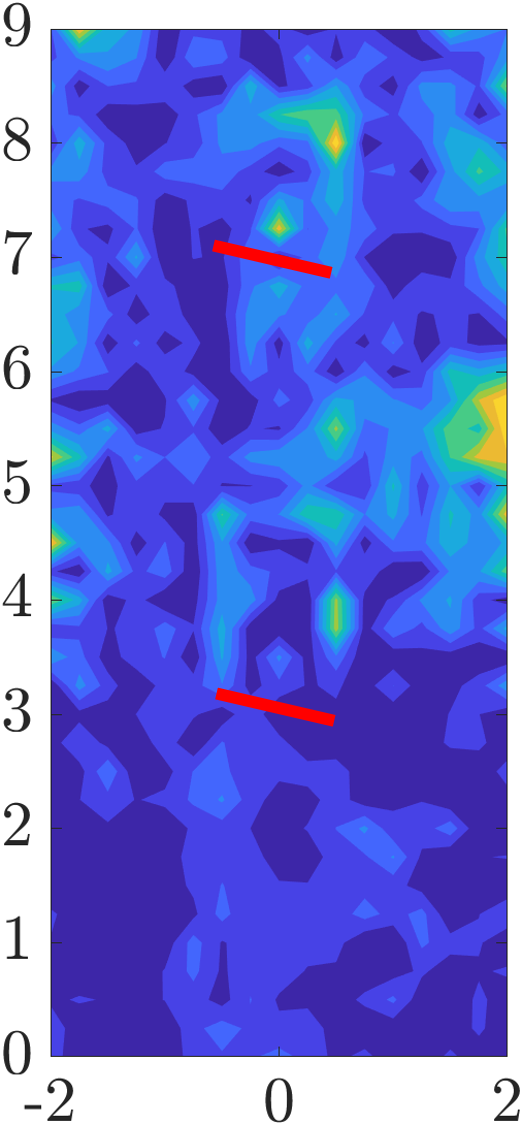}
       \\[-.1cm]
       \hspace{0cm}
       {$z$}
       \end{tabular}
       &&
              \hspace{0.3cm}
        \begin{tabular}{c}
       \includegraphics[width=0.15\textwidth]{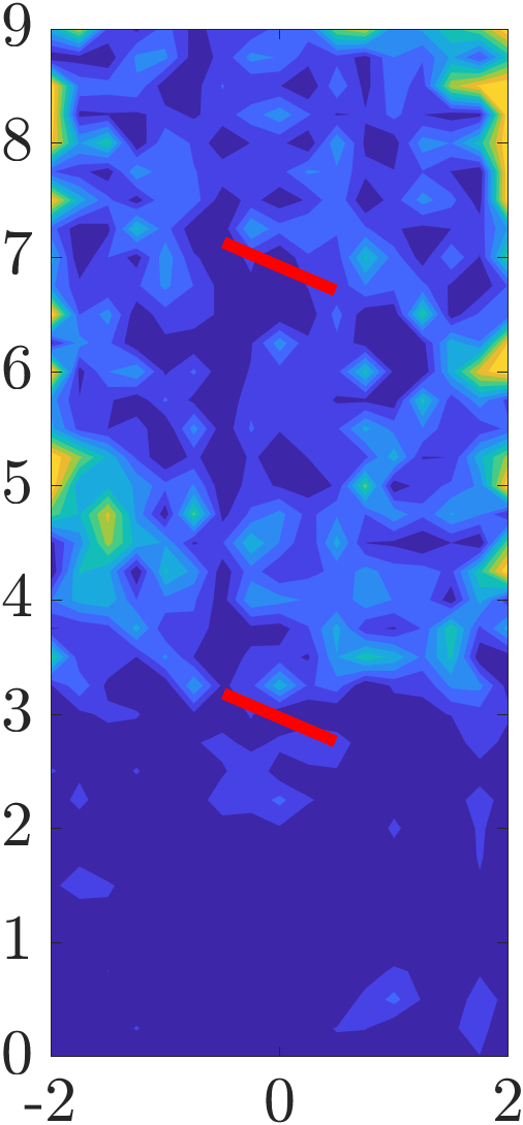}
       \\[-.1cm]
       \hspace{0.1cm}
       {$z$}
       \end{tabular}
       &
        \hspace{-0.3cm}
        \begin{tabular}{c}
       \includegraphics[width=0.195\textwidth]{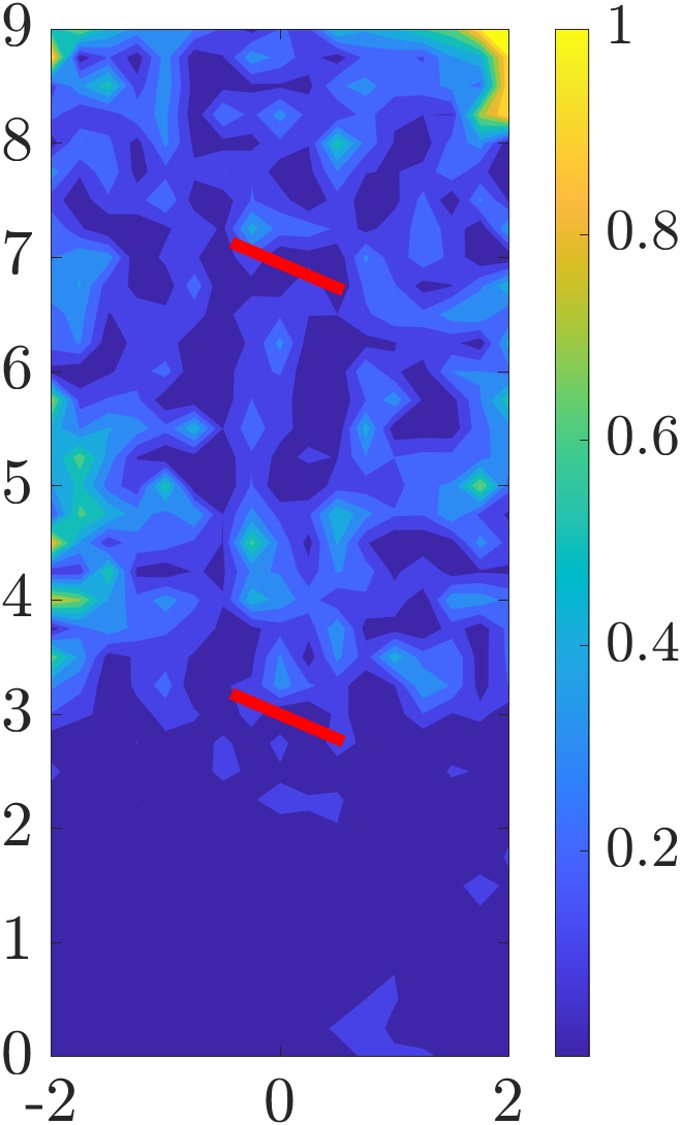}
       \\[-.1cm]
       \hspace{-.7cm}
       {$z$}
       \end{tabular}
       \end{tabular}
       \vspace{-.1cm}
        \caption{Colormaps of relative $\mathrm{err}_\mathrm{rel}$ (top row) and normalized $\mathrm{err}_\mathrm{norm}$ (bottom row) errors achieved by the EKF over the hub-height plane at $t=1536$ sec for (a,c) $15^\degree$ and (b,d) $30^\degree$ uniform yaw angles. In each sub-figure, the left plots correspond to the use of Type A prior models and the right plots correspond to the use of Type B prior models. The red lines mark turbine {rotors} and the wind blows from bottom to top.}
        \label{fig.yaw_comparisons}
        \end{center}
\end{figure*}

\begin{figure*}
\begin{center}
        \begin{tabular}{cccc}
        \hspace{-1.2cm}
        \subfigure[]{\label{fig.rank_gamma}}
        &&
        \hspace{0.35cm}
        \subfigure[]{\label{fig.performance}}
        &
        \\[-.45cm]
        \hspace{-.9cm}
	\begin{tabular}{c}
        \vspace{.5cm}
        \rotatebox{90}{\normalsize number of sensors}
       \end{tabular}
               &
       \hspace{-0.1cm}
	    \begin{tabular}{c}
    \includegraphics[width=0.35\textwidth]{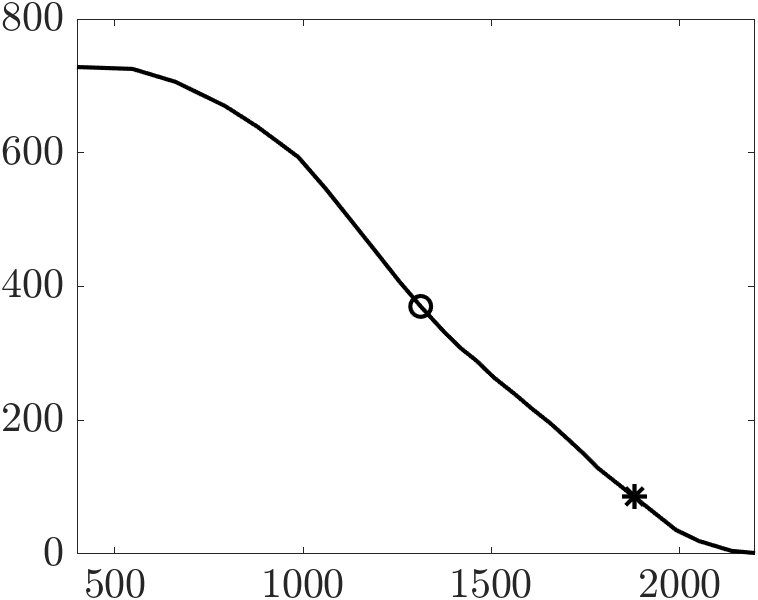}
       \\[0.1cm]
       \hspace{-.4cm}
       {\normalsize $\beta$}
       \end{tabular}
       &\hspace{0.6cm}
    \begin{tabular}{c}
        \vspace{.5cm}
        \rotatebox{90}{\normalsize $(f-f_c)/f_c$}
       \end{tabular}&
       \hspace{-.15cm}
        \begin{tabular}{c}
 \includegraphics[width=0.35\textwidth]{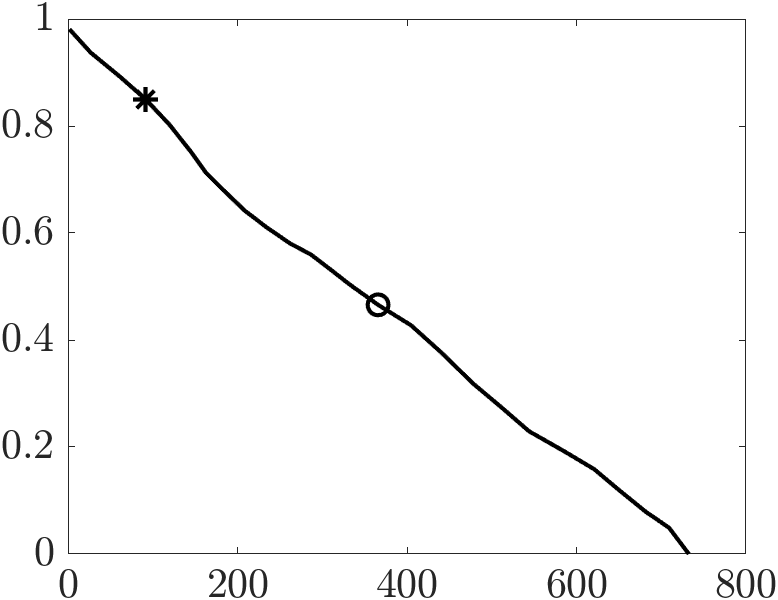}
       \\[0.1cm]
       \hspace{.4cm}
       {\normalsize number of sensors}
       \end{tabular}
       \end{tabular}
       \vspace{-.1cm}
        \caption{{(a) Number of retained sensors as a function of $\beta$; and (b) performance loss vs the number of retained sensors in estimating the entire 2D velocity field. Here, $f\DefinedAs \trace(V_d X\,+\, X\,L\,R\,L^T)$ is the performance index resulting from the solution of problem~\eqref{eq.sensel} and $f_c$ corresponds to the attainable performance when all sensors are available ($\beta=0$). The $\circ$ and $\ast$ symbols on both plots correspond to cases with $\beta=1312$ and $1879$, which retain $370$ and $90$ pressure sensors, respectively.}}
        \label{fig.gamma_analysis}
\end{center}
\end{figure*}

\begin{figure*}
\begin{center}
        \begin{tabular}{cccc}
        \hspace{-2.1cm}
        \subfigure[]{\label{fig.gamma1312}}
        &&
        \hspace{0.8cm}
        \subfigure[]{\label{fig.gamma1879}}
        &
        \\[-.45cm]
        \hspace{-1.8cm}
	\begin{tabular}{c}
        \vspace{.5cm}
        \rotatebox{90}{$z$}
       \end{tabular}
               &
       \hspace{0.15cm}
	    \begin{tabular}{c}
    \includegraphics[width=0.35\textwidth]{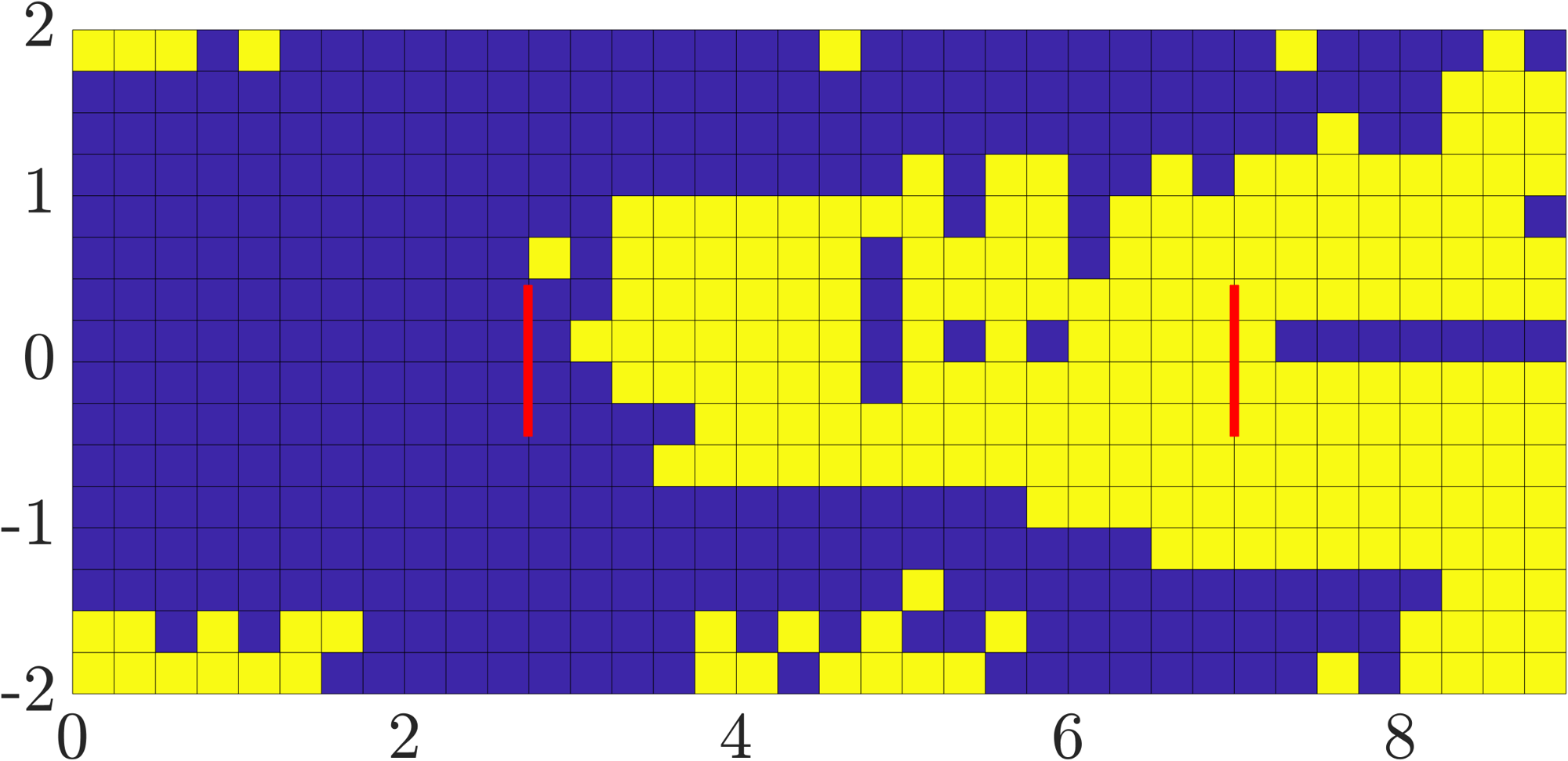}
       \\[-.1cm]
       \hspace{-.6cm}
       {$x$}
       \end{tabular}
       &&
       \hspace{.45cm}
        \begin{tabular}{c}
 \includegraphics[width=0.35\textwidth]{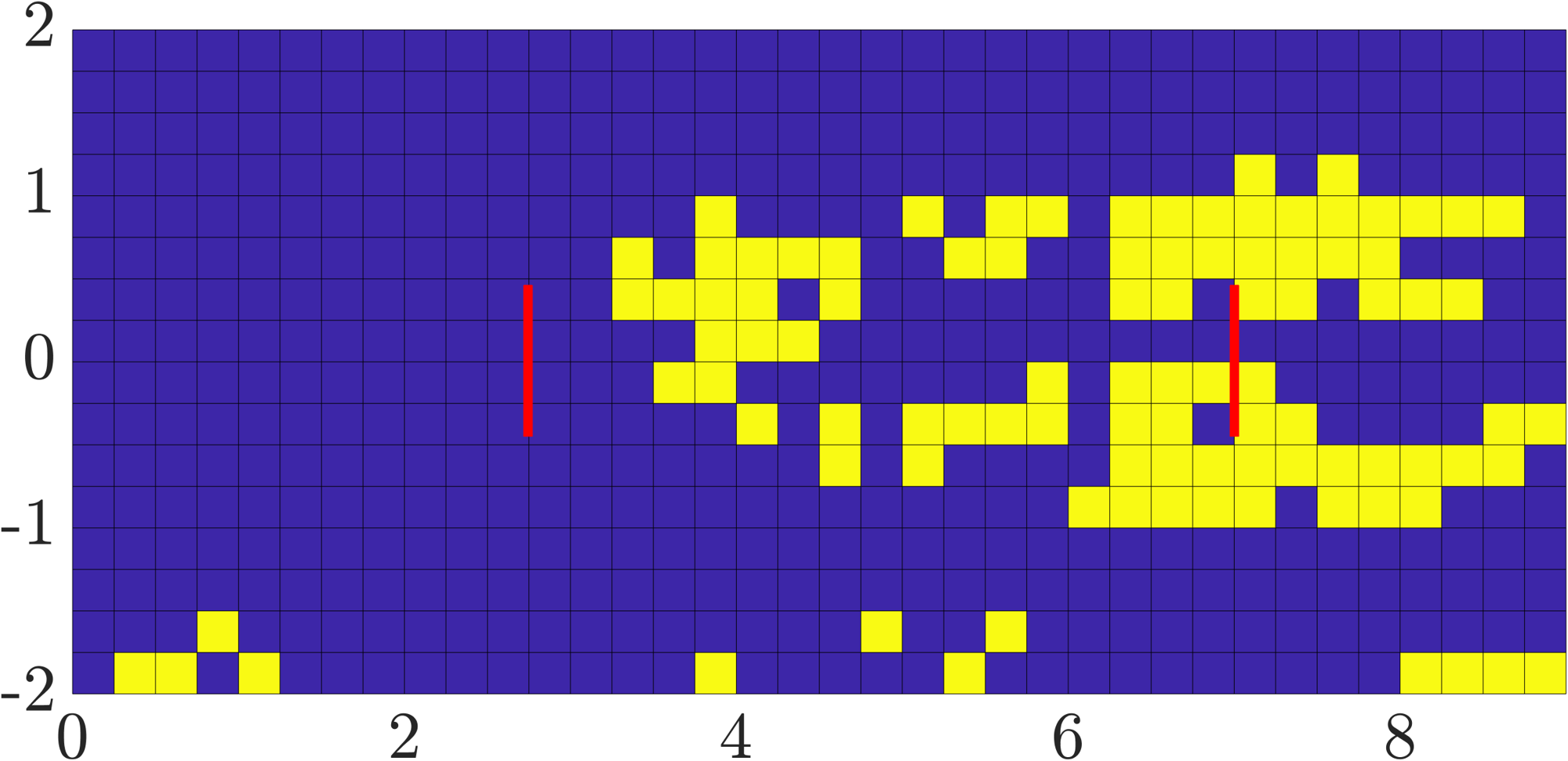}
       \\[-.1cm]
       \hspace{-.6cm}
       {$x$}
       \end{tabular}
       \end{tabular}
       \vspace{-.1cm}
        \caption{{Sparsity patterns for ground-level pressure sensors obtained from solving problem~\eqref{eq.sensel} with (a) $\beta=1312$ and (b) $\beta=1879$.
        Yellow and blue squares denote retained and dropped sensors, respectively. The red lines mark turbine rotors and the wind blows from left to right.}}
        \label{fig.SenSel}
\end{center}
\end{figure*}

\begin{figure*}
\begin{center}
        \begin{tabular}{cccc}
        \hspace{-1.3cm}
        \subfigure[]{\label{fig.EKF_2D_rel_norm_sensel_beta390}}
        &&
        \hspace{0.45cm}
        \subfigure[]{\label{fig.EKF_2D_inst_vel_sensel_beta390}}
        &
        \\[-.45cm]
        \hspace{-0.9cm}
	\begin{tabular}{c}
        \vspace{.5cm}
        \rotatebox{90}{$z$}
       \end{tabular}
       &
       \hspace{-0.1cm}
	    \begin{tabular}{c}
\includegraphics[width=0.35\textwidth]{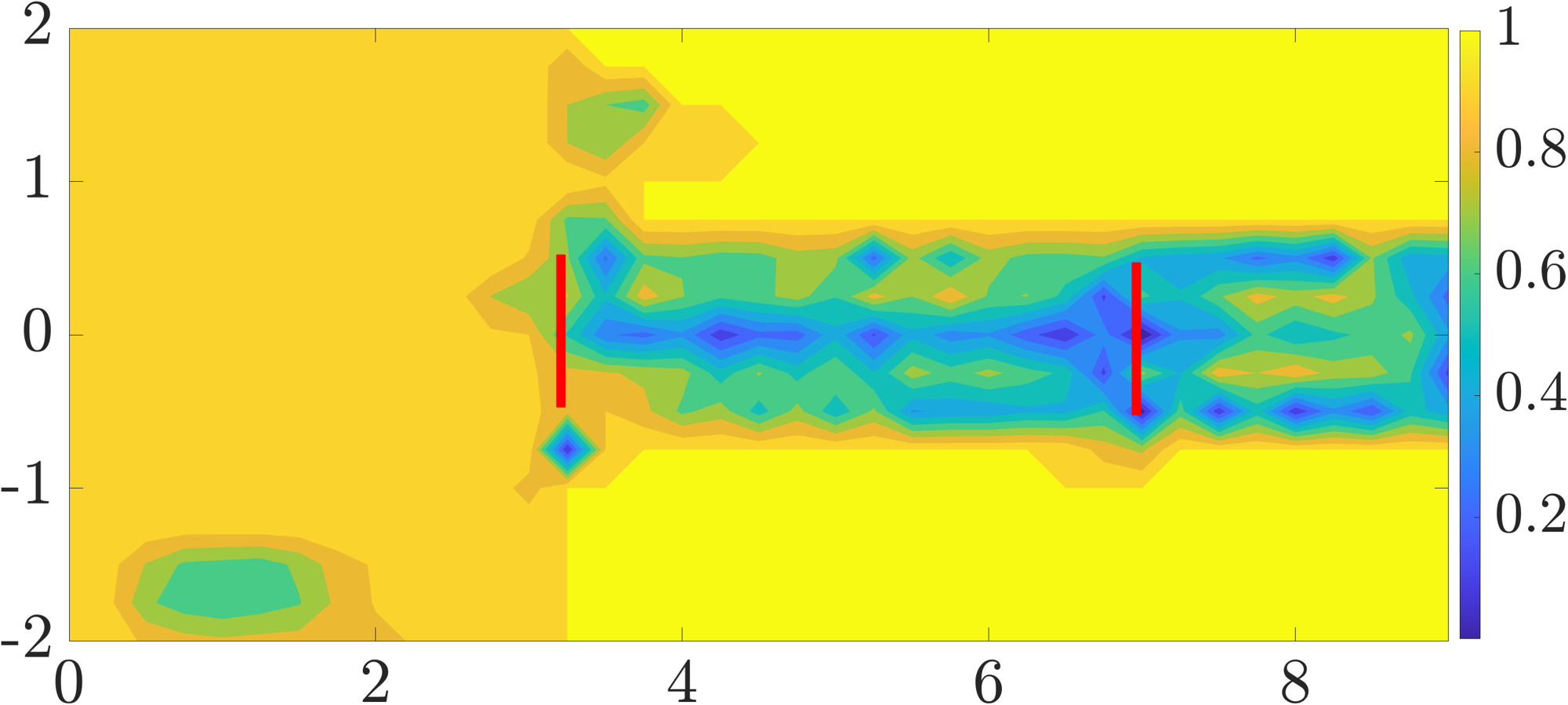}
       \end{tabular}
       &&
       \hspace{.25cm}
        \begin{tabular}{c}
       \includegraphics[width=0.35\textwidth]{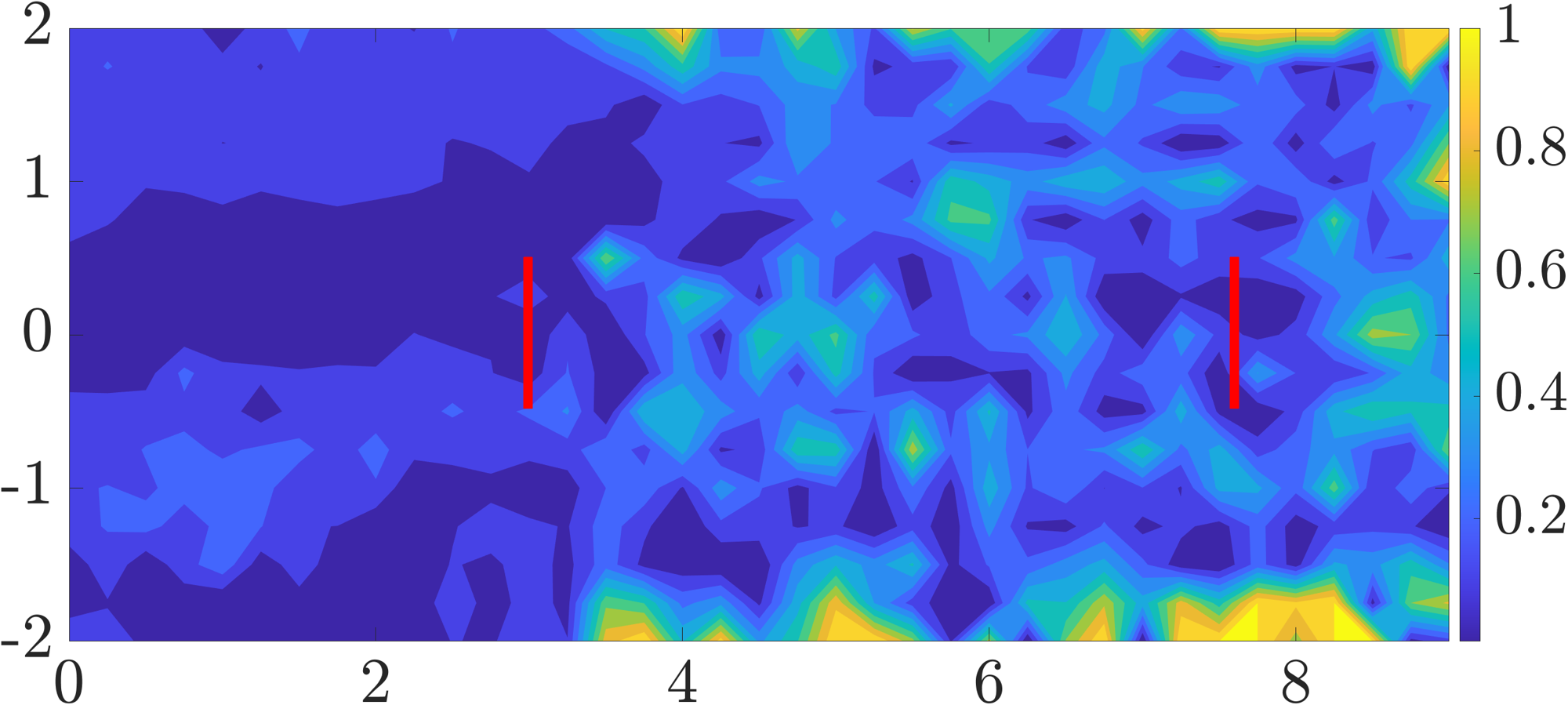}
       \end{tabular}
       \end{tabular}
       \\[-0.1cm]
       \begin{tabular}{cccc}
        \hspace{-1.3cm}
        \subfigure[]{\label{fig.EKF_2D_rel_norm_sensel_beta410}}
        &&
        \hspace{0.45cm}
        \subfigure[]{\label{fig.EKF_2D_inst_vel_sensel_beta410}}
        &
        \\[-.45cm]
        \hspace{-0.9cm}
	\begin{tabular}{c}
        \vspace{.5cm}
        \rotatebox{90}{$z$}
       \end{tabular}
       &
       \hspace{-0.1cm}
	    \begin{tabular}{c}
       \includegraphics[width=0.35\textwidth]{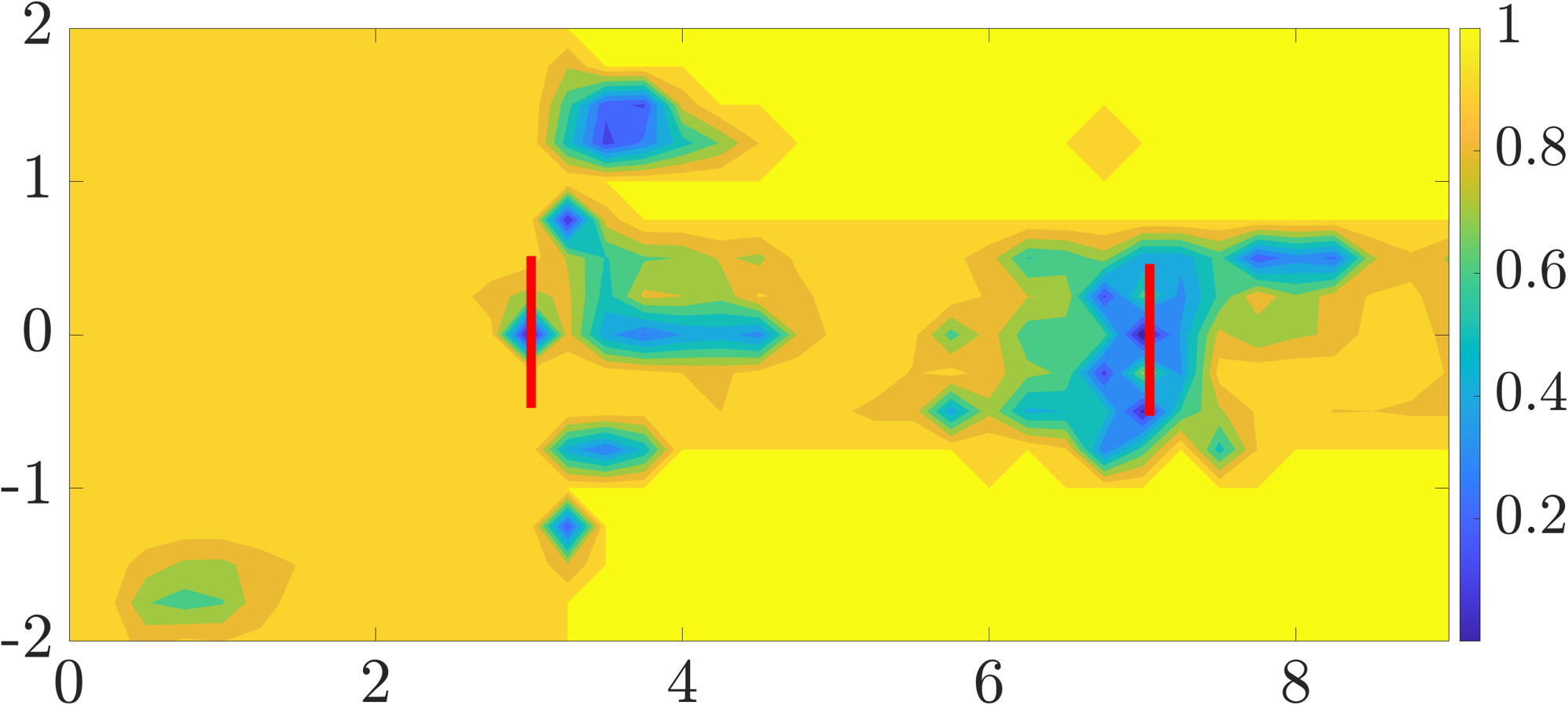}
       \\[-.1cm]
       \hspace{0.4cm}
       {$x$}
       \end{tabular}
       &&
       \hspace{0.25cm}
        \begin{tabular}{c}
       \includegraphics[width=0.35\textwidth]{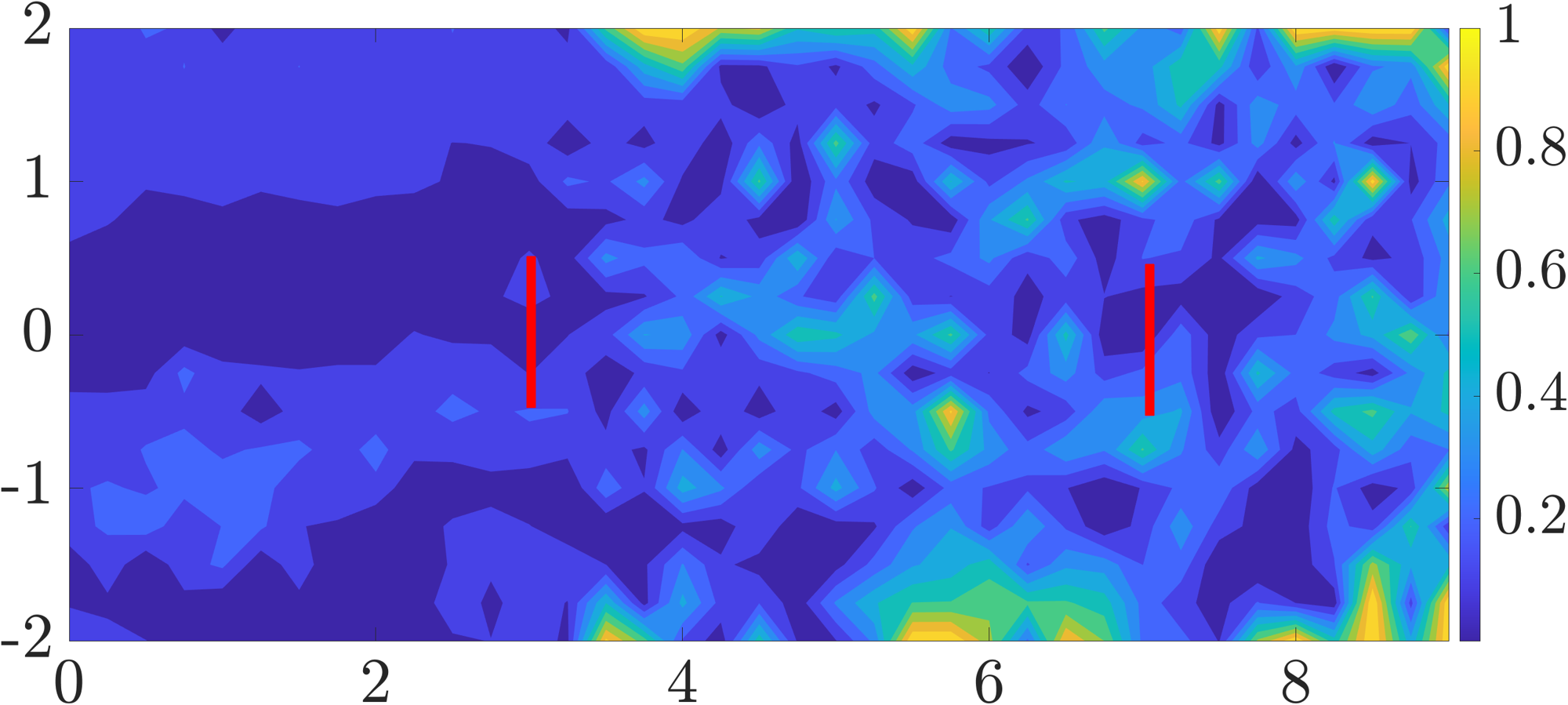}
       \\[-.1cm]
       \hspace{0.4cm}
       {$x$}
       \end{tabular}
       \end{tabular}
        \caption{{Colormaps of relative error $\mathrm{err}_\mathrm{rel}$ (a,c) and normalized $\mathrm{err}_\mathrm{norm}$ error (b,d) achieved by the EKF over the hub-height plane at $t=1536$ using the pressure sensor layouts shown in Figs.~\ref{fig.gamma1312} and~\ref{fig.gamma1879}, which were identified by problem~\eqref{eq.sensel} with $\beta=1312$ and $1879$, respectively. Turbine rotors are marked by thick red lines and the wind direction is from left to right.}}
        \label{fig.sensel_plots_err_comp}
\end{center}
\end{figure*}

Figure~\ref{fig.yaw_comparisons} compares the result of EKF built around Type A and Type B models with respect to the statistical and instantaneous error metrics $\mathrm{err}_\mathrm{rel}$ and $\mathrm{err}_\mathrm{norm}$ over the 2D plane at the hub height of the 2-turbine cascade. Two uniform yaw angles of $\gamma=15^\degree$ and $30^\degree$ have been considered. For the smaller yaw angle ($\gamma=15^\degree$), Figs.~\ref{fig.yaw_comparisons}(a,c) show that {the EKF estimations} are comparable for both types of prior models. {This is, however, not the case for the larger yaw angle ($\gamma=30^\degree$) as the EKF built around the Type B prior model achieves a slightly better estimation, especially in the wake of the leading turbine (Figs.~\ref{fig.yaw_comparisons}(b,d)); see, e.g., the regions with $\mathrm{err_{rel}}\approx80\%$ for type A model in the wake regions of both turbines (Fig.~\ref{fig.yaw_comparisons}(b)) and the absence of such high errors for the same model when the yaw angle is only $15^\degree$ (Fig.~\ref{fig.yaw_comparisons}(a)). The quality of estimation downwind of the first turbine} can be critically important in controlling the second turbine. {In spite of the robustness of the stochastic dynamical models to small variations in the turbines' yaw, they need to be updated (retrained) if the yaw misalignment becomes larger than $15^\degree$.}  We note that while the dynamic modifier $-B K_f$ in model~\eqref{eq.modified-dyn} is identified via an offline optimization routine, subsequent updates to this term in accordance with yaw-induced variations in velocity correlations can be conducted in a timely manner (even online) by initializing the optimization routines from the solution of non-yawed configurations.

\section{Sensor placement}
\label{sec.sensel}

The placement and spatial concentration of pressure sensors across a given wind farm can greatly influence the performance of Kalman filters that are designed to estimate the incoming wind. The necessity to preserve acceptable performance while minimizing costs renders the optimal placement of sensors an integral part of any flow estimation strategy. Following the sparsity-promoting framework of~[\onlinecite{linfarjovTAC13admm}] and a convex SDP reformulation~\cite{polkhlshc13,dhijovluoCDC14}, we cast the sensor placement problem as one of optimally selecting a subset of potential sensors that achieve the least steady-state estimation error, i.e.,
\begin{align} 
\non
	\minimize\limits_{L, \, X}
	& 
	~~
 \trace(V_d X\,+\, X\,L\,R\,L^T)
	\, + \, 
	\beta\, \ds{\sum_{i=1}^{n}} w_i \norm{L\,\mre_i}_2
	\\
 \non
	\subject 
	&
~~
	(A_f-L\,C)^T  X \,+\, X (A_f-L\,C) \,+\, C^TC  \;=\; 0
	\\
 \label{eq.sensel}
	&
~~
	\, X\succ  0
\end{align}
for the Kalman gain $L \in \bbR^{2n \times m}$ and the observability Gramian $X\in \bbR^{2n\times 2n}$ of the error dynamics
\begin{align*}
    \dot {\tilde{\bv}}(t) \;=\; \left(A_f - L\,C\right)\tilde{\bv}(t) \,+\, B\,\bw(t) \,-\, L\,\eta (t)
\end{align*}
where $\tilde{\bv} \DefinedAs \bv - \hat{\bv}$, $\bw$ and $\eta$ are zero-mean white processes with covariances $\Omega$ and $R$, respectively, $n$ is the number of states in Eq.~\eqref{eq.modified-dyn}, and $m$ is the number of available measurements. The composite objective function consists of the $\cH_2$ norm of the estimation error $\tilde{\bv}$, which quantifies variance amplification from process ($\bw$) and measurement noise ($\eta$) to the error, and a regularizing term that promotes column-sparsity of the Kalman gain $L$. In problem~\eqref{eq.sensel}, $V_d \DefinedAs B\,\Omega\, B^T$, $\beta>0$ specifies the importance of sparsity,
$w_i$ are nonzero weights, $\mre_i$ is the $i$th unit vector in $\bbR^m$, and $C = C_0$ from Eq.~\eqref{eq.Ck}, which corresponds to linearization around initial conditions $\bv_0$.
The desire to promote column-sparsity in matrix $L$ arises from $L$ multiplying the output matrix $C$ from the left in the first optimization constraint. As a result, when the $i$th column of $L$ is identically equal to zero, the $i$th measurement is not used. Therefore, designing a Kalman filter that uses a subset of available sensors is equivalent to designing a column-sparse $L$~\cite{polkhlshc13}.  
We utilize the proximal gradient algorithm developed in Refs.~[\onlinecite{zarjovCDC18}] and [\onlinecite{zarmohdhigeojovTAC20}] to solve convex optimization problem~\eqref{eq.sensel} and identify optimal sparsity patterns (i.e., sensing architectures) that guarantee the least estimation error. While assuming an invariant output matrix $C$ is inaccurate for the proposed Kalman filtering algorithms besides the LKF, our numerical experiments show that subsequent solutions of problem~\eqref{eq.sensel} with updated $C_k$ do not result in significant modulations to the resulting sparsity patterns. 

To analyze the effect of sensor placement throughout the two-turbine wind farm considered in the previous sections, we solve problem~\eqref{eq.sensel} with different values of the parameter $\beta$. The turbines are assumed to be facing the wind (no yaw misalignment).
When $\beta=0$, the solution to problem~\eqref{eq.sensel} can be obtained by solving the observer Riccati equation arising from the corresponding KKT conditions. This typically leads to an unstructured matrix $L$ (i.e., $L$ has no zero entries) and therefore all available measurements will be used. On the other hand, as $\beta$ increases, more columns of the solution $L$ become zero leading to sparser sensing architectures but lower filter performance. 
As evident from Fig.~\ref{fig.rank_gamma}, $\beta$ adjusts the rate of column sparsity, e.g., for $\beta < {547}$ the number of sensors is not affected. For {$547<\beta<2130$} the number of sensors monotonically decreases, and for $\beta \geq {2130}$, sparsity promotion drops all sensors. Figure~\ref{fig.performance} shows the relative performance loss as a function of the number {of} retained sensors. The performance loss has been computed relative to when all sensors are used. Clearly, a reduction in the number of sensors yields a lower performance. However, the graceful performance degradation relative to the optimal Kalman filter that uses all available sensing capabilities favors the utility of our sensor selection strategy in addressing practical engineering limitations. It is noteworthy that in the absence of any pressure sensors on the ground, the anemometer readings from the nacelles only achieve {2\%} of the total attainable performance; see Fig.~\ref{fig.performance}.

Figure~\ref{fig.SenSel} shows the distribution of ground-level pressure sensors resulting from the solution of problem~\eqref{eq.sensel} with {$\beta=1312$} and {$1879$}, which result in {$370$} and {$90$} retained sensors across the wind farm, respectively. These cases are marked by $\circ$ and $\ast$ symbols on Fig.~\ref{fig.gamma_analysis}.
As evident from this figure, increasing $\beta$ results in the selection of a smaller subset of pressure sensors within the wake region of the two turbines, which hints at the dynamical significance of the fluid mechanics within these regions. Figure~\ref{fig.sensel_plots_err_comp} shows the relative and normalized errors achieved by the EKF for the two cases where {370} and {90} pressure sensors are used in addition to the two anemometers on the turbine nacelles. {When $370$ sensors are retained, the relative error is lower in the wake region of the first turbine, where more sensors have been retained. This is because of the higher concentration of pressure sensors in the wake region.} The normalized error is also smaller in the same region. We note that this level of performance, {especially in estimating the wind before impinging the second turbine, is achieved with less than $50\%$} of the ground-level pressure sensors; see discussion in Sec.~\ref{sec.non-yawed}.

\section{Concluding remarks}
\label{sec.concluding_remark}

{We have studied} the efficacy of various Kalman filtering algorithms in estimating changes in hub-height wind velocity due to atmospheric variations that are simulated by high-fidelity LES. Our estimation algorithms rely on a stochastic dynamical model of hub-height velocity and a projection strategy for mapping pressure changes from the hub height of the wind turbines to the ground. For the first, we {used} the stochastically forced linearized NS equations around 2D velocity profiles of the average wind speed provided by static engineering wake models. To model background turbulence, we {proposed} white- and colored-in-time process noise models to excite statistical responses from our linear filters that reproduce parts of the statistical signature of atmospheric turbulence at hub height in accordance with a high-fidelity LES. The white-in-time forcing model ensures that our estimated velocity field matches the total kinetic energy of the flow and the colored-in-time forcing model is designed via inverse modeling to match the streamwise and spanwise velocity correlations. 

{We examined the} performances of conventional {Kalman filtering} algorithms in estimating {the streamwise velocity field at the hub-height of} a two-turbine cascade under non-yawed and uniformly yawed conditions. {As expected, we observed better performance in using UKF over the EnKF and the EnKF over the EKF. Nevertheless, given the computational cost of the UKF, we identified} the EKF and the EnKF as potential candidates for further development of such forecasting tools{, with the latter offering both high accuracy and computational speed if implemented on parallel computing architectures. We also demonstrated how the correlation between the sign of velocity fluctuations at the nacelle and nearby points can be used as an application-specific heuristic for improving the ability to track wind variations ahead of leading wind turbines. In addition to baseline conditions, we examined the performance of our estimation framework in the presence of uniform yaw misalignment. We attributed the reasonably good performance of our estimators to misalignments of up to $15^\degree$ to the physics-based nature of our stochastic dynamical prior models. For larger misalignment angles, we offered a solution that involved retraining the prior model of hub-height turbulence. Finally, we complemented} our proposed estimation framework with a convex optimization-based sensor selection strategy that enables us to drop pressure sensors with the least contribution to the performance of the Kalman filter. This approach identifies sensors within the near-wake region of leading turbines as being the most significant in detecting atmospheric changes using ground-pressure sensors.

\subsection{Practical application of proposed framework}

Our results serve as a proof of concept for the use of ground-level pressure measurements as an economical alternative or complement to LiDAR measurements in providing short-term estimates of the incoming wind. The development of the proposed forecasting framework would require time series data for training prior models and pressure projection kernels that are tuned to the particular terrain and atmospheric conditions in a wind farm. Data from high-fidelity LES codes that are coupled with meso-scale codes such as WRF~\cite{wrf-ver4} and are tuned to simulate real wind conditions over wind farms (see, e.g.,~[\onlinecite{sangarcirzhaiunleo20}]) would aid this effort. The second-order statistics used in training our stochastic model may alternatively be computed from LiDAR measurements. Finally, we anticipate that the stochastic models and projection kernels would require additional parameterization and periodic tuning to ensure good estimation over different atmospheric stability conditions after deployment.

\subsection{Limitations and future directions}

Our work in developing a computationally efficient estimation framework aims at supporting real-time wind turbine control. While we propose the use of stochastic dynamical models of wind farm turbulence for achieving this goal, others have proposed the use of coarse-grained 3D LES~\cite{baumey19,janmey23}, which is motivated by the lower grid-sensitivity of influential large-size coherent motions.
The ability of our lower-order estimators in capturing dominant spatio-temporal features (e.g., the power-spectral density) of the turbulent flow and their generalizability across various atmospheric conditions as part of a comparison with such higher fidelity nonlinear alternatives calls for in-depth examination in future work. In pursuit of an alternative middle ground, nonlinear terms could also be computed from the estimated velocity at each time step and fed back into the linearized dynamics~\cite{dinill24}.

Restricting the spatial domain of our models to the 2D plane at hub height comes with computational benefits, but it can also result in inaccurate predictions of the rotor equivalent wind speed, especially if wind farms are located on rough terrain. Furthermore, in the absence of {a wall-normal velocity component}, such models are agnostic to vertical interactions and momentum transport and cannot benefit from the strength of pressure correlations between the ground and planes below hub height{~\citep{linrodleozarTorque24}}. One way to address this problem is to use 3D extensions of stochastic dynamical models, that extend beyond the hub height and account for the entire vertical range affected by the rotor blades~\citep{linrodleozarTorque24}. 
Such models may be based on the linearized NS equations around 3D curled wake profiles~\citep{marannflechu19,basshashagaymen22} that capture wake deflection due to yaw misalignment in addition to ground effects. 

Another aspect that we have not addressed involves changes in atmospheric stability conditions over diurnal cycles, which could result in changes to pressure correlations between the ground and hub height and thereby affect the validity of pressure projection kernels. A detailed analysis of stratification effects under different atmospheric stability conditions calls for additional high-fidelity simulations that account for potential buoyancy forces that influence flow momentum. such results, when incorporated into the framework, may improve estimation. 

As evident from the analysis of the relative statistical error $\mathrm{err}_\mathrm{rel}$, the Kalman filters succeed in tracking the energy of velocity fluctuations, especially in regions where prior models are trained to match steady-state flow statistics. However, as discussed in Sec.~\ref{sec.KF_comparison}, tracking instantaneous atmospheric variations is much more difficult, which could be related to weak observability; see Appendix~\ref{sec.appendixC}. As such, analyzing the efficacy of our forecasting approach in predicting transient responses to gusts and sudden wind speed ramp-ups, goes beyond the scope of this paper. This direction, which could require time-varying approaches to estimation, is left to future research.

\appendix

\section{Decomposition of $Z$ into $BH^*+HB^*$}
\label{sec.appendix1}

Due to its Hermitian structure, the solution $Z$ to problem~\eqref{eq.CC} can be written as 
\begin{align*}
    Z \;=\; S\,+\,S^*,
\end{align*} 
where $S = B\,H^*$. In this appendix, we detail the steps required for obtaining matrices $B$ and $H$; see Ref.~[\onlinecite{zarchejovgeoTAC17}] for further details. Let $\pi(Z)$ and $\nu(Z)$ represent the number of positive and negative eigenvalues of matrix $Z$, respectively. There exists an invertible matrix $T$ that can bring $Z$ into a canonical form:
\begin{align*}
    \hat{Z} \;\DefinedAs\; T \, Z \, T^* \;=\; 2 \begin{bmatrix} I_\pi & 0 & 0 \\ 0 & -I_\nu & 0 \\ 0 & 0 & 0 \end{bmatrix}.  
\end{align*}
Here, $I_\pi$ and $I_\nu$ are identity matrices of size $\pi(Z)$ and $\nu(Z)$. When $\pi(Z) \leq \nu(Z)$, the choice of
\begin{align*}
    \hat{S} \;=\; \begin{bmatrix} I_\pi & -I_\pi & 0 & 0\\ I_\pi & -I_\pi & 0 & 0\\ 0 & 0 & -I_{\nu - \pi} & 0 \\ 0 & 0 & 0 & 0 \end{bmatrix} 
\end{align*}
not only satisfies $\hat{Z} = \hat{S} + \hat{S}^*$, but enables the decomposition $\hat{S}=\hat{B}\hat{H}^*$ using
\begin{align*}
    \hat{B} \;=\; \begin{bmatrix} I_\pi & 0 \\ I_\pi & 0 \\ 0 & I_{\nu - \pi} \\ 0 & 0 \end{bmatrix},
    \quad
    \hat{H}\;=\; \begin{bmatrix} I_\pi & 0 \\ -I_\pi & 0 \\ 0 & -I_{\nu - \pi} \\ 0 & 0 \end{bmatrix}.
\end{align*}
When $\pi(Z) = \nu(Z)$, $I_{\nu - \pi}$ and the corresponding rows and columns in matrices $\hat{B}$ and $\hat{H}$ are empty. 
On the other hand, when $\pi(Z)>\nu(Z)$, $\hat{S}$ can take the same form, but with
\begin{align*}
    \hat{B} \;=\; \begin{bmatrix} I_{\pi-\nu} & 0 \\ 0 & I_\nu \\ 0 & I_\nu \\ 0 & 0 \end{bmatrix},
    \quad
    \hat{H} \;=\; \begin{bmatrix} I_{\pi-\nu} & 0 \\ 0 & I_\nu \\ 0 & -I_\nu \\ 0 & 0 \end{bmatrix}.
\end{align*}
Finally, matrices $B$ and $H$ can be determined as $B=T^{-1}\hat{B}$ and $H=T^{-1}\hat{H}$.

\section{Large-eddy simulations}
\label{sec.appendix2}

A cascade of two NREL-$5$ MW wind turbines~\cite{jonbutmussco09} is simulated using the LES code UTD-WF~\cite{sancirrotleo15,sancarareleo17,Ciri17,ciri2018}. The LES code utilizes the immersed boundaries to represent turbine towers and nacelles and the rotating actuator disk model to mimic the blades. The computational domain is $32d_0\times 10.24 d_0 \times 10d_0$ in the streamwise, spanwise, and vertical directions, respectively. The distance from the inlet to the upstream turbine is $9d_0$. 
No-slip boundary conditions are enforced at the bottom of the computational domain, as well as on the towers and nacelles, a free-slip boundary condition is applied to the top of the computational domain, and a periodic boundary condition is applied to both sides of the domain in the spanwise direction. The grid achieves its highest resolution at the hub height where it is uniformly spaced in all three directions with $\Delta x = \Delta y = \Delta z = 0.025d_0$. Away from the turbines, the grid is stretched in the vertical direction.

The LES was conducted under neutral atmospheric stability conditions with atmospheric boundary layer conditions mimicked by superimposing the turbulent velocity field from a precursor simulation to a mean velocity profile expressed by a power law. The free-stream velocity $U_\infty$ is chosen to be about $80\%$ of a rated wind speed 
of $11.4\,m/s$~\cite{jonbutmussco09}. The superposition of the mean flow and the turbulence from the precursor simulation ultimately results in a hub-height turbulence intensity of $8\%$ impinging the first turbine in the cascade.  

The LES results were interpolated on the coarser grid associated with the stochastic dynamical model before time-averaged and root mean square profiles of the velocity fluctuations, which were used in training our prior model, were computed. Statistical quantities were computed using $750$ instantaneous snapshots of the 3D velocity field.
These {snapshots} were taken over a period of about $27$ minutes of {real-time} operation ($70$ non-dimensional time units based on a reference length and speed of $126 \,m$ and $11.4\,m/s$) with a temporal resolution of about 2 seconds ($0.2$ non-dimensional time units).

\section{Observability analysis}
\label{sec.appendixC}

To analyze the infinite-horizon observability of the system, we consider 3 cases: (i) hub-height wind estimation using hub-height velocity measurements; (ii) wind estimation using pressure measurements from the hub height in addition to velocity measurements from nacelle-mounted anemometers; and (iii) wind estimation using pressure measurements from the ground in addition to velocity measurements from nacelle-mounted anemometers. The second case evaluates the innovation due to measurements of pressure in lieu of velocity and the third case evaluates the innovation due to measurements of pressure from the ground, which require projection from the hub height to the ground.
The observability of the pair $(C,A_f)$ in each case can provide insights into the expected quality of estimation. While the state matrix $A_f$ 
corresponds to the modified dynamics~\eqref{eq.modified-dyn}, the output matrix $C=C_0$ from Eq.~\eqref{eq.Ck}, which corresponds to linearization around the initial condition.

\begin{figure*}
\begin{center}
        \begin{tabular}{cccccc}
        \hspace{-.5cm}
        \subfigure[]{\label{fig.sval_val}}
        &&
        \hspace{-0.7cm}
        \subfigure[]{\label{fig.sval_vel_u}}
        &&
        \subfigure[]{\label{fig.sval_vel_w}}
        &
        \\[-.45cm]
        \hspace{-0.2cm}
	\begin{tabular}{c}
        \vspace{.5cm}
        \rotatebox{90}{$\lambda_i$}
       \end{tabular}
       &
       \hspace{-0.1cm}
	    \begin{tabular}{c}
\includegraphics[width=0.2\textwidth]{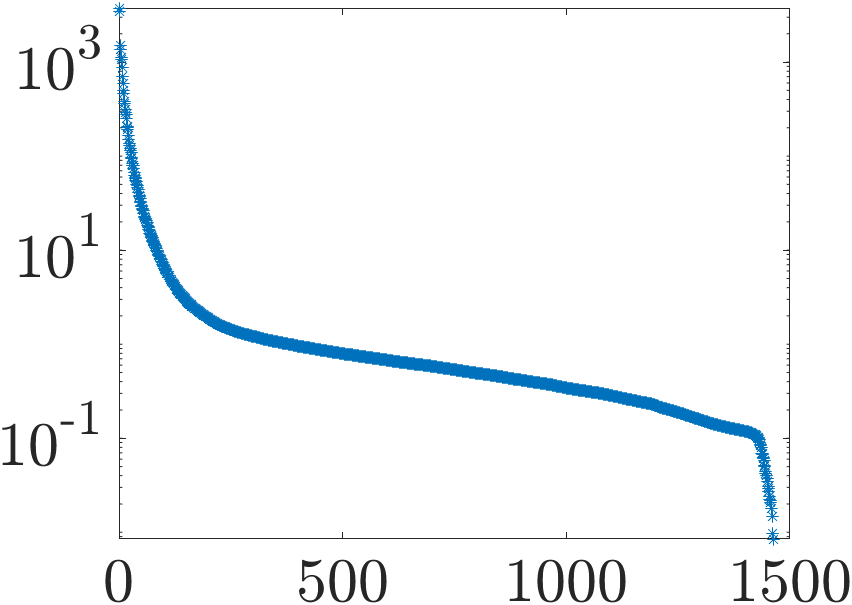}
       \end{tabular}
       &
       \hspace{-0.2cm}
	\begin{tabular}{c}
        \vspace{.5cm}
        \rotatebox{90}{$z$}
       \end{tabular}
       &
       \hspace{-0.2cm}
        \begin{tabular}{c}
       \includegraphics[width=0.35\textwidth]{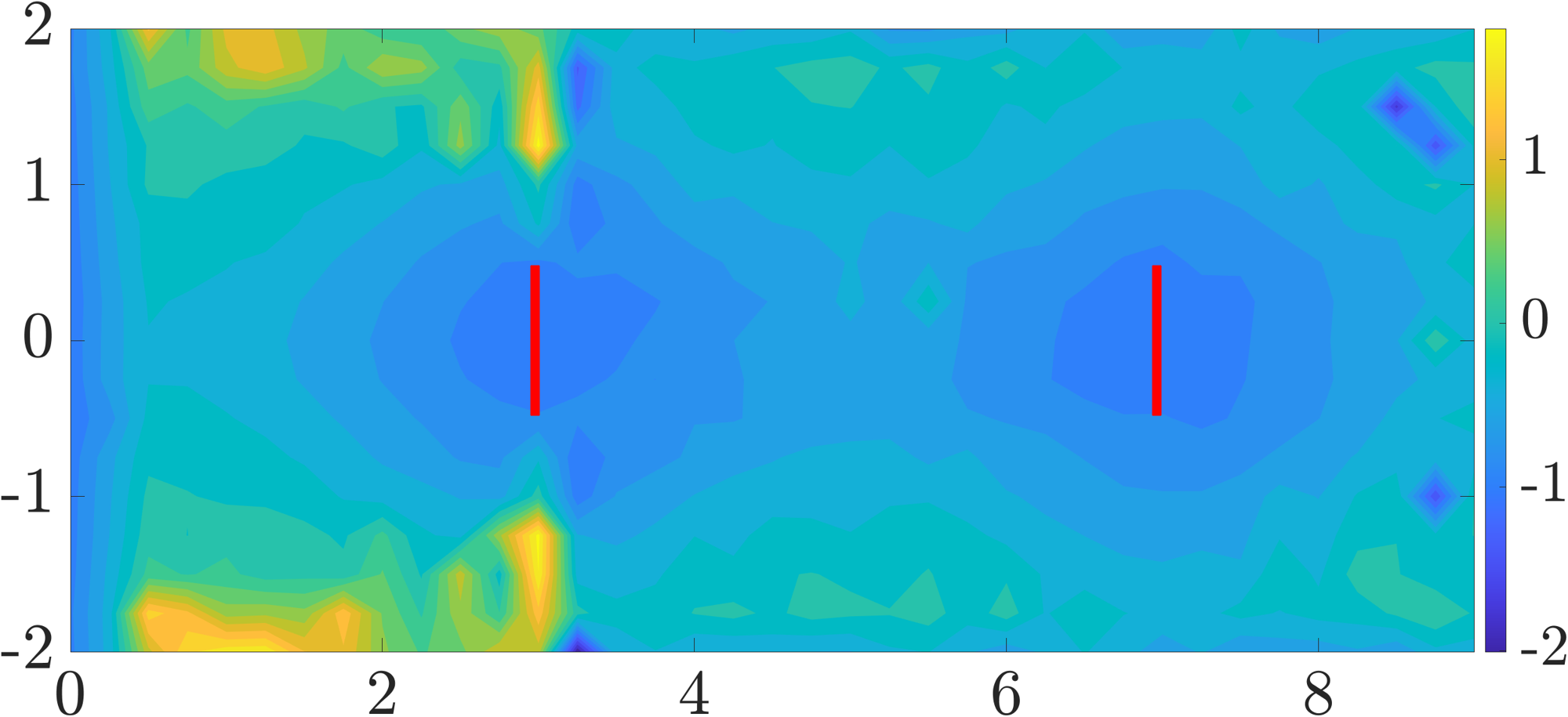}
       \end{tabular}
       &&
       \hspace{.05cm}
        \begin{tabular}{c}
       \includegraphics[width=0.35\textwidth]{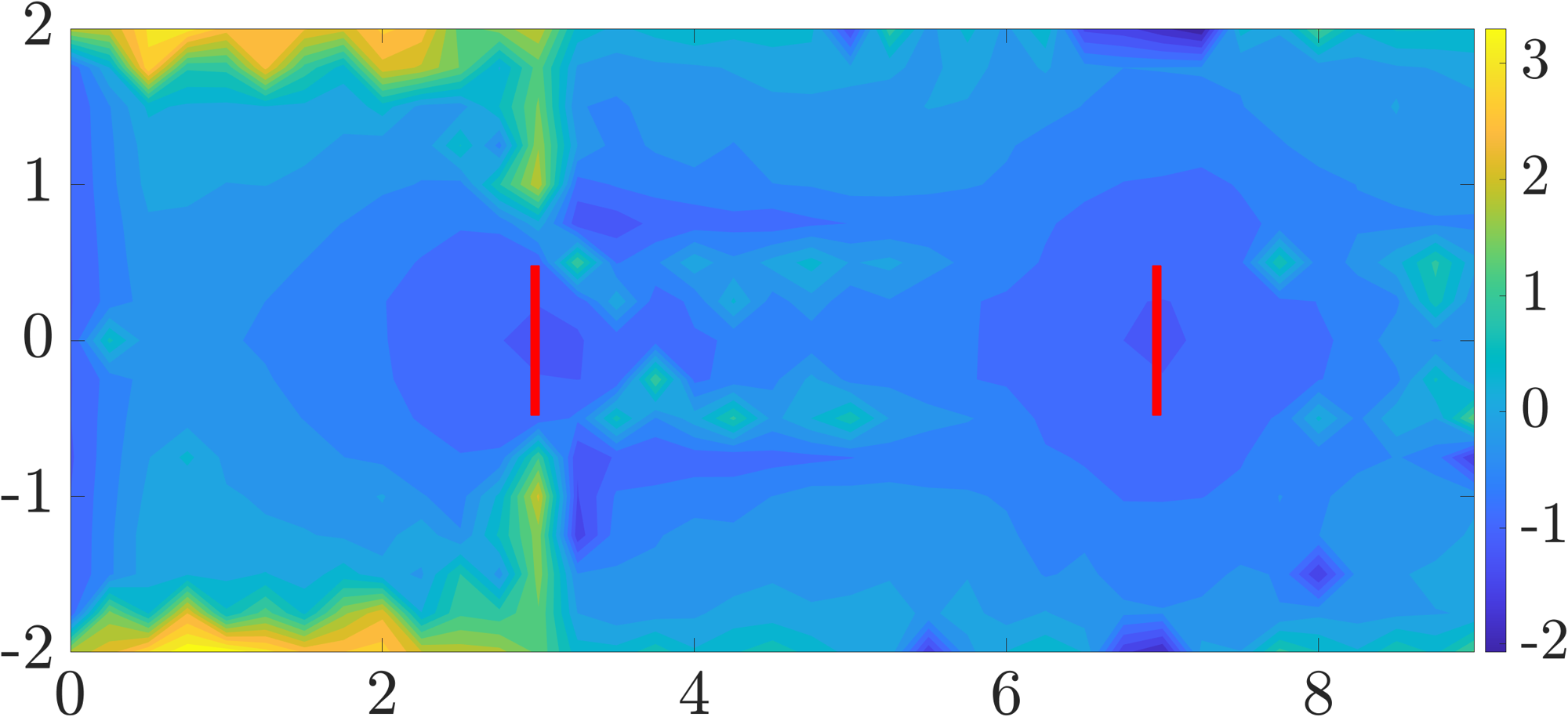}
       \end{tabular}
       \end{tabular}
       \\[-0.1cm]
       \begin{tabular}{cccccc}
        \hspace{-.5cm}
        \subfigure[]{\label{fig.sval_HHpress}}
        &&
        \hspace{-0.7cm}
        \subfigure[]{\label{fig.sval_HHpress_u}}
        &&
        \subfigure[]{\label{fig.sval_HHpress_w}}
        &
        \\[-.45cm]
        \hspace{-0.2cm}
	\begin{tabular}{c}
        \vspace{.5cm}
        \rotatebox{90}{$\lambda_i$}
       \end{tabular}
       &
       \hspace{-0.1cm}
	    \begin{tabular}{c}
\includegraphics[width=0.2\textwidth]{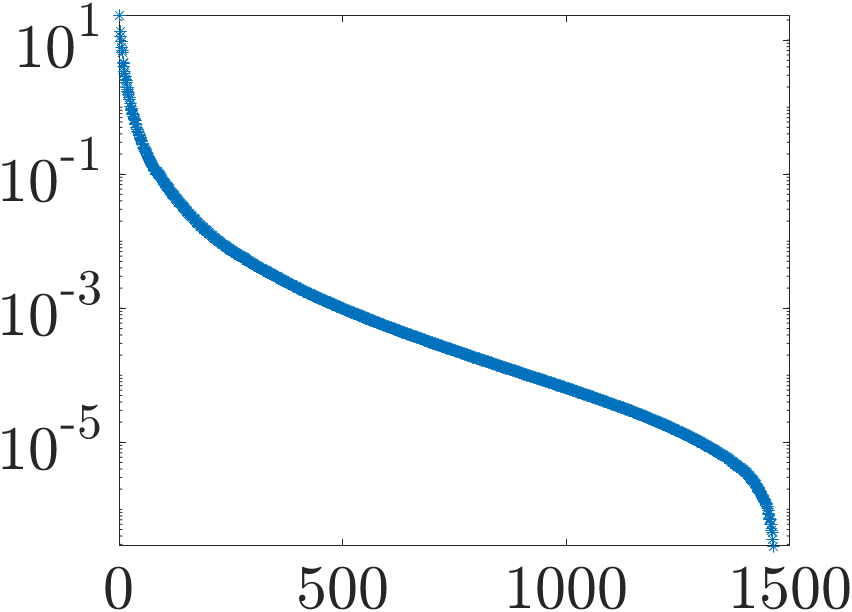}
       \end{tabular}
       &
       \hspace{-0.2cm}
	\begin{tabular}{c}
        \vspace{.5cm}
        \rotatebox{90}{$z$}
       \end{tabular}
       &
       \hspace{-0.2cm}
        \begin{tabular}{c}
       \includegraphics[width=0.35\textwidth]{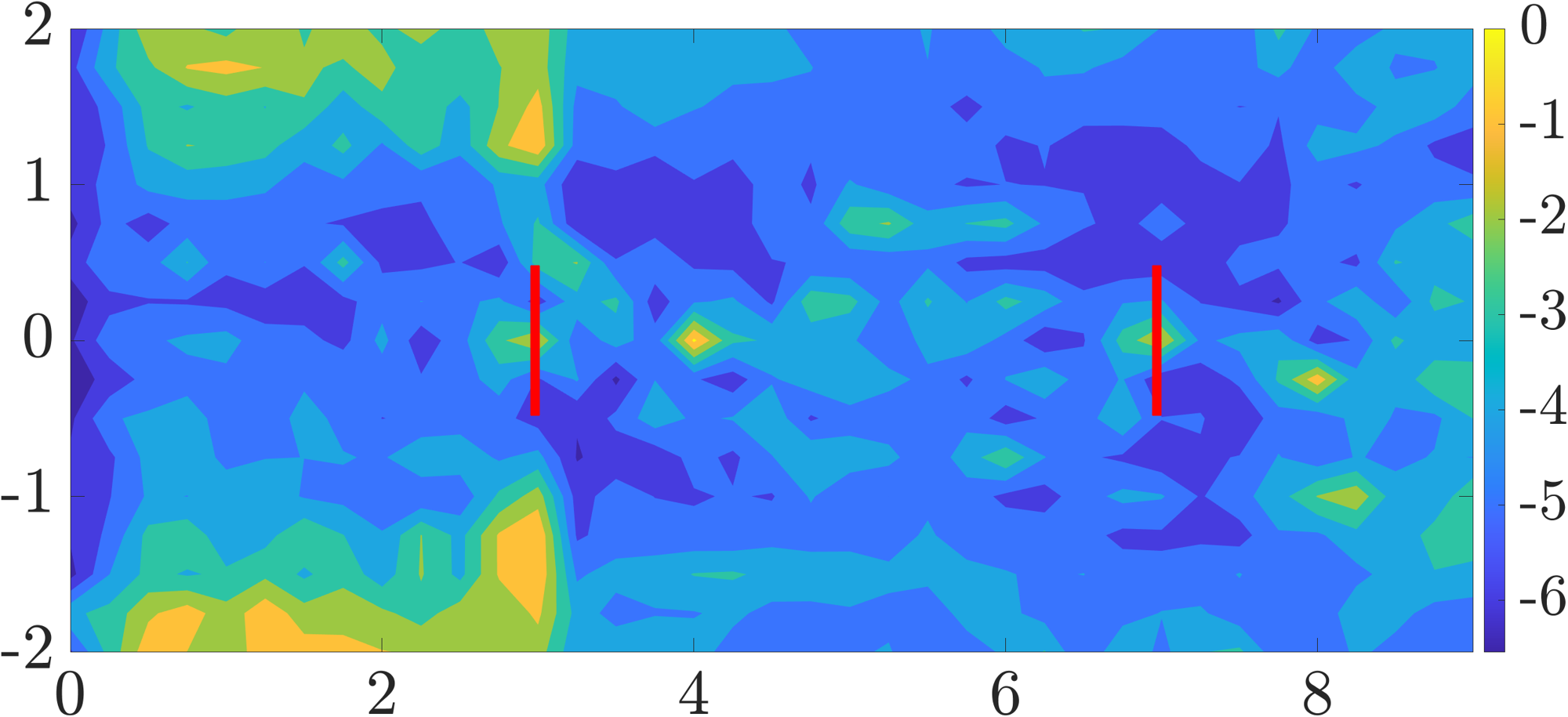}
       \end{tabular}
       &&
       \hspace{.05cm}
        \begin{tabular}{c}
       \includegraphics[width=0.35\textwidth]{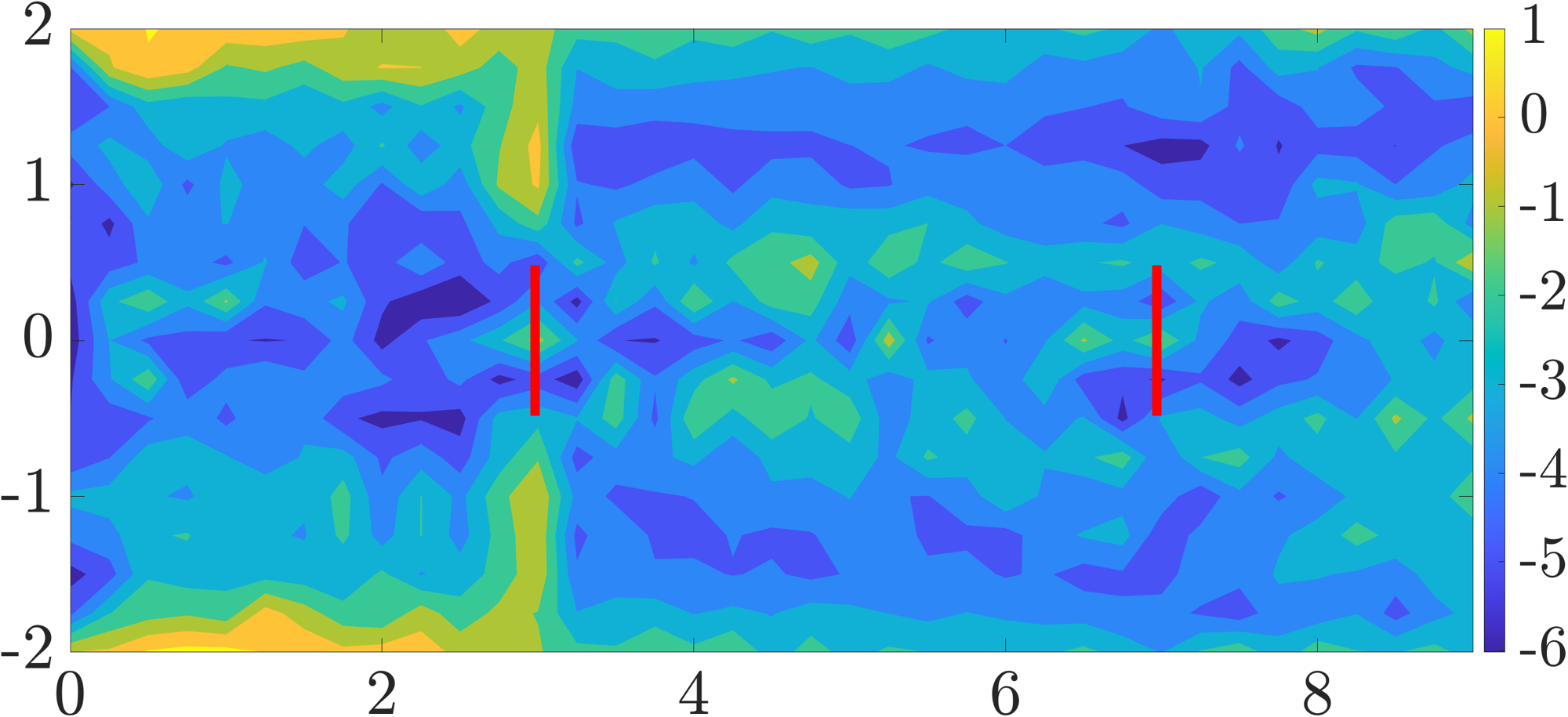}
       \end{tabular}
       \end{tabular}
       \\[-0.1cm]
       \begin{tabular}{cccccc}
        \hspace{-.5cm}
        \subfigure[]{\label{fig.wwLESmap_tmp}}
        &&
        \hspace{-0.7cm}
        \subfigure[]{\label{fig.wwLNSmap_tmp}}
        &&
        \subfigure[]{\label{fig.wwLNSmap_tmp2}}
        &
        \\[-.45cm]
        \hspace{-0.2cm}
	\begin{tabular}{c}
        \vspace{.5cm}
        \rotatebox{90}{$\lambda_i$}
       \end{tabular}
       &
       \hspace{-0.1cm}
	    \begin{tabular}{c}
       \includegraphics[width=0.2\textwidth]{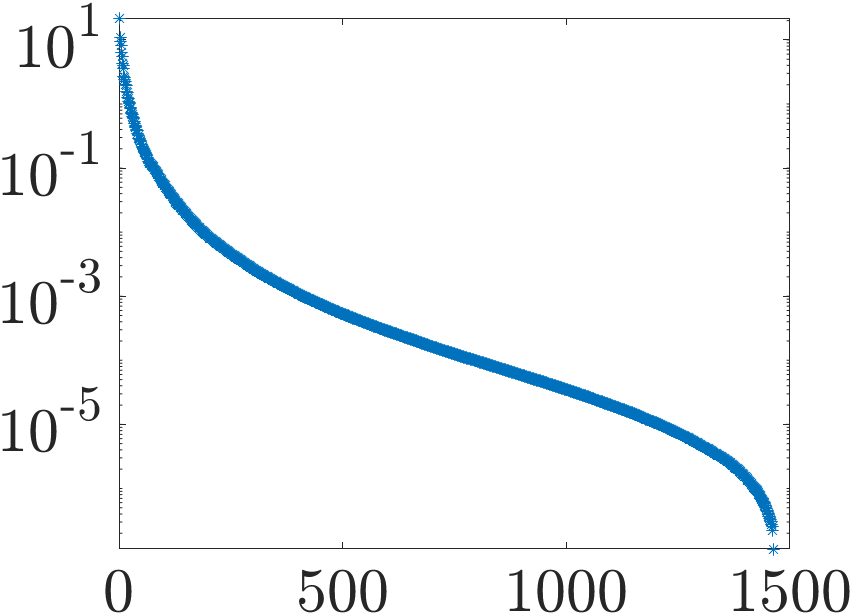}
       \\[.1cm]
       \hspace{0.4cm}
       {$i$}
       \end{tabular}
       &
       \hspace{-0.2cm}
	\begin{tabular}{c}
        \vspace{.5cm}
        \rotatebox{90}{$z$}
       \end{tabular}
       &
       \hspace{-0.2cm}
        \begin{tabular}{c}
       \includegraphics[width=0.35\textwidth]{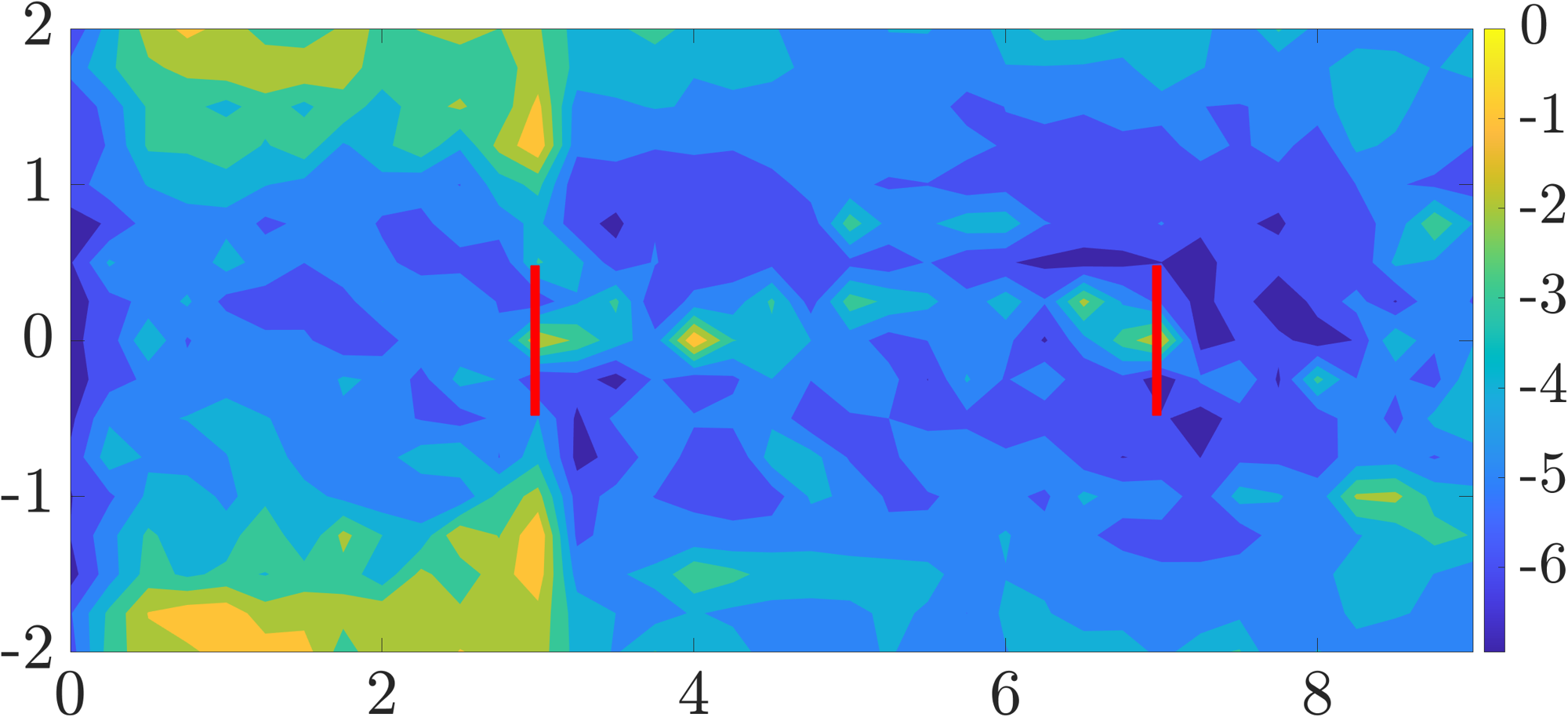}
       \\[-.1cm]
       \hspace{0.4cm}
       {$x$}
       \end{tabular}
       &&
       \hspace{0.05cm}
        \begin{tabular}{c}
       \includegraphics[width=0.35\textwidth]{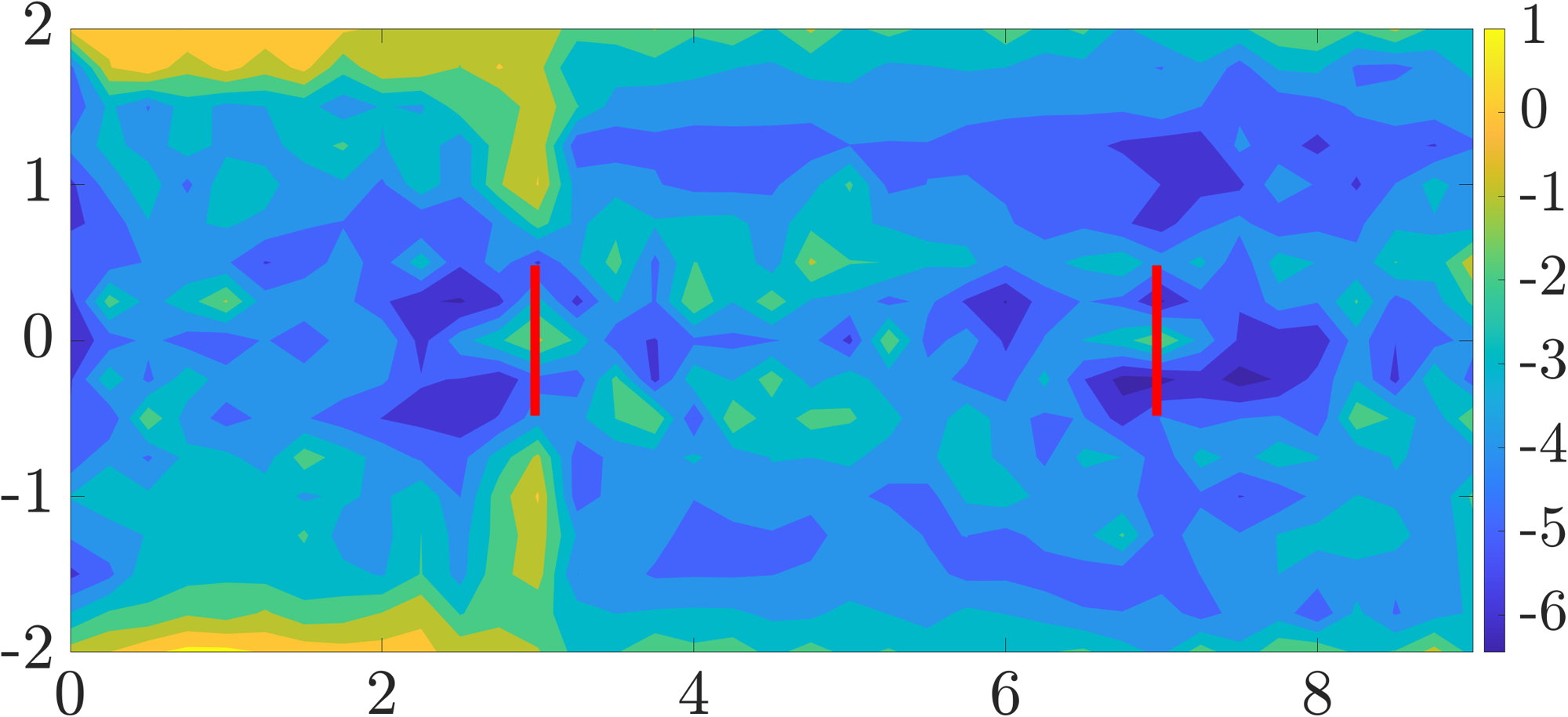}
       \\[-.1cm]
       \hspace{0.4cm}
       {$x$}
       \end{tabular}
       \end{tabular}
        \caption{Eigenvalues of the observability Gramian $W_o$ (a,d,g) and the corresponding colorplots for the level of observability in the streamwise (b,e,h) and spanwise (c,f,i) velocity components over the 2D plane at hub height. First row: case (i), where hub-height velocity is taken as the measurement; Second row: case (ii), where hub-height pressure and nacelle velocities are taken as measurements; and third row: case (iii), where ground pressure and nacelle velocities are taken as measurements.}
        \label{fig.observability_colormap}
\end{center}
\end{figure*}

As $A_f$ is Hurwitz, the observability Gramian $W_o$ solves the Lyapunov equation
\begin{align*}
    A_f^T\, W_o \,+\, W_o\, A_f \;=\; -C^T C.
\end{align*}
The first column of Fig.~\ref{fig.observability_colormap} shows the eigenvalues of $W_o$ for the three cases discussed above. While the Gramians are not rank deficient, the large condition number in the second (Fig.~\ref{fig.sval_HHpress}) and third cases (Fig.~\ref{fig.wwLESmap_tmp}), which consider pressure sensing, is an indicator of weak observability~\cite{moo81}, that can explain the inability to track; see Table~\ref{table:Grammian}. In the absence of a clear-cut in the eigenvalues that would allow us to identify the unobservable subset from the null space of $W_o$, we evaluate the level of observability across the 2D plane at hub height. Following a sequence of elementary column operations on the matrix of eigenvectors, we associate each eigenvalue to the spatial location corresponding to the pivot entry in its eigenvector{; depending on whether the pivot entry is found in the first or second half of the eigenvector, we associate it with either the streamwise or spanwise velocity components corresponding to a particular spatial location dictated by the discretized state-space. An observability map can thereby be sequentially populated by the eigenvalues of $W_o$ and the identified spatial locations.}


The colorplots shown in the second and third columns of Fig.~\ref{fig.observability_colormap} reflect {the ordering of the eigenvalues of the observability matrices} over the 2D plane for the streamwise and spanwise velocity components, respectively. These colormaps demonstrate the highest levels of observability in the upwind regions to the sides of the computational domain. While the dynamically significant wake regions have a lower degree of observability, the eigenvalues associated with these regions are not the weakest, which is in agreement with our Kalman filtering results (cf.~e.g., Fig.~\ref{fig.comparisons}(b,d)). In particular, as shown in Figs.~\ref{fig.observability_colormap}(e,h), the center-point one diameter behind the leading turbine, which was highlighted in Sec.~\ref{sec.non-yawed} for its potential to provide approximately 54 seconds of preview wind information to the second turbine, exhibits a high degree of observability. It is also evident that the velocity measurements at the nacelles of the two turbines significantly improve the observability at those locations; see small regions of higher observability in Figs.~\ref{fig.observability_colormap}(e,f,h,i) around $(x,z)=(3,0)$ and $(x,z)=(7,0)$. Finally, Figs.~\ref{fig.observability_colormap}(h,i) indicate a reduction in observability behind the second turbine, which can be attributed to low levels of coherence between the pressure fields on ground and at the hub-height within those regions (cf.~Fig.~\ref{fig.LCM}). In agreement with the results of this observability analysis, Table~\ref{table:Grammian} shows the normalized error $\mathrm{err}_\mathrm{norm}$ (Eq.~\eqref{eq.NormalizedError}) (integrated over the horizontal domain and the simulation time) to also increase over the 3 cases studied in this appendix.

\begin{table}[ht]
\caption{Space-time-averaged normalized error $\mathrm{err}_\mathrm{norm}$ for different cases considered in observability analysis.} 
\centering 
\begin{tabular}{c| c c c} 
\hline\hline 
\\[-1.8ex]
Case~ & ~$\lambda_{max}$ & ~~$\mathrm{cond}(W_o)$ & ~~$\ds{\int_{\bx,t}} \mathrm{err}_\mathrm{norm}$ \\ [0.5ex] 
\hline 
\\[-0.30cm]
(i)~ &~ $3.76\times 10^3$ &~~ $4.43\times 10^5$ &~~ $5.54\times 10^{-2}$ 
\\ 
(ii)~ &~ $3.08\times 10^1$ &~~ $8.35\times 10^7$ &~~ $1.82 \times 10^{-1}$ \\
(iii)~ &~ $3.10\times 10^1$ &~~ $1.87\times 10^8$ &~~ $2.42\times 10^{-1}$ \\ [.6ex] 
\hline 
\end{tabular}
\label{table:Grammian} 
\end{table}

\vsp
\begin{acknowledgments}
This work was supported in part by the National Science Foundation under awards 1916715 and 1916776 (I/UCRC for Wind Energy, Science, Technology, and Research) and from the members of WindSTAR I/UCRC. Any opinions, findings, and conclusions or recommendations expressed in this material are those of the author(s) and do not necessarily reflect the views of the National Science Foundation or the sponsors. The Office of Information Technology Cyberinfrastructure Research Computing (CIRC) at The University of Texas at Dallas and the Texas Advanced Computing Center are acknowledged for providing computing resources.
\end{acknowledgments}

\section*{Data Availability Statement}
The data that support the findings of this study are available from the corresponding author upon reasonable request.

\vspace{3ex}


%

\end{document}